\UseRawInputEncoding
\documentclass[fleqn,usenatbib]{mnras}

\usepackage{newtxtext,newtxmath}

\usepackage[T1]{fontenc}
\usepackage{ae,aecompl}

\usepackage{graphicx}	
\usepackage{amsmath}	
\usepackage{float}

\usepackage[T1]{fontenc}
\usepackage{ae,aecompl}

\usepackage{times}
\usepackage{amsmath}
\usepackage{color}
\usepackage{bm}
\usepackage{graphicx}
\usepackage{tabu}
\usepackage{epsfig}
\usepackage{longtable,tabu}
\usepackage{url}
\usepackage{xcolor}
\usepackage{ulem}
\usepackage{supertabular}
\usepackage{caption}
\usepackage[multidot]{grffile}
\usepackage{subcaption}
\def \CHp{\ifmmode{\rm CH^+}\else{$\rm CH^+$}\fi}
\def \Lya{\ifmmode{\rm Ly \alpha}\else{$\rm Ly \alpha$}\fi}
\def \Cp{\ifmmode{\rm C^+}\else{$\rm C^+$}\fi}
\def \twC{\ifmmode{\rm ^{12}C}\else{$\rm ^{12}C$}\fi}
\def \thC{\ifmmode{\rm ^{13}C}\else{$\rm ^{13}C$}\fi}
\def \thCHp{\ifmmode{\rm ^{13}CH^+}\else{$\rm ^{13}CH^+$}\fi}
\def \twCHp{\ifmmode{\rm ^{12}CH^+}\else{$\rm ^{12}CH^+$}\fi}
\def \HH{\ifmmode{\rm H_2}\else{$\rm H_2$}\fi}
\def \twCO{\ifmmode{\rm ^{12}CO}\else{$\rm ^{12}CO$}\fi}
\def \HCOp{\ifmmode{\rm HCO^+}\else{$\rm HCO^+$}\fi}
\def \amm{\ifmmode{\rm NH_3}\else{$\rm NH_3$}\fi}
\def \wat{\ifmmode{\rm H_2O}\else{$\rm H_2O$}\fi}
\newcommand{\msol}{\hbox{\,${\rm M}_\odot$}}

\newcommand{\lsol}{\hbox{\,${\rm L}_\odot$}}
\def\kms{\hbox{\,km\,s$^{-1}$}}

\def\cc{\ifmmode{\,{\rm cm}^{-3}}\else{$\,{\rm cm}^{-3}$}\fi}
\def\cq{\ifmmode{\,{\rm cm}^{-2}}\else{$\,{\rm cm}^{-2}$}\fi}
\def\Lya{\ifmmode{\,{\rm Ly}\alpha}\else{${\rm Ly}\alpha$}\fi}
\def\escqa{\ifmmode{\,{\rm erg \,cm}^{-2} {\rm s}^{-1} {\rm arcsec}^{-2}}\else{${\rm erg\,cm}^{-2} {\rm s}^{-1} {\rm arcsec}^{-2}$}\fi}
\def\ergs{\ifmmode{\,{\rm erg \, s}^{-1}}\else{${\rm erg\, s}^{-1}$}\fi}
\def\ergccs{\ifmmode{\,{\rm erg \, cm^{-3}\,s^{-1}}}\else{${\rm erg\,cm^{-3}\, s^{-1}}$}\fi}

\title[Where infall meets outflows: the CH$^+$ and Ly$\alpha$ perspectives]{Where infall meets outflows: turbulent dissipation probed by CH$^+$ and Ly$\alpha$ in the starburst/AGN galaxy group SMM J02399$-$0136 at z$\sim$2.8}
\author[A. Vidal-García et al.]{A. Vidal-García$^1$,\thanks{E-mail: alba.vidal@ens.fr}
E. Falgarone$^{1}$, 
F. Arrigoni Battaia$^2$,
B. Godard$^{1,3}$,  
R. J. Ivison$^{4}$,  
\newauthor 
M. A. Zwaan$^{4}$,  
C. Herrera$^5$,
D. Frayer$^6$,   
P. Andreani$^4$,
Q. Li$^7$,
R. Gavazzi$^8$
\\
\\
$^{1}$ LPENS, Ecole Normale Supérieure, Université PSL, CNRS, Sorbonne Université, Université Paris-Diderot, Paris, France\\
$^2$  Max-Planck-Institute f\"ur Astrophysik, Karl-Schwarzschild-Strasse 1, D-85748 Garching bei M\"unchen, Germany\\
$^{3}$ Observatoire de Paris, Ecole Normale Supérieure, Universit\'e PSL, Sorbonne Université, CNRS LERMA, Paris, France\\
$^{4}$ European Southern Observatory, Karl-Schwarzschild-Strasse 2, D-85748 Garching bei M\"unchen, Germany\\
$^5$ Institut de radioastronomie millimétrique, 106 rue de la Piscine, Saint Martin d'Hères, France\\
$^6$ Green Bank Observatory, PO Box 2, Green Bank, WV, 24944, USA\\
$^7$ Kavli Institute for Astronomy and Astrophysics, Peking University, Beijing, People's Republic of China\\
$^8$ Sorbonne Universit\'e, CNRS, UMR7095, Institut d'Astrophysique de Paris, F-75014, Paris, France\\
}

\date{Accepted 2021 May 15. Received 2021 May 3; in original form 2020 February 10}

\pubyear{2018}

\begin{document}
\label{firstpage}
\pagerange{\pageref{firstpage}--\pageref{lastpage}}
\maketitle

\begin{abstract}
We present a comparative analysis of the \CHp(1-0) and \Lya\ lines, observed with the Atacama Large Millimeter Array (ALMA) and Keck telescope respectively, in the field of the submillimetre-selected galaxy (SMG) SMM\,J02399$-$0136 at $z\sim2.8$, which comprises a heavily obscured starburst galaxy and a broad absorption line quasar, immersed in a large \Lya\  nebula.
This comparison highlights the critical role played by turbulence in channeling the 
energy across gas phases and scales, splitting the energy trail between hot/thermal and cool/turbulent phases in the circum-galactic medium (CGM).
The unique chemical and spectroscopic properties of \CHp\ are used to infer the existence of a massive ($\sim 3.5 \times 10^{10}$ \msol), highly turbulent reservoir of diffuse molecular gas of radius  $\sim 20$\,kpc 
coinciding with the core of the \Lya\ nebula.  The whole cool and cold  CGM is shown to be inflowing towards the galaxies at a velocity $\sim$ 400 \kms. Several kpc-scale shocks are  detected  tentatively in \CHp\ emission. Their specific location in space and velocity with respect to the high-velocity \Lya\ emission suggests that they lie at the interface of the inflowing CGM and the high-velocity \Lya\ emission, and signpost the feeding of CGM turbulence by AGN- and stellar-driven outflows.  The mass and energy budgets of the CGM require net mass accretion at a rate commensurate with the star formation rate (SFR). 
From this similarity, we infer that the merger-driven burst of star formation and black-hole growth are ultimately fuelled by large-scale gas accretion.
\end{abstract}
\begin{keywords}
galaxies: high-redshift -- galaxies: starburst -- galaxies: intergalactic medium -- galaxies: formation -- turbulence -- molecular processes
\end{keywords}



\section{Introduction}

\begin{table*}
  \centering
  \caption{Known components of SMM\,J02399$-$0136.}
  \begin{tabular}{lccll} \\ 
  \hline \noalign {\smallskip}
Name & $\alpha_{\rm J2000}$ (h\,m\,s)& $\delta_{\rm J2000}$ ($^\circ\,'\,''$)& Description & References\\ 
\hline \noalign {\smallskip}
ELAN  &&&Enormous Ly\,$\alpha$ nebula around the entire system & $1,2$ \\
L1   &02:39:51.815 $^a$&$-$01:35:58.39&Dusty BAL QSO & $1,3,5$\\
L1N  &02:39:51.838$^c$ &$-$01:35:57.19&Marginally extended and somewhat bluer northern companion to the QSO & $1,4$ \\
W1$^d$  &  & &Feature extending westwards for several arcsec from BAL QSO, seen in CO(3--2) &  $5$\\
L2   &02:39:52.00$^b$ &$-$01:35:57.5 &Low-surface-brightness blue structure, east of BAL QSO & $1,4$\\
L2SW &02:39:51.951$^a$&$-$01:35:58.92&Highly obscured gas-rich starburst galaxy near BAL QSO & $4$\\
L3   &02:39:51.59$^b$ &$-$01:36:04.8 &Faint Ly\,$\alpha$ blob & $2$\\ 
\hline \noalign {\smallskip}
  \end{tabular}
 \noindent
 
 $1 -$ \cite{Ivison1998}. $2 -$ \cite{Li2019}. $3 -$ \cite{Vernet2001}. $4 -$ \cite{Ivison2010a}. $5 -$ \cite{Frayer2018}. \\
 $^a$ Coordinates for L1 and L2SW are measured from the ALMA continuum imaging of \cite{Frayer2018}, accurate to $\pm0.001$\,s and $\pm0.002$\,s in $\alpha_{\rm J2000}$, respectively, and $\pm0.02''$ in $\delta_{\rm J2000}$. $^b$ For L2 and L3, the positions come from \cite{Li2019}, corrected for the difference between their position for L1 and that of \cite{Frayer2018}.  $^c$ For L1N, the position comes from the Abell\,370 Frontier Fields {\it Hubble Space Telescope} imaging of \cite{Lotz2017}, corrected for the difference between the position for L1 and that of \cite{Frayer2018}. $^d$ The position of W1 cannot be given accurately because it is an extended and elongated source (see Fig. 5 in \citet{Frayer2018}).
  \label{tab:components}
\end{table*}

\begin{figure}
\begin{center}
	\includegraphics[width=\columnwidth]{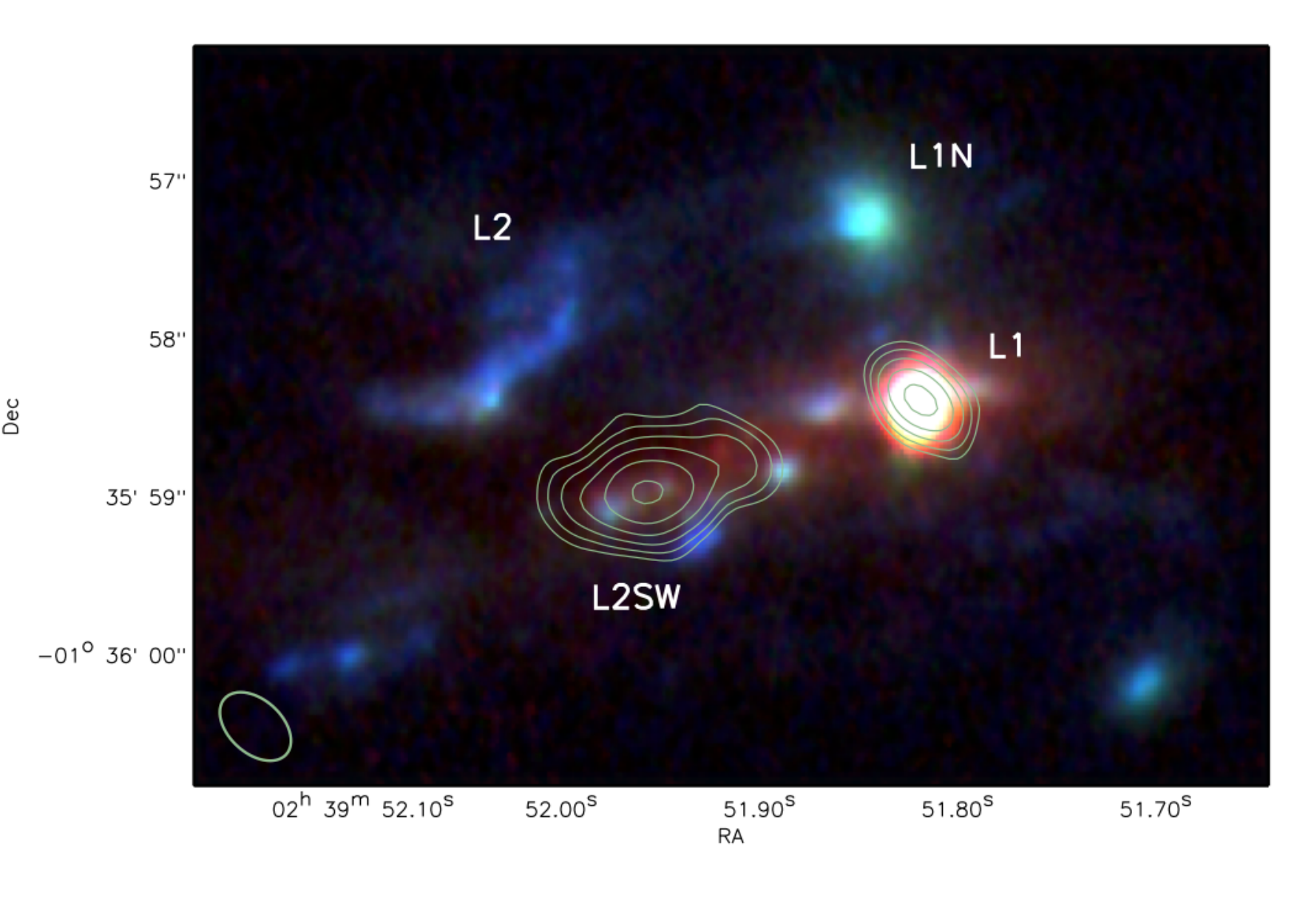}
    \caption{  Overview of the galaxy group, SMM\,J02399$-$0136. The key components are labelled: L2SW for the dusty starburst, L1 for the BAL QSO, following the notation of \citet{Ivison2010a}. ALMA rest-frame 360-$\mu$m continuum image with contours starting at $3\sigma$, spaced by $\sqrt 2$, where the peak is 1.58\,mJy\,beam$^{-1}$ and the local r.m.s. noise, $\sigma$, is 55\,$\mu$Jy\,beam$^{-1}$.  The Frontier Fields {\it HST} frames \citep{Lotz2017} used to make the underlying false-colour composite were shifted 0.1\,arcsec West and 0.1\,arcsec South to align with the available {\it Gaia} stars.  The astrometry is then be good to better than a small fraction of the ALMA synthesised beam, shown inset. The \Lya\ blob, L2, is out of the field and the extended CO structure, W1, is not shown.}
    \label{fig:HST}
    \end{center}
\end{figure}

In cosmological simulations, the growth of galaxies in the early Universe results from the accretion
of pristine gas in virialized dark matter halos, modulo ejection of matter by stars and active galactic 
nuclei  (AGN) \citep[e.g.][]{Schaye2015,Hopkins2018,Madau2014,Somerville2015,Tacconi2020}. 
This gas regulation involves complex exchanges of mass and energy that take place 
in the circum-galactic medium (CGM) around growing galaxies.  
So, the CGM naturally holds many of the clues to
the physics of galaxy evolution \citep{Tumlinson2017,Hafen2019,Wright2020}.
The challenges raised by feedback, and by AGN-driven winds in particular, are so prodigious that they
have only started to be touched by new generations of numerical simulations \citep[e.g.][]{Biernacki2018,Richings2018b,Costa2020,Dave2020}.

Observationally, a new era is opening up \citep[see the review of][]{Hodge2020}. While ejection is observed via ubiquitous ionised and neutral 
galactic winds \citep[e.g.][]{Rupke2018,Veilleux2020,Faisst2020} and contributes to the self-regulation of cosmic star formation, 
evidence of accretion is slowly building up \citep{Bouche2013,Zabl2019,Schroetter2019,Bielby2020,Walter2020,Fu2021}. 

One new avenue is the direct detection of extended halos ($\sim$ 10 kpc) of [CII] or CO emission around high-redshift galaxies 
 \citep{Ivison2010b,Ivison2011,Carilli2013,Tumlinson2017} 
but the dynamics of the gas traced by line emission of extended halos cannot be unambiguously ascribed to inflows or outflows \citep[see e.g.][]{Fujimoto2019}.
Only absorption lines provide the straightforward space-velocity information to 
disentangle inflows from outflows, without providing the physical origin of inflows, though. 

As bright sub-millimetre sources, dusty starburst galaxies at high redshift 
provide unique background sources to probe the environment of growing
galaxies using highly sensitive absorption spectroscopy in the sub-millimetre domain, 
just as quasars do in the visible domain.
The power of absorption spectroscopy is even greater against gravitationally lensed starburst galaxies that allow far more
sensitive  detections.
The sub-millimetre domain happens to be that of the fundamental transitions of light hydrides that, by linking hydrogen with heavy elements, are the very first steps 
of chemistry in space \citep{Gerin2016}.
Light hydrides have very high dipole moments so that the critical density of their fundamental transition is orders of magnitude higher than that of CO(1--0), the usual tracer 
of molecular gas. In diffuse molecular gas, their $J$=1 level is therefore weakly populated, so that molecular gas of low density ($n_{\rm H} < 10^3$\cc) can be uniquely detected by absorption of the fundamental transition of light hydrides against sub-millimetre background sources 
\citep[e.g.][for OH$^+$ lines]{Riechers2013,Indriolo2018,Berta2021}.

Among all hydrides, \CHp, one of the three first molecular species ever detected in space \citep{Douglas1941}, 
has unique chemical and spectroscopic
properties, such that it highlights the sites of dissipation of mechanical energy \citep{Draine1986,Gredel1993,Federman1996,Godard2009,Godard2014}.
Seen in absorption, it reveals not only the presence of low density molecular gas but also the trails of the supra-thermal energy in this medium. 
Turbulent transport in a biphase medium comprising cold ($T\sim 100$\,K) and warm ($T\sim 10^4$\,K)  neutral phases
may also contribute to \CHp\ production in diffuse gas \citep{Valdivia2017}. 
In either case, absorption of \CHp(1-0)  traces the small pockets of "lukewarm" gas in a cold medium of low density. In the rest of the paper, the term "cool molecular gas" 
is adopted for the gas phase seen in  \CHp(1-0) absorption \footnote{We note here that cold for \Lya\ observers and cosmologists is $T \sim 10^4$\,K, while 
it is $T \sim 10$\,K or below for submillimetre astronomers, and $T \sim 10^2$\,K for radioastronomers, which is
the temperature of the cold neutral medium (CNM) in thermal pressure balance with the warm neutral medium (WNM) at $T \sim 10^4$\,K. The cool molecular gas traced by \CHp(1-0) absorption is warmer 
than the bulk of star-forming molecular clouds at $T \sim 10-40$\,K.}.

  With ALMA,  the \CHp($J=1$--0) line has been detected in all the lensed
submillimetre-selected galaxies (SMGs) targeted so far at 
redshifts $z= 1.7$-- 4.2 \citep[][herafter Paper I, and Vidal-Garcia et al., in prep.]{Falgarone2017}.
Most of these are lensed starburst galaxies discovered by {\it Herschel} \citep{Eales2010,Oliver2010}.  
\CHp\ absorption lines, found in 15 of the 18 SMGs observed, 
are unexpectedly deep (opacities 0.25--1.20) and broad (average FWHM $ \sim$ 400\,\kms). 
The emission lines, when fully visible, are extremely broad  (FWHM $\sim 1500$\,\kms) and are understood as originating in myriad molecular shocks \citep{Godard2019} presumably powered by hot galactic winds penetrating the CGM (Paper I).

In Paper I, we  assumed that the width of the absorption lines is dominated by turbulence and show that the \CHp(1-0) absorption lines reveal  previously unseen massive and 
highly turbulent reservoirs (TR) of diffuse molecular gas ($M_{\rm TR} \sim$  $10^{10}$\,\msol) extending far outside ($> 10$\,kpc)  the compact starburst cores of radii $\approx 1$\,kpc.
As absorption lines alone cannot provide the extent of the absorbing medium along the line-of-sight, 
the {\bf original} method followed to determine the radius (and mass) of the diffuse gas traced by \CHp\ absorption 
involves several assumptions (Paper I and Appendix \ref{A}): \\
(1) the high occurrence of absorption detections suggests  not only that the surface filling factor of the diffuse gas against the background continuum source is high but also that
the turbulent reservoirs of diffuse molecular gas and the starburst phase are coeval. We thus assume that the TR lifetime is of the order of the duration of the starburst phase, $t_{\rm SB}\sim 50-100$ Myr \citep{Tacconi2008,Greve2005},\\
(2) following results obtained in the Milky Way \citep{Godard2014}, the \CHp\ abundance in this diffuse molecular gas is assumed to be proportional to the turbulent energy transfer rate per unit volume, $\epsilon \sim \overline{\rho} \overline{v}_{\rm turb}^3/r_{\rm TR}$, that depends on the average gas mass density $\overline{\rho}$ within the TR volume, on the mean turbulent velocity inferred from the observed absorption linewidth,  $\overline{v}_{\rm turb}= 0.7\Delta v_{\rm abs}$, and on the unknown radius $r_{\rm TR}$, causing a degeneracy between the \CHp\ abundance, $X(\CHp)$, and $r_{\rm TR}$, \\
(3) we break this degeneracy with the finding that, in the few cases where the stellar mass of the starburst galaxy is known, the radius $r_{\rm esc}$ at which the escape velocity 
of the galaxy surrounded by its massive CGM is equal to the mean turbulent velocity is such that $r_{\rm esc} \sim  \overline{v}_{\rm turb} \times  t_{\rm SB}$ within a factor of two, 
in spite of stellar masses differing by a factor of 10 and very different gas velocity dispersions. 
 This is why both the TR lifetime and the starburst phase duration, $t_{\rm SB}$, are about equal to the dynamical time of the TR large-scale turbulence, $t_{\rm dyn} \sim r_{\rm TR}/ \overline{v}_{\rm turb}$.  

A validation of these unrelated assumptions can only be achieved by direct imaging of the CGM. 
This was the primary goal of the present work, in which we compare our \CHp(1-0) absorption results with the \Lya\ line emission in 
 SMM\,J02399$-$0136 at $z\sim2.8$, previously detected \citep{Ivison1998,Vernet2001} and recently imaged  at high spectral resolution, $R \approx 4000$, with the Keck/KCWI \citep{Li2019}.
 As often, this primary goal has been largely exceeded and the \Lya\ -- \CHp\ comparison in space, and even more importantly in velocity-space, has opened up new perspectives that we also report in this paper.

SMM\,J02399$-$0136, the first SMG ever found, is gravitationally lensed  by the  foreground galaxy cluster Abell 370 with a magnification factor in the range $\mu=2.0-3.2$\footnote{Magnification values obtained from the  {\it Hubble Space Telescope} ({\it HST}) Frontier Fields program \citep{Lotz2017} which includes different models of the lense maps for Abell 370. https://archive.stsci.edu/prepds/frontier/lensmodels/} \citep{Magnelli2012,Johnson2014,Richard2014,Diego2016}. Its total intrinsic bolometric luminosity is $L_{\rm bol} = 1.2 \times 10^{13}$ \lsol\ \citep{Ivison2010a}.
 It is a complex and vast galactic nursery (see Table~\ref{tab:components} and Fig.\ref{fig:HST})  that harbors the starburst galaxy  detected in the FIR dust continuum emission (L2SW), a broad absorption line quasar (BAL QSO) 
  also seen in the FIR continuum (L1)  and faint companions visible in the rest-frame UV (L2 and L1N) and undetected in the FIR. 
A new \Lya\ emitter, L3, has been recently discovered in the field \citep{Li2019}.
  
The starburst galaxy provides 58 \% of the total IR luminosity \citep[log$_{10}$ $L_{\rm IR}=12.94\pm 0.05$, corrected for gravitational amplification --][]{Magnelli2012}, as judged by the relative fractions of rest-frame 122-$\mu$m continuum emission from L1 and L2SW in naturally weighted ALMA images \citep{Ferkinhoff2015}.  The implied SFR for L2SW is around 870\,M$_\odot$\,yr$^{-1}$ for a \citet{Salpeter1955} stellar initial mass function (IMF) covering 0.1--100\,M$_\odot$ \citep[e.g.][]{Kennicutt1998}, ignoring the considerable uncertainty related to the IMF. 
 The molecular gas mass inferred from CO observations is $M_{\HH} = (2.3 \pm 0.3) \times 10^{11} (\alpha_{\rm CO}/2.1)$\,\msol, where the CO-to-H$_2$ conversion factor is taken to the average $\alpha_{\rm CO} = 2.1$\,M$_\odot$ (K km\,s$^{-1}$\,pc$^2$)$^{-1}$ over its various gas-rich components  \citep{Frayer2018}.  It extends over $\sim 25$ kpc in the source plane ~\citep{Frayer1998,Ivison2010a}.  
A new massive elongated structure  (W1) extending 13 kpc westward of L1 has been recently discovered by \citet{Frayer2018}. 
It is detected in CO(1-0) single-dish and ALMA CO(3-2) observations, but undetected in higher-$J$ CO lines with ALMA. The nature of this low-excitation component, with a velocity red-shifted with respect to that of L1 is undetermined.

  For $H_0=67.4$\kms\,Mpc$^{-1}$,
  $\Omega_{\rm M}=0.315$ and $\Omega_\Lambda=0.685$ \citep{Planck-Collaboration2020}, 1\,arcsec corresponds to a proper size of 8.0\,kpc at $z$ =2.8  (uncorrected for lensing magnification which acts mainly East-West, \citealt{Frayer2018}). 
  Throughout the paper, we adopt a correction of $\times$ 2.25 along the shear direction and $\times$ 1.06 perpendicular to that, giving a total of $\times$ 2.38 \citep{Ivison2010a}.
  The correspondences 3.55 kpc/arcsec, 7.55 kpc/arcsec and  3.36 kpc/arcsec are therefore used in the East-West, North-South and NE-SW (and NW-SE) directions, respectively. 

The paper is organised as follows:  Sect. 2 is dedicated to the improved determination of the galaxy redshifts, the ALMA and IRAM-30m \CHp(1-0) results are presented in Sect. 3, the summary of the Keck/KCWI \Lya\ results 
is given in Sect. 4, the comparison of the \CHp(1-0) and \Lya\ data is conducted in Sect. 5. The discussion, in Sect. 6, includes the many CO observations of this galaxy group and an estimation of the energy budget. The paper ends in Sect. 7 with a summary of our findings and open questions.

\begin{figure*}
\begin{center}
	\includegraphics[width=\textwidth]{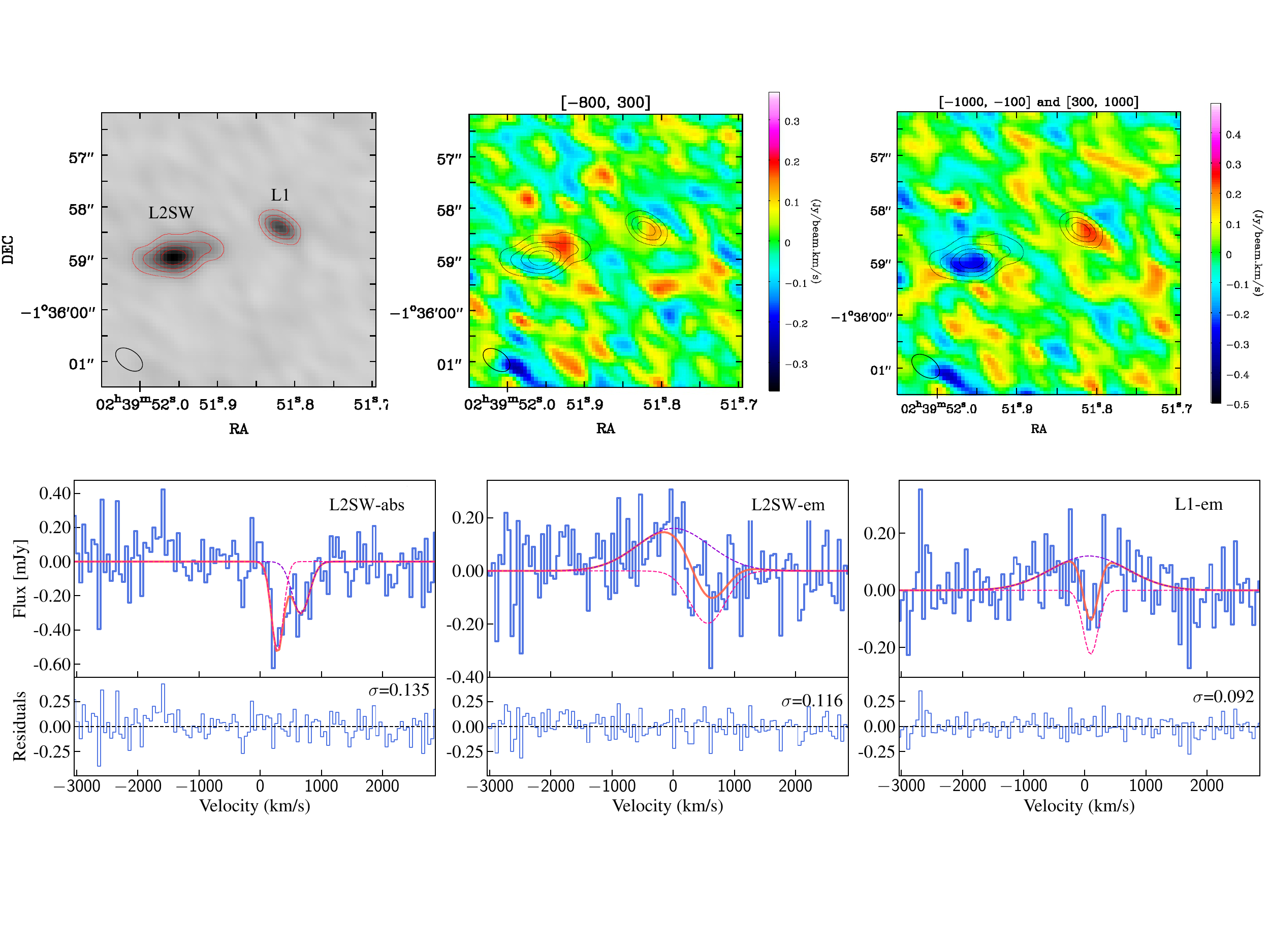}
    \caption{{\it Upper row:} {\it Left:} ALMA rest-frame 360-$\mu$m continuum image of SMM\,J02399$-$0136 (gray scale and red contours at 20, 40, 60 and 80$\%$ of the peak at 1.58 mJy/beam) with $\sigma=55$ $\mu$Jy/beam. The SMG is labelled L2SW and the BAL QSO L1, following the notations of~\citet{Ivison2010a}.  The synthesised beam is shown in the bottom left corner. 
    {\it Middle:} Map of the \CHp\ line integrated area (i.e. moment-0) over the velocity range [-800,300] \kms. The r.m.s. noise level over the displayed area is $\sigma_{\rm m0} =$ 0.06 Jy/beam \kms. {\it Right:} \CHp\ moment-0 map over the velocity range [-1000,-100] and [300,1000] \kms.   The r.m.s. noise level over the displayed area is $\sigma_{\rm m0} =$ 0.08 Jy/beam \kms. The black contours show the dust continuum emission of the galaxies.
    {\it Lower row:}  Three  \CHp(1-0) continuum-subtracted ALMA spectra, with Gaussian fits of the emission and absorption. The origin of the velocity scale is  at $z_{\rm ref}=2.8041$.The dashed lines show individual Gaussian
    components, in absorption and/or emission. The residual spectra and the r.m.s. noise level are also displayed. 
   The spectra are obtained over the L2SW extended continuum peak (80\% contour level) ({\it left}),  and over the extended regions at levels $\geq \sigma_{\rm m0}$ (yellow range)  $\sim 0.5$ arcsec west of L2SW ({\it middle}) and extending over $\sim 1$ arcsec south-west of L1 ({\it right}).
    }
    \label{fig:3specGal}
    \end{center}
\end{figure*}

\section{ Improved redshift measurements for the L1 and L2SW galaxies}

The redshifts used in the various observations of SMM\,J02399$-$0136 are different: $z=2.803$ in \cite{Ivison1998}, \cite{Frayer1998} and the present \CHp\ observations, $z=2.808$ in \cite{Genzel2003} and \cite{Frayer2018} and $z=2.8048$ for the \Lya\ observations \citep{Li2019}. They correspond to redshift determinations achieved with different lines and angular resolutions. 
The velocity difference corresponding to the two extreme redshifts is $\sim$ 300 \kms, with a remarkable trend:  the larger the beam and  the lower the line excitation, the higher the redshift \footnote{Similarly, two redshifts differing by $\sim 300$\kms\ have been measured for the starburst galaxy G09v1.40 \citep{Bussmann2013,Yang2017} likely due to a large scale outflow detected in \CHp\ emission \citep{Falgarone2017}. }. 
In the present paper, all the velocity scales are computed with respect to a reference redshift $z_{\rm ref}=2.8041 \pm 0.0004$ inferred from an ALMA CO(7-6) image at higher resolution (0.48"$\times$0.46") than that published 
(0.72"$\times$ 0.61") in \cite{Frayer2018}, obtained using Briggs weighting rather than natural weighting to maximise the resolution. 
The CO(7-6) high excitation line is a better tracer of the dense hot starburst cores than the lower excitation CO(3-2) line, even observed at high angular resolution with ALMA:  the CO(3-2) channel maps in \cite{Frayer2018} clearly show the existence of an extended low excitation component at $v \sim 340$\,\kms\ that contaminates the L1 and L2SW line profiles.
From the new 1mm continuum map, the positions of the galaxies, 
based on the radio phase calibrators,  
are given in Table~\ref{tab:components}.
In the  CO(7-6) line data cube, the centre velocity of the galaxies shifts to the red for larger apertures.  
For an aperture associated with the beam size, the velocities (on the $z=2.808$ scale, that of the CO(7-6) ALMA data) of the 
mid-points of the half-power strengths centred on the 1mm continuum
positions are $v_{\rm L1}=-297\pm30$ \kms\ and $v_{\rm L2SW}=-320\pm20$ \kms. 
The velocity difference between L1 and L2SW being only 23 \kms, the reference redshift for the two galaxies, $z_{\rm ref}$, is inferred from the observed average velocity -308.5 \kms\ of the two galaxies in the $z=2.808$ scale.
It provides the best reference frame for the galaxies themselves because it combines the highest angular resolution with the kinematics of the CO(7-6) lines that peak at the same positions as the 1mm continuum sources. 

All the velocities reported in the paper are measured in this reference frame. The absolute velocity estimates are uncertain by only $\pm 15$ \kms, corresponding to the uncertainty on $z_{\rm ref}$.


\section{The \CHp(1-0) perspective}

\subsection{ALMA observations and results}\label{sub:subalma}

\begin{figure*}
\begin{center}
		\includegraphics[width=\textwidth]{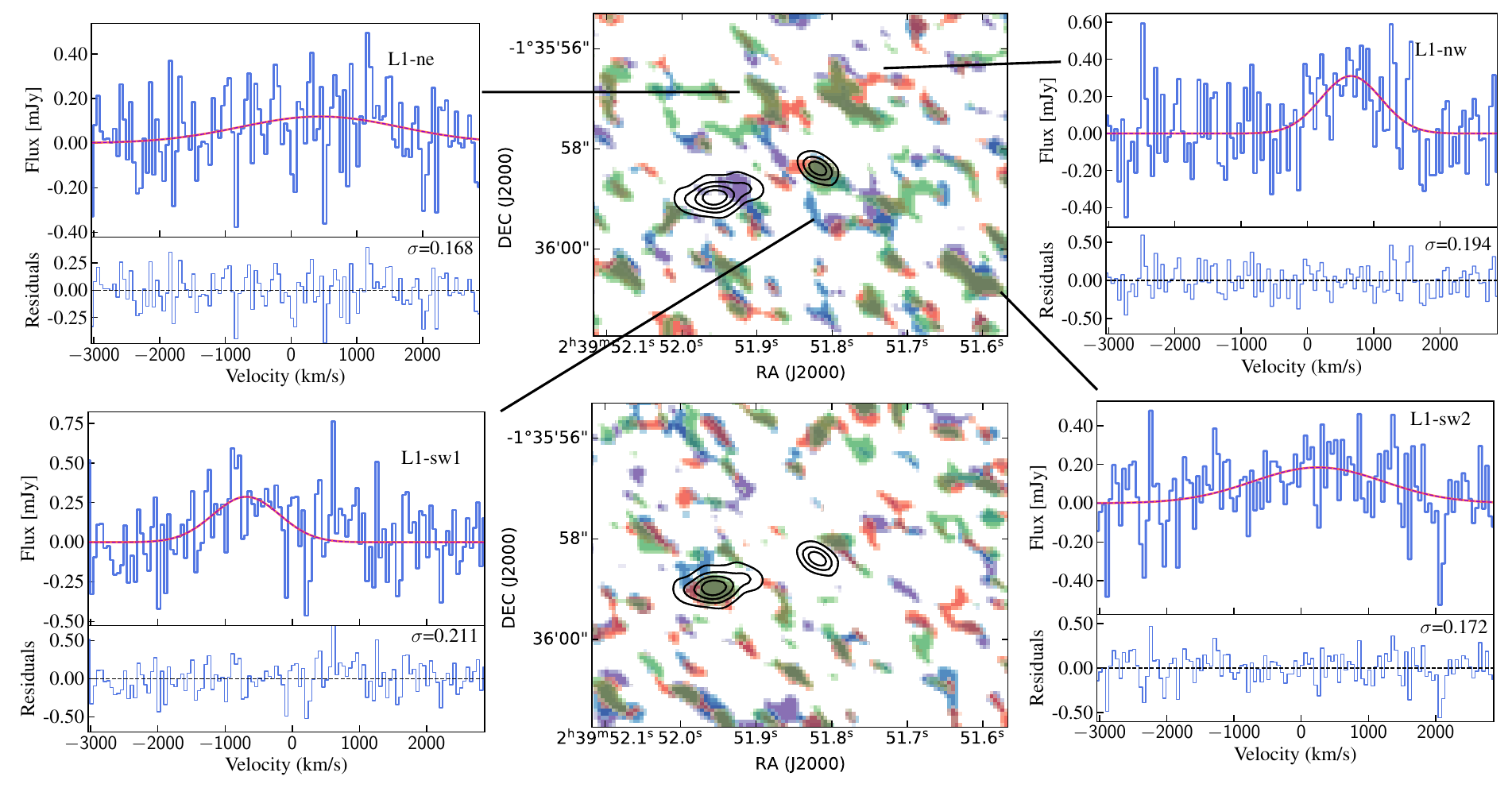}
    \caption{ {\it Central panels:} Overlay of four \CHp\ moment-0 maps displaying only the pixels brighter than $\sigma_{\rm m0}$ ({\it  top}) and weaker than 
      $- \sigma_{\rm m0}$ ({\it  bottom}).   The velocity coverages of the four moment-0 maps are: [-1350,250 ] \kms\ (purple), [-1000,-100] and [300,1000] \kms\  (blue), [0,1000] \kms\ (red) and [500,1500] \kms\ (green). The r.m.s. noise level over the displayed area are $\sigma_{\rm m0} =$ 0.06 and 0.08 Jy/beam \kms\ for the narrower (1000 \kms) and broader (1600 \kms) velocity coverage, respectively.  {\it Outer panels:} Four \CHp(1-0) continuum-subtracted ALMA spectra of the indicated extended areas encompassing the structure shapes in the different moment-0 maps. Gaussian fits of the emission lines and residual spectra are displayed. The origin of the velocity scale is at $z_{\rm ref}=2.8041$.  The black contours mark the continuum emission of L1 and L2SW, as in Fig. \ref{fig:3specGal}.
}
    \label{fig:4specL1}
    \end{center}
\end{figure*}

SMM\,J02399$-$0136  is one of eight SMGs observed for the first time in ALMA Cycle 4 as part of project
2016.1.00282.S. The \CHp(1-0) line ($\nu_{\rm rest}$ = 835.137\,GHz, \cite{Amano2010}) 
at redshift $z \sim$ 2.8 is shifted into ALMA Band 6 at frequency 219.5 GHz. 
Observations were done with 42 antennas in configuration C40-4. The minimum and maximum baselines were 15~m and 704~m, respectively, providing a  maximum recoverable size of $\sim 5$ arcsec. J0238+1636, J0239-0234 and J0006-0623  were observed as bandpass, phase and flux calibrators.
Data reduction was performed within the Common Astronomy Software Application (CASA) version 5.4.0.
We followed the standard ALMA pipeline data reduction. Visual inspection of the different calibration steps, i.e. bandpass, phase, amplitude, and flux, showed satisfactory calibration. We additionally flag antenna DV23 during a short period of time because of bad amplitude.
Imaging was done in the same CASA version. We cleaned the data using the CASA task {\it tclean}, with the hogbom algorithm. We spectrally regridded the data to a channel width of 50\kms. To deconvolve the image, we used a mask contouring the sources which was defined in the continuum image. The restored synthesised beam size is 0.59"$\times$0.36" ($PA$=54 deg). The phase centre is  RA: 2h 39m 51.86s, Dec: -1deg 35' 58.81" (J2000).

The ALMA rest-frame 360 $\mu$m continuum image of SMM\,J02399$-$0136 is shown in Figs.~\ref{fig:HST} and \ref{fig:3specGal}. Only the starburst galaxy, L2SW, and the BAL QSO, L1, are detected.
The radius of the starburst galaxy L2SW in the source plane has been inferred from a simplified lens modelling  to be $r_{\rm L2SW} =1.8\pm 0.1$ kpc (see Appendix \ref{B}) consistent with the radii 
measured for SMGs at redshifts $z\sim$ 1-- 3 for which the mid-IR emission is dominated by star formation ~\citep[e.g.][]{Ikarashi2017}.

The only three \CHp\ lines detected in the immediate vicinity of L1 and L2SW are displayed in Fig.~\ref{fig:3specGal}. In the direction of the L2SW peak of dust continuum emission, 
only absorption is observed: it is redshifted with respect to the galaxies average velocity.  An inverse P-Cygni profile is observed one synthesised beam westward of this peak. In the direction of L1, 
there is no absorption detected against the peak continuum emission but the upper limit is high, $\tau_0<0.7$, because the continuum source is weak. 
Absorption is tentatively detected one synthesised beam west of the peak, mostly  against the \CHp\ line emission. 
The regions over which absorption and emission are detected appear  at the $\sim 3 \sigma_{\rm m0}$ level  in the moment-0 maps of Fig.~\ref{fig:3specGal} computed over two different velocity intervals ($\sigma_{\rm m0}$ is the r.m.s. noise level of the moment-0 maps).
In L1 and L2SW, the \CHp\ emitting regions are more extended, at that level, than the synthesised beam and spatially offset from the continuum peaks, an offset already reported in Paper I for other starburst galaxies.

\begin{table*}
\small
\begin{center}
\caption[]{\bf{Characteristics of the \CHp\ line profiles and inferred properties of the  turbulent reservoir around L2SW and L1,  {\bf computed for $t_{\rm SB}=66$~Myr}. All the velocities are given  with respect to $z_{\rm ref}=2.8041$. 
}}
\bigskip
\smallskip
\begin{tabular}{llllllllllll}
\hline \noalign {\smallskip}
Source   &   &  $S_{\rm cont}^a$  &$v_{\rm abs}$   & $\Delta v_{\rm abs}$  & $v_{\rm em}$  & $\Delta v_{\rm em}$  & $\tau_0$  &  $N(\CHp)$   & $r_{\rm TR}$   &  X(\CHp) & $M_{\rm TR}$  \\
              &  & mJy &\kms  &  \kms               & \kms                   &   \kms      &               &  10$^{14}$\cq &  kpc    & $10^{-8}$     & $10^{10}\msol$   \\
\hline \noalign {\smallskip}
L2SW     & LV$^b$ & 1.5            &  280$\pm$40      & 210$\pm$70 &     --   &   --                                    &  0.45   &   2.9  &       & 0.7   &      \\
          &   HV$^b$&                     &  660$\pm$70      & 290$\pm$160  &  --   &  --                                    &  0.26   &  2.3   &       &  1.3  &         \\
       & LV+HV$^c$      &             &  430$\pm$40       & 580$\pm$100  &   --      &  --                                  &  0.29   &  5.0   &  23 &    5.2    &  3.5$^d$   \\
L2SW-em & & 0.7                        & 600$\pm$180 & 750$\pm$440 & 40$\pm$380   & 1280$\pm$640  &   0.37   &  8.3  &    30 & 8.8   &   4.4$^d$\\
L1-em  & & 0.4               &  80$\pm$90  &  290$\pm$130  & 60$\pm$240 & 1350$\pm$440                &  0.62   &  5.4  &   12  & 1.3   &   3.3$^e$ \\
Average $^f$ &  &  &                  380$\pm$200 & 490$\pm$470 &   55 $\pm$ 450  & & &   6.2 & 22 & & 3.7 \\
\hline \noalign {\smallskip}
\end{tabular}
\label{tab:tabfitvalues}
\end{center}
$^a$ Rest-frame 360$\mu{\rm m}$ continuum flux of the galaxies.  
$^b$ Low (LV) and high (HV) velocity components in Fig.~\ref{fig:3specGal} (bottom left).
$^c$ Fit with one single component that allows the determination of the global TR properties. $^d$ Mass estimated for $r_{\rm SMG}=1.8{\rm kpc}$, the intrinsic radius of L2SW 
inferred from lens modelling (Appendix B). $^e$ Mass estimated for $r_{\rm SMG}=0.8{\rm kpc}$ (Appendix B). $^f$ Weighted averages for $v_{\rm abs}$, 
$\Delta v_{\rm abs}$ and $v_{\rm em}$ of the single component fit for L2SW, L2SW-em and L1-em.
\end{table*}

The line profiles are decomposed into Gaussian components: one or two components in absorption and one in emission. 
 The parameters in Table~\ref{tab:tabfitvalues} are inferred from the fits made simultaneously to the emission
and absorption lines: $\Delta v_{\rm em}$ and $\Delta v_{\rm abs}$ are
the FWHM of the emission and absorption lines,  $v_{\rm abs}$ (resp.  $v_{\rm em}$) is the offset velocity of the absorption (resp. emission) Gaussian
with respect to the redshift  $z_{\rm ref}=2.8041$ (Sect. 2).
 The uncertainties on the parameters have not been evaluated with the bootstrap method  
 to take into account the correlation between these parameters because the emission lines are only tentatively detected. 

 A remarkable feature in the moment-0 maps of Fig.~\ref{fig:3specGal} is the presence  in the L1 environment of elongated structures 
above the $\sigma_{\rm m0}$ level, whose shape  depends on the velocity range over which the line moment-0 is computed.   
To highlight the velocity-dependence of their shape, Fig.~\ref{fig:4specL1} displays the overlays of all the extended structures brighter than $\sigma_{\rm m0}$  in four different moment-0 maps of the close environment of L1 and L2SW. The choice of their velocity coverage optimises the signal-to-noise ratio of the moment-0 values (i.e. coverage of 1000 and 1600 \kms\ much larger than the resolution to increase the signal but not as large as the lines themselves to limit the noise level), while the choice of their velocity reflects the distribution of the structures in velocity.  The overlays of extended structures weaker than $- \sigma_{\rm m0}$ in the same moment-0 maps are also shown.
\CHp\ spectra integrated over areas encompassing the structure contours in the four different moment-0 maps are also displayed in Fig.~\ref{fig:4specL1}. 
These structures are possible regions of \CHp\  line emission located within  $\sim 10$~kpc from the BAL QSO, L1.

A search for emission lines further away from the galaxies, conducted in velocity-space and space over the 1/3 primary beam of the ALMA observations (43"/3$\approx 14$"), has led to the discovery of  possible additional structures of \CHp\ emission.
The criteria used to identify structures in different moment-0 maps are: {\it(i)} line-integrated intensity brighter than the $\sigma_{\rm m0}$ level, where  $\sigma_{\rm m0}$ is the r.m.s. noise level 
of the moment-0 values over the searched area, {\it(ii)} more extended, at that level, than the synthesised beam, i.e. $\Omega > 0.17 {\rm arcsec}^2$ and 
{\it(iii)} localised within the 1/3-primary beam area where the noise is minimum (i.e. at a distance less than 7" from the phase centre).  
No lower limit on the linewidth was included in the selection criteria.
The search was performed with CASA  and is by no means a systematic search of all the \CHp\ emission spots. This is the goal of an on-going study (Hayatsu et al., in prep.). 

\begin{figure*}
\begin{center}
	\includegraphics[width=\textwidth]{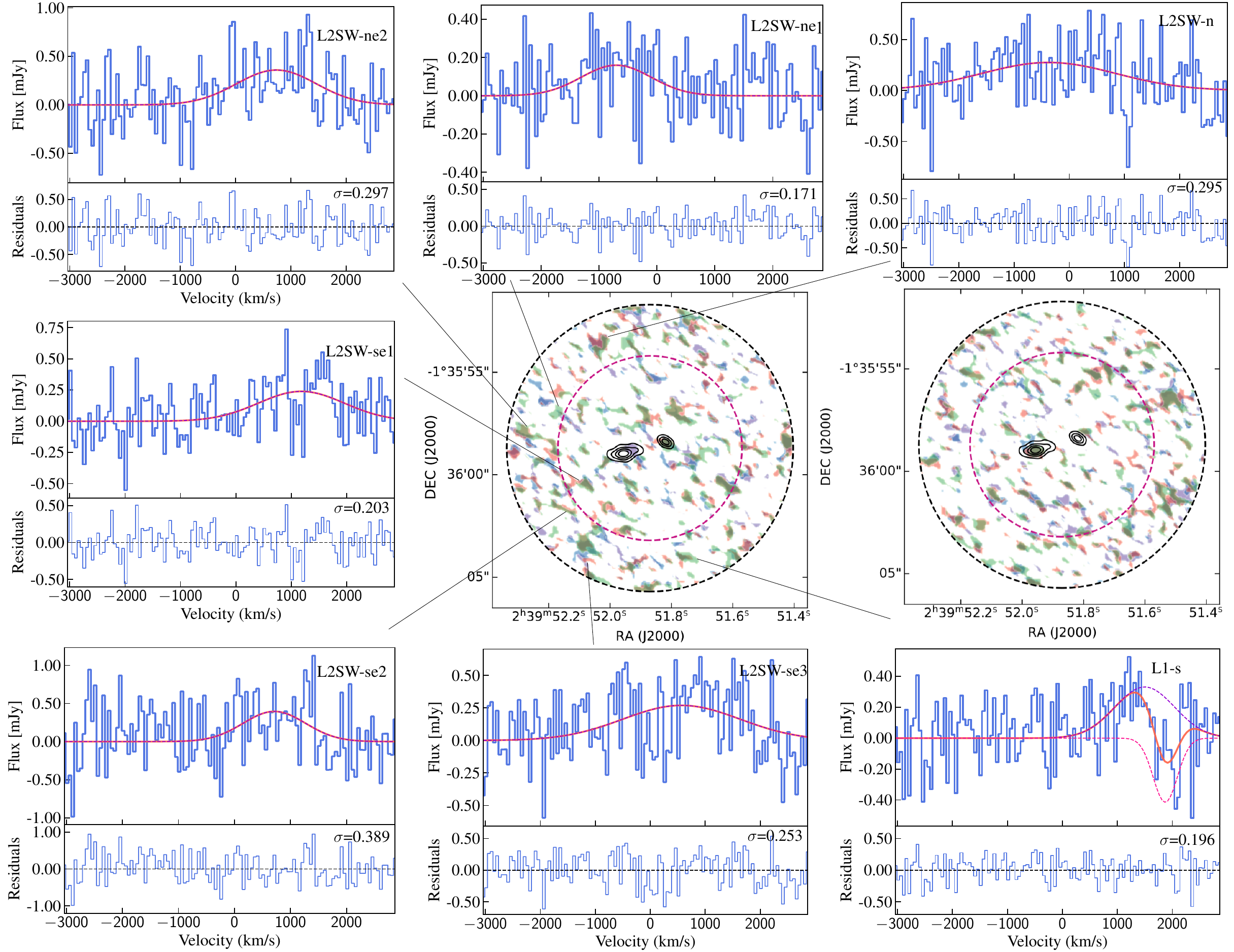}
    \caption{ Same as Fig.~\ref{fig:4specL1} over one third of the ALMA primary beam, shown as a dashed black circle.
   The r.m.s. noise levels of the four moment-0 maps over the 1/3 primary beam of ALMA are similar to those in the field of  Fig.~\ref{fig:4specL1}. The 
   distance of 4.5 arcsec to the phase centre is given by the dashed magenta circle.}
  \label{fig:7specFar}
  \end{center}
\end{figure*}

However, the number of negative extended structures challenges the statistical significance of the positive structures. 
While the number of positive structures exceeds that of negative structures close to the phase centre (Fig.~\ref{fig:4specL1}), it is less clear cut in the  wider field, out to the 1/3 response point of the primary beam (Fig.~\ref{fig:7specFar}).
We show in Appendix \ref{C} that the bright and extended negative structures could be due to combined sidelobes of multiple regions of weak emission and that  
the number of pixels in positive extended structures larger than one synthesised beam exceeds that in negative ones within 4.5 arcsec from the centre 
of phase (Fig.~\ref{fig:stat}). 
This positive excess suggests that a fraction of the positive structures in the inner part of the ALMA field of view are not due to  
residual sidelobes, although this excess does not tell which are the real emission structures and which are the sidelobes among the positive ones. 
We also note that, as the total power is missing in our ALMA data, in particular its spatial distribution over a few arcsec, the power in the structures of negative moment-0 is overestimated. 

 The \CHp\ spectra of seven additional candidates are displayed in Fig.~\ref{fig:7specFar} 
with Gaussian fits to their  line profiles and residual spectra. 
Spectra taken within the same $\sigma_{\rm m0}$ contour as the selected moment-0 structures but localised in their vicinity are displayed 
in Fig.~\ref{fig:9specnoise} to illustrate the high quality of the baselines in the inner part of the ALMA field. 

These structures of possible \CHp\  line emission are complex because 
the lines are weak and extremely broad, and  emission and absorption coexist over various velocity ranges, depending on the position. This is why, 
although there are several other prominent positive extended structures in the moment-0 maps of Fig.~\ref{fig:7specFar},  the seven spectra selected are those for which simultaneous Gaussian fits to the emission and absorption could converge.

 \begin{table*}
\small
\begin{center}
\caption[]{\bf{Characteristics of the tentative \CHp\ emission detections.  }}
\bigskip
\smallskip
\begin{tabular}{lllllllll}
\hline \noalign {\smallskip}
Name       & Flux density            &    $\sigma$ $^a$ & $v_{\rm em}$ $^b$ & $\Delta v_{\rm em}$ $^b$ & S/N $^c$ & d $^d$& d $^e$ & $\Omega$ $^f$ \\         
                &  mJy                       &   mJy                                             &   \kms             & \kms            &   & arcsec  & kpc  & arcsec$^2$\\
\hline \noalign {\smallskip}
L1-em  &   0.12 $\pm$ 0.05     &        0.09     &     60  $\pm$   240  &   1350  $\pm$440  & 2.8  &1.1  &  3.9 & 0.34\\
L2SW-em &      0.16 $\pm$ 0.12    &           0.12  &  40 $\pm$380    &    1280$\pm$640   &3.0  &  1.1 & 3.9 & 0.38\\
L1-ne &    0.12 $\pm$ 0.11 &  0.17    &   530 $\pm$ 1200 & 2900$\pm$3100&  2.7               & 1.7& 5.7 & 0.46\\
L1-nw &    0.31$\pm$0.17 & 0.19   & 740 $\pm$ 290 & 1070$\pm$690 &  3.8             & 2.5  & 8.4 & 0.42\\ 
L1-sw1  & 0.29$\pm$0.17 & 0.21 & -600$\pm$330 & 1160 $\pm$ 775 &   3.3 & 2.0 & 6.7 & 1.03\\
L1-sw2  &  0.18 $\pm$ 0.11 & 0.17 & 340 $\pm$ 720  &  2290$\pm$ 1700 & 3.6 & 4.5 & 15.1 & 0.32\\ 
L2SW-ne1  & 0.16 $\pm$ 0.15 & 0.17 &  -620$\pm$ 655 & 1420 $\pm$ 1540 & 2.5 & 3.8 &  12.8 & 0.44\\
L2SW-se1  &  0.24 $\pm$ 0.13 &  0.20 & 1270 $\pm$ 500 & 1820$\pm$ 1210 & 3.6 &  3.2 & 10.8 & 0.54\\
Average  &    0.25  &  --     &    170 $\pm$ 540 $^g$& 1330$\pm$270 $^g$&  --  &  & 8.4 & 0.49 \\ 
\hline \noalign{\smallskip}
L1-s   $^h$ & 0.34 $\pm$ 0.24     &  0.20  &1590$\pm$ 766 & 1275$\pm$ 940 & 2.8 & 5.6 & 42.3  &  0.48\\
L2SW-n   & 0.27 $\pm$ 0.15  & 0.30 &-285 $\pm$ 545 & 2860 $\pm$ 1315 & 3.4 & 5.4& 25.4 & 1.85\\
L2SW-ne2   &  0.36 $\pm$ 0.13 &  0.30 & 825 $\pm$ 320 & 1700 $\pm$ 750 & 3.5 &  5.0 & 17.8 & 1.03\\
 L2SW-se2   & 0.48$\pm$ 0.15  & 0.40 &  805 $\pm$ 215 & 1390 $\pm$ 500 & 3.2 &  5.0 & 16.8 & 0.67\\
 L2SW-se3     & 0.27$\pm$ 0.11 & 0.25 & 655 $\pm$ 525 & 2510$\pm$ 1270 & 3.8 & 5.3 & 17.8 &1.00  \\
 Average$^i$  &    0.25  &  --     &    412 $\pm$ 550 $^g$& 1450$\pm$390 $^g$&  --  &  & 28.8 & 0.75 \\ 
 \hline \noalign{\smallskip}
\end{tabular}
\label{tab:11spots}
\end{center}
$^a$  Spectrum noise computed for a spectral resolution $\delta v = 50$ \kms,  
$^b$ Line centroid and FWHM inferred from the Gaussian fits shown  in Figs.~\ref{fig:4specL1} and \ref{fig:7specFar},
$^c$ Line signal-to-noise ratio computed over the FWHM, 
$^d$ Projected distance of the structure from the phase centre, expressed in arcsec,
$^e$ Same distance, expressed in kpc, corrected from lensing shear by adopting the three kpc/arcsec correspondences in the source plane discussed in Sect. 1,  
$^f$  Solid angle of the structure, to be compared to the ALMA synthesised beam area, 0.17 arcsec$^2$,
$^g$ Weigthed averages, 
$^h$ For this source, absorption was fitted simultaneously with the emission. The absorption parameters are the line centroid $v_{\rm abs}= 1955 \pm 210 $ \kms\ and FWHM $\Delta v_{\rm abs}= 520 \pm 675$ \kms, $^i$ Average values for the 13 structures.
\end{table*}

The characteristics of these candidate \CHp\  emission regions and those in the close vicinity of L1 and L2SW are given in Table~\ref{tab:11spots}. 
The first eight structures are the most plausible  
because they are located within 4.5 arcsec from the phase centre and the r.m.s. noise level 
of their residuals is $<$ 0.2 mJy. 
The others are less plausible, being further from the phase centre, with noisier spectra but the comparison of their \CHp\ emission to that of  \Lya\  suggests that three of them may be real (see Sect. 5). 
Note that, although the peak intensity is of the order of the r.m.s. noise level in 50 \kms\ channels, the signal-to-noise 
ratios computed over their full linewidth are all $\sim$ 3 because they are extremely broad ($\overline{\Delta v}_{\rm em} \sim 1300$ \kms).  

Four straightforward results regarding the lines of the eight most plausible candidates appear in this Table:\\
(1) their average centroid velocity, ${\overline v}_{\rm em} = 170$ \kms, is redshifted with respect to the reference frame of the galaxies, \\
(2) their centroid velocities are highly scattered, with a standard deviation $\sigma_{v_{\rm em}}$ = 540\kms, \\
(3) their average FWHM, ${\overline \Delta v}_{\rm em}=1330$ \kms, is extremely large for molecular lines, with FWHM up to $\sim$ 3000 \kms\ and their standard deviation is small, \\
(4) on average, they are $\sim$ 3 $\times$ more extended than the ALMA synthesised beam area, mostly because of their elongation. 
  
Several of these structures are elongated over more than 1 arcsec, along directions that are not aligned with the lensing shear direction, roughly aligned with the L2SW--L1 direction \citep{Frayer2018}. In a few cases they are barely resolved in the transverse direction.
  
\subsection{IRAM-30m observations }\label{sub:subiram} 

The \CHp(1-0) observations
were carried out in June and November 2019 at the frequency $\nu=219.584$\,GHz corresponding to the rest frame frequency $\nu_0=835.079$\,GHz of the \CHp($J$=1-0) transition\footnote{This value is the first to have been determined experimentally by \citet{Pearson2006}. It differs by only 58.5 MHz, or 21\kms\  from that determined later by \citet{Amano2010}. }
 at redshift $z=$2.803. Two EMIR receivers \citep{Carter2012} were operated, E0 at 90.9 GHz for the $^{12}$CO($J$=3-2) line and E2 for the \CHp\ line. The wobbler was operating at 
 0.5 Hz with a throw of $\pm 30"$.  The weather conditions were good  with a SSB system temperature varying between 160 and 220 K. 
 At this frequency, the half-power beam width is 11.2" (close to the 1/3 ALMA primary beam) and the flux/brightness conversion for point sources is $S{\rm [Jy]}= 9.6 \,T_{\rm A}^*{\rm [K]}$. 
A spectral resolution of 64 MHz provided a velocity resolution of 87 \kms.  
The spectra were reduced and analysed with the CLASS package of the GILDAS software\footnote{http://www.iram.fr/IRAMFR/GILDAS}.

 \begin{figure}
	\includegraphics[width=\columnwidth]{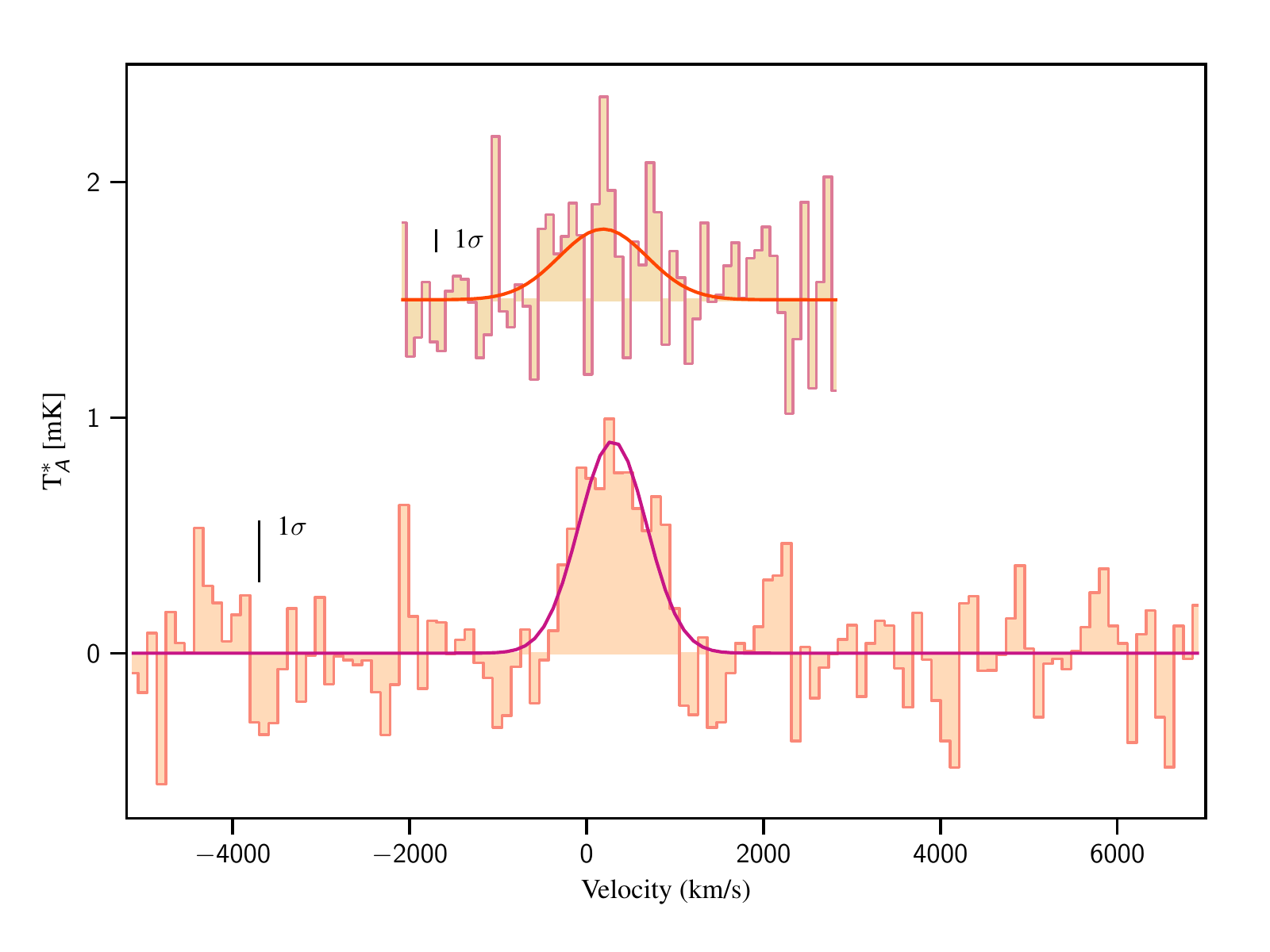}
    \caption{IRAM-30m observations:  the \CHp(1-0) line {\it (top)} and \twCO(3-2) line {\it(bottom)} and their Gaussian fits. The origin of the velocity scale is $z_{\rm ref}=2.8041$.}
    \label{fig:IRAM}
\end{figure}

A broad \CHp\ line has been tentatively detected with a peak intensity $T_{\rm A}^* = 0.3 \pm 0.1 $\,mK in resolution elements of 87 \kms, a centroid velocity $v_{\rm \CHp(1-0)} = 100 \pm 170$ \kms\ and a FWHM 
$\Delta v_{\rm \CHp(1-0)} = 1300 \pm  500$\kms\  (Fig.\ref{fig:IRAM}). 
The \twCO(3-2) line,  $T_{\rm A}^* = 0.8 \pm 0.2 $\,mK, is centred at  $v_{\rm CO(3-2)} = 210 \pm 60$ \kms,  and narrower, 
 $\Delta v_{\rm CO(3-2)} = 800 \pm 100$ \kms, than the \CHp\  line (Fig.\ref{fig:IRAM}). The offset velocities of the two lines with respect to $z_{\rm ref }$ are similar within the error bars.

\begin{figure}
\includegraphics[width=\columnwidth]{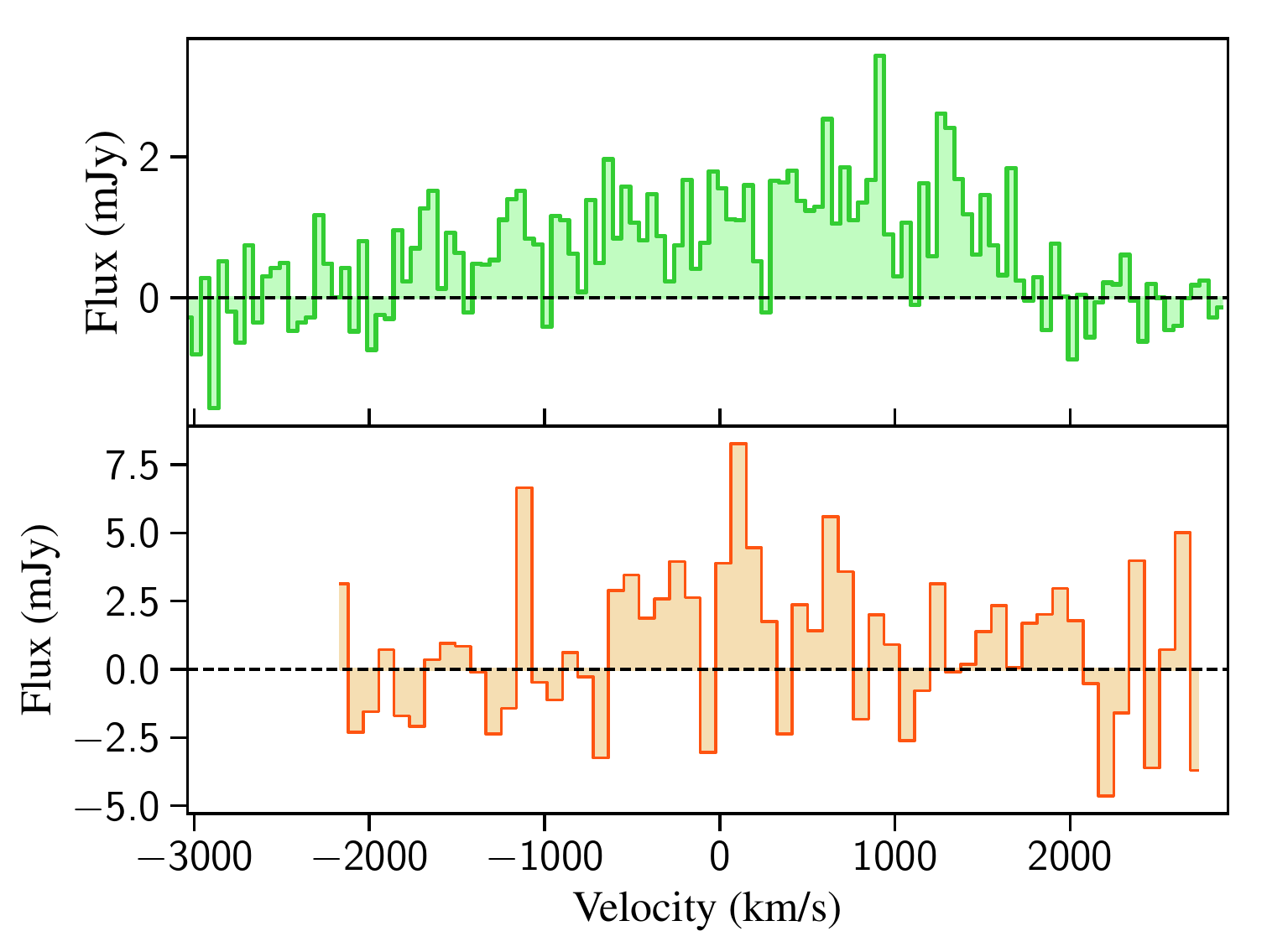}
    \caption{Comparison of the sum of all the \CHp\ emission line spectra detected by ALMA {\it (top)} with the tentative IRAM-30m detection {\it (bottom)}. The origin of the velocity scale is $z_{\rm ref}=2.8041$. }
    \label{fig:SumALMA}
\end{figure}

The sum of the eight most plausible \CHp\ line fluxes identified in the  ALMA data (Table\,\ref{tab:11spots}), weighted by the IRAM-30m beam profile at the relevant location, 
is shown in Fig.\,\ref{fig:SumALMA}, top.
This sum is smaller than the flux  measured with the single-dish IRAM-30m telescope (Fig. \ref{fig:SumALMA}, bottom). 
The line integrated fluxes over [-1000, 1000] \kms\  are  1.4$\pm$0.4 Jy\kms\ and 3.5$\pm$1.4 Jy\kms\ for ALMA and IRAM-30m, respectively. Over a broader interval,  [-1900, 2200] \kms,
 these integrated fluxes are 2.2$\pm$0.5 Jy\kms\ and 4.7$\pm$1.6 Jy\kms, indicating that the ensemble of scattered \CHp\ emission structures detected by  ALMA 
 in the inner 4.5 arcsec of the field of view contributes to about 50\% of the power detected by IRAM-30m. 
 
 This comparison suggests that these eight structures are the brightest of a broad distribution of weaker structures 
that contribute to the single-dish detection but are resolved out by ALMA and/or too weak to be detected. This is  also in line with our statement that the \CHp\ emission structures we report are not a complete census of the \CHp\ emission spots in the dataset.
 The line shapes of the two spectra in Fig. \ref{fig:SumALMA} are notably different which is likely due to the combination of a complex radiative transfer of the \CHp\ line (i.e. locally  the profiles are P-Cygni and inverse P-Cygni) 
 and ALMA filtering.
 
Last, we note that the total power detected by the IRAM single-dish telescope, 3.5 Jy\,\kms\ over 2000 \kms\  would correspond, if distributed uniformly, to an integrated line brightness of 4.7 mJy \kms\ beam$^{-1}$ in an ALMA moment-0 map at the resolution 0.59"$\times$0.36", integrated over 1600 \kms.  With 
a surface filling factor of 8\% in the ALMA 1/3 primary beam,  it would amount to $\sim 60$ mJy \kms\ beam$^{-1}$, that is the $\sigma_{\rm m0}$  level 
of the moment-0 maps. Recovering the total power and its spatial distribution at a few arcsec scale 
would possibly shift the moment-0 brightness distributions towards more positive values.

\subsection{Properties of the turbulent CGM inferred from the \CHp\ lines}\label{sub:turb}

\subsubsection{The \CHp\ absorption lines}

The  absorption line optical depth, computed at the velocity $v_{\rm abs}$ of the deepest absorption, 
is derived from the Gaussian  fits:
\begin{equation}
   \tau_0=-\ln\,\frac{S(v_{\rm abs})}{S_{\rm cont}+S_{\rm em}(v_{\rm abs})},
 \label{eq:tau}
  \end{equation}
 with $S_{\rm cont}$ and $S_{\rm em}(v_{\rm abs})$  are the dust continuum flux  
  and the line emission flux at velocity $v_{\rm abs}$, respectively, and $S(v_{\rm abs})$ is the flux resulting from the absorption.
  The optical depth estimates of the absorption lines (Table~\ref{tab:tabfitvalues}) are not affected by differential lensing because 
the background dust emission photons and those -- at the same frequency -- that escaped absorption by \CHp\ molecules  have experienced the same lensing. 
The velocity resolution is coarse and we cannot rule out the
possibility that portions of the line might be saturated. The estimated values of $N(\CHp)$
could  therefore be lower limits, so in Table~\ref{tab:tabfitvalues} we provide them without a compiled
error. However, the r.m.s. uncertainty on $\tau_0$ in the spectra of Fig.\ref{fig:3specGal}  is $\sim$ 0.15 so that the relative uncertainty on $\tau_0$
is between 25 and 40\%.
 
The \CHp\ column densities are the 
average  values in a solid angle subtended by the area, $\pi
r_{\rm SMG}^2$, of the background source assumed to be uniformly covered by the absorbing screen:
\begin{equation}
N(\CHp)=3\times10^{12}\cq\ \tau_0 \,(\Delta v_{\rm abs}/1\kms) 
\label{eq:NCHp}
\end{equation}

The absorption lines in the direction of L2SW are {\it redshifted with respect to the galaxies systemic velocity} at $z_{\rm ref}=2.8041$. 
These redshifted absorption lines are evidence of several inflowing gas streams onto the SMG at velocities, $v_{\rm in} = v_{\rm abs}$. 
The projected inflow velocities in the direction of L2SW are large, $v_{\rm in,L2SW} = 280 - 660$ \kms\ (see Table~\ref{tab:tabfitvalues} ). In the direction of L1, the absorption is 
also redshifted but the offset velocity is much smaller,  $v_{\rm in,L1} \sim 80$ \kms. 

Following Paper I, we estimate the radius  of the reservoir of diffuse molecular gas at the origin of the \CHp\ absorption by 
connecting the \CHp\  formation rate with the turbulent dissipation rate,
 assuming that the turbulent dynamical time $r_{\rm TR}/ \overline{v}_{\rm turb,TR}$ is of the order of the duration 
of the starburst phase $t_{\rm SB}$:
\begin{equation}
r_{\rm TR} =  \overline{v}_{\rm turb,TR} t_{\rm SB}.
\label{eq:radius}
\end{equation}
where $\overline{v}_{\rm turb,TR}=340$\kms\ is the mean turbulent velocity provided by the weighted-average width of the absorption lines $\overline{v}_{\rm turb,TR} = 0.7 \,\overline{\Delta v}_{\rm abs}$ (Table~\ref{tab:tabfitvalues}).

 The starburst phase duration,  $t_{\rm SB}$,  is often estimated as the consumption time of the molecular gas mass due to star formation, $t_{\rm SB}=M_{\rm \HH}/{\rm SFR}$. Values of  $t_{\rm SB}$ consistent with those of  \cite{Tacconi2008}, usually around 50--100 Myr, have appeared in the literature as far back as \cite{Frayer1999} and \cite{Greve2005}. 
 In the case of SMM\,J02399$-$0136, the starburst galaxy itself, L2SW, comprises approximately 25\% of the total molecular gas mass \citep[see][]{Frayer2018} yielding a bespoke value of $t_{\rm SB}\approx 66$\,Myr for its SFR of about 870 \msol\ yr$^{-1}$ (Sect. 1). We caution here that $t_{\rm SB}$ uncertainties stated in the literature have usually ignored the considerable systematics related to gas mass and SFR --- those due to $\alpha_{\rm CO}$ and the IMF, for example. 
 
 We recall in Appendix\,\ref{A} the method followed to 
infer the \CHp\ abundance from the fact that the \CHp\ formation in the diffuse component is primarily driven by turbulent dissipation. 
Once $r_{\rm TR}$ is known, the degeneracy between  the radius and the \CHp\ abundance is broken and the mass
of the diffuse molecular CGM (TR) can be determined. 

In Table~\ref{tab:tabfitvalues}, the different values of the radius $r_{\rm TR}$, abundance X(\CHp)  and mass $M_{\rm TR}$  of the turbulent reservoir 
obviously refer to the same entity, probed by different sight-lines. They are close to each other because the absorption optical depths are similar, being bracketed 
by unity on the upper side and $\sim 0.2$, the sensitivity of the observations on the brightest continuum source. The dynamic range on $M_{\rm TR}$ is therefore limited. 
The values in Table~\ref{tab:tabfitvalues}  are computed  for $t_{\rm SB} = 66$\,Myr, a molecular fraction ${f}_{\HH}=1$ and the local density of the absorbing medium, $n_{\rm H}=50$\cc.
This value of the local density is inferred from the Milky Way studies \citep{Godard2009}. An upper limit on the local gas density of $n_{\rm H} \lesssim 10^3$\cc\ is provided by the fact that the \CHp\ line 
is seen in absorption (see Appendix~\ref{D}). 

The last line of Table~\ref{tab:tabfitvalues}  gives the average values of the three estimates of the TR properties derived from the three lines-of-sight towards L2SW, L2SW-em and L1-em. 
The average of the absorption offset velocities, $v_{\rm abs}$, provides a mean inflow velocity, $\overline{v}_{\rm in}= 380\pm 200$\kms\ 
with respect to L1 and L2SW at $z_{\rm ref}=2.8041$. 

The \CHp\ absorption lines in SMM\,J02399$-$0136 therefore highlight the presence of a massive turbulent reservoir of diffuse molecular 
gas, $M_{\rm TR}= 3.7 \times 10^{10} \, t_{\rm 66}^2  \msol $, where $t_{\rm 66}=t_{\rm SB} (66{\rm Myr})^{-1}$,
inflowing towards the galaxies at $ \overline{v}_{\rm in}=380$\kms, a velocity comparable 
to its mean turbulent velocity $\overline{v}_{\rm turb,TR}=340$\kms. 
Its inferred radius $r_{\rm TR} \sim 22 \, t_{\rm 66} {\rm kpc}$ is independent of the lensing magnification since it is inferred from absorption lines.

\subsubsection{The \CHp\ emission lines}

The detected line flux densities (Table~\ref{tab:11spots}) correspond to line brightness temperatures $\sim$ 10\,mK in the source frame, assuming a uniform brightness over the structure solid angle (see Appendix \ref{D}).
This assumption provides a lower limit for the density in the gas component of the CGM where \CHp\ is formed and radiates. 
In Appendix \ref{D}, we give  the results of non-LTE radiative transfer for the \CHp($J$=1-0)  emission using the RADEX code \citep{van-der-Tak2007}. They provide the gas density and temperature conditions required to form a 10\,mK emission line, for two relevant \CHp\ column densities and a local velocity dispersion of 20 \kms.
Densities $n_{\rm H} \gtrsim 10^4$\cc\ are required for temperatures $<10^4$K and $N(\CHp)=10^{13}$ \cq. 

As the \CHp\ radical has an extremely short lifetime, $t _{\CHp} \sim 1$ yr (Appendix \ref{A}),  these structures, rich in \CHp\ and located at distances of several $\times$ 10 kpc from the galaxies, cannot have been stripped by galactic winds from the galaxies' interstellar medium, because the advection time, even at velocities $\sim 1000$\kms\ would be $>10$ Myr.  Therefore \CHp\ must have formed in situ, where it is observed. 
Given the large \CHp\ linewidths and the large densities required for line emission, the \CHp\ seen in emission most likely formed  in shocks. 
However, the \CHp\ emission lines are extremely broad, with an average FWHM $\sim 1330$ \kms\ (Table \ref{tab:11spots}), as in several other SMGs (Paper I). 
 \CHp\  cannot originate in shocks at such high velocities
since this would dissociate the \HH\  molecule needed for the \CHp\ formation and sputter dust grains necessary for the re-formation of \HH.
Most likely they originate in lower velocity ($v_{\rm sh} \approx 20$\kms) UV-irradiated molecular shocks (Paper I and \cite{Godard2019}).  A particularly interesting result 
of  \cite{Godard2019} is that the \CHp\ column density in the post-shocked layer of these irradiated shocks increases linearly with the UV-field irradiation over a broad range of pre-shock densities. 

The large velocity dispersion of the ensemble of individual low-velocity molecular shocks (LVMS) is at the origin of the broad \CHp\  linewidths. We ascribe it to turbulence generated in the post-shock layer of 
high-velocity shocks (HVS) in the CGM (see Sect. 6.2).  
 The dynamical energy transfer between HVS and LVMS  might be similar to that observed in  the 50-kpc long shock in the Stephan's Quintet driven by a galaxy collision with a gas tidal stream. Within this large-scale shock, the [CII] and CO line velocity dispersions are $\sim 1000$\kms\  \citep{Guillard2012,Appleton2013,Appleton2017} spanning the velocity difference between the intruder galaxy and the tidal stream of atomic gas.   
 
  The physics of such powerful energy transfers between $\sim$10kpc-scale shocks feeding a turbulent cascade in the multiphase post-shock layers is complex. It is the focus of several recent thorough investigations
\citep[e.g.][]{Gaspari2013,Gaspari2017,Li2020,Gronke2020b,Lochhaas2020,Voit2020}.

 We show below the added value of a joint investigation of \CHp\ molecular lines and \Lya\ observations in this field.


 \section{The  \Lya\  perspective  }

\begin{figure}
	\includegraphics[width=\columnwidth]{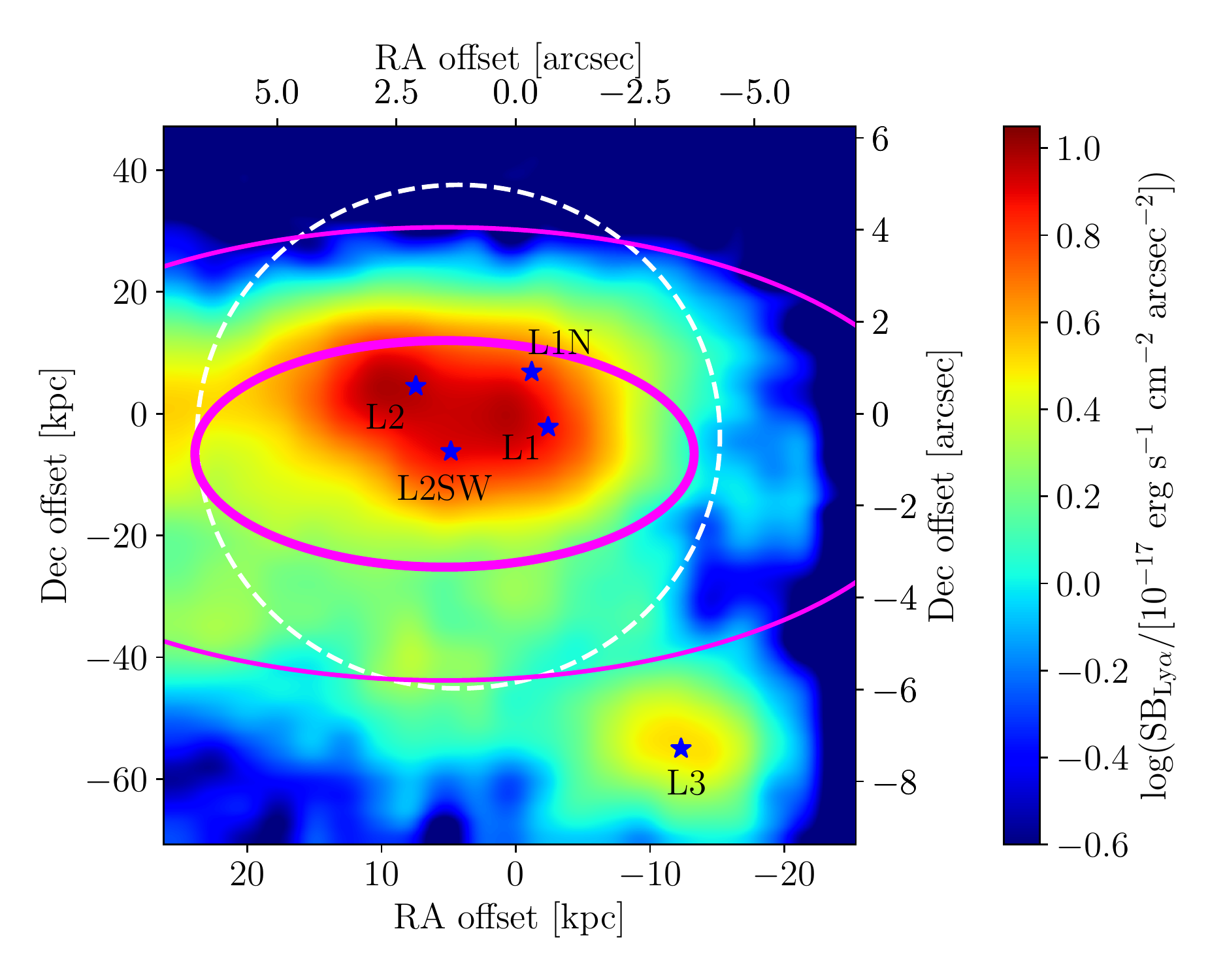}
    \caption{Continuum-subtracted \Lya\ emission  of the gas surrounding SMM\,J02399$-$0136 adapted from \citet{Li2019}. Note that we have taken the lensing shear direction into account, so that the kpc/arcsec correspondence is not the same along the RA and Dec axis (see Sect. 1).  The magenta ellipses therefore show the size of the turbulent reservoir of diffuse molecular gas seen in \CHp(1-0) absorption against the L2SW and L1, estimated following the argument of Paper I for $t_{\rm SB} = 50$\,Myr (thick) and 100Myr (thin) that bracket the 66~Myr value found for SMM\,J02399$-$0136. The dashed white circle shows the HPBW=11.2" of the IRAM-30m telescope at 220 GHz centred at RA(2000)=02:39:51.89,
Dec(2000)= - 01:35:58.9.  }
    \label{fig:Lyamap}
\end{figure}

\begin{figure}
		\includegraphics[width=\columnwidth]{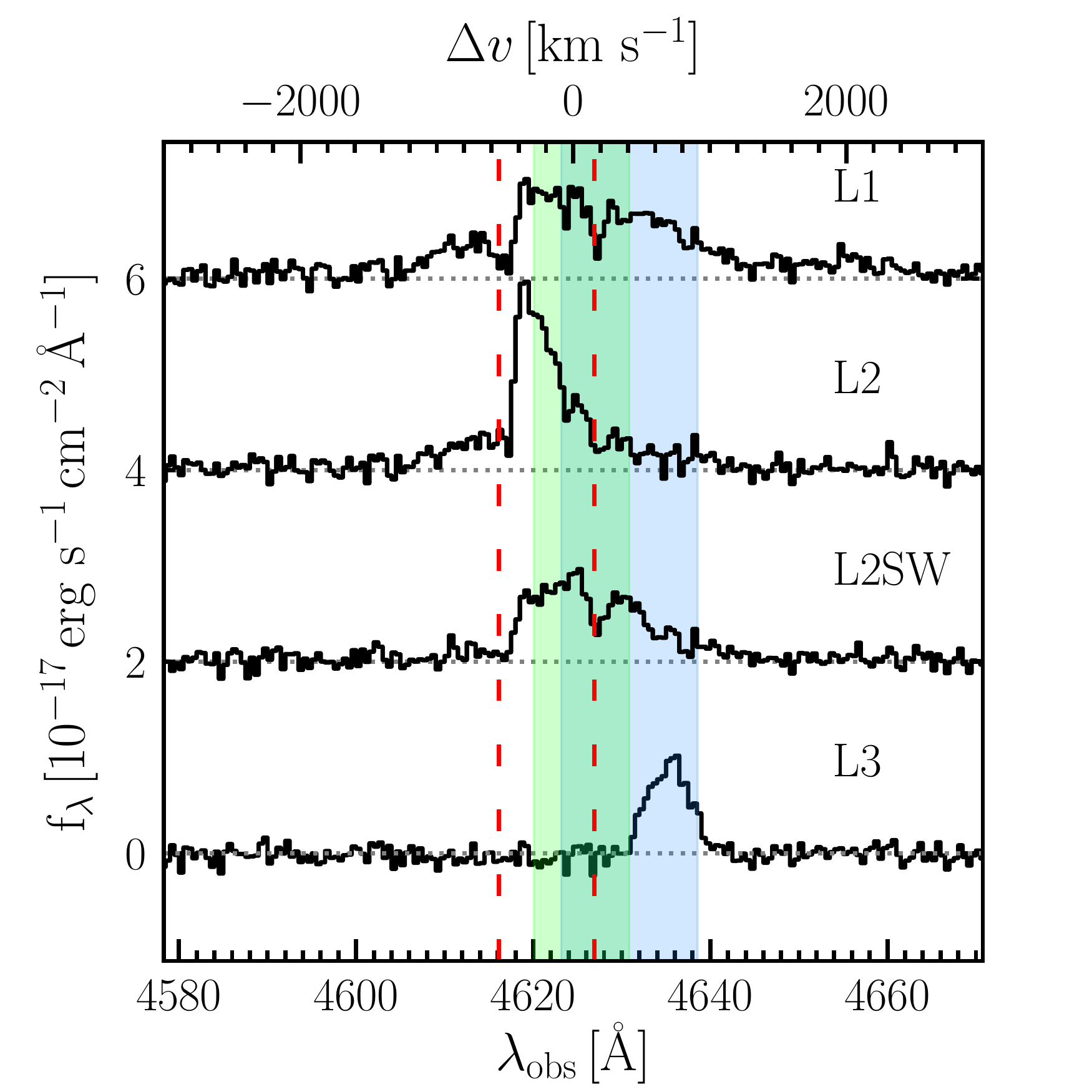}
    \caption{Ly$\alpha$ spectra observed in the direction of L1, L2, L2SW  and L3 by \citet{Li2019}.  The origin of the velocity scale, 
    computed for $z_{\rm ref}=2.8041$. 
    The velocity range highlighted in blue (resp. green) is that of the full velocity coverage of the 
    absorption in the direction of L2SW (resp. in the west of L1).  Note that the two velocity ranges overlap. The dashed red lines show the uniformity 
    across the field of the velocities of the narrow \Lya\ absorption troughs at $v \sim-550$ and $\sim$150 \kms.  Adapted from \citet{Li2019}.}
    \label{fig:Lyaspec}
\end{figure}

SMM\,J02399$-$0136  was known from the earliest spectroscopic observations with CFHT to be a \Lya\ emitter and the total extent of the emission was found to be 13 arcsec or $\sim 100$\,kpc \citep{Ivison1998,Vernet2001}. These galaxies
have recently been observed with the  Keck/KCWI  \citep{Li2019}  and an extended  \Lya\ nebula  has been imaged over $\sim  80$ kpc (Fig.~\ref{fig:Lyamap}). However, it is not quite as large and luminous as the \Lya\ nebulae
referred to as ELANe by \citet{Cai2017} and \cite{Arrigoni-Battaia2018}. The bright spot, L3, previously unknown, is the focus of \citet{Li2019} who propose that it is a dark cloud made visible by fluorescence excited by the BAL QSO, L1.
The spectra observed at the position of the starburst galaxy L2SW, the BAL QSO L1, the blue companion L2, and the dark cloud L3 are shown in Fig.~\ref{fig:Lyaspec}. 
The lines are highly non-Gaussian and asymmetrical 
close to the QSO. Towards L3, the line is narrow (FWHM=353 \kms) and close to Gaussian \citep{Li2019}.
Two narrow absorption troughs at $v \sim 150$ \kms\ and $-550$ \kms\ are also detected at the three positions L1, L2, L2SW and at the same velocities (Fig.~\ref{fig:Lyaspec}).

\begin{figure*}
		\includegraphics[width=\columnwidth]{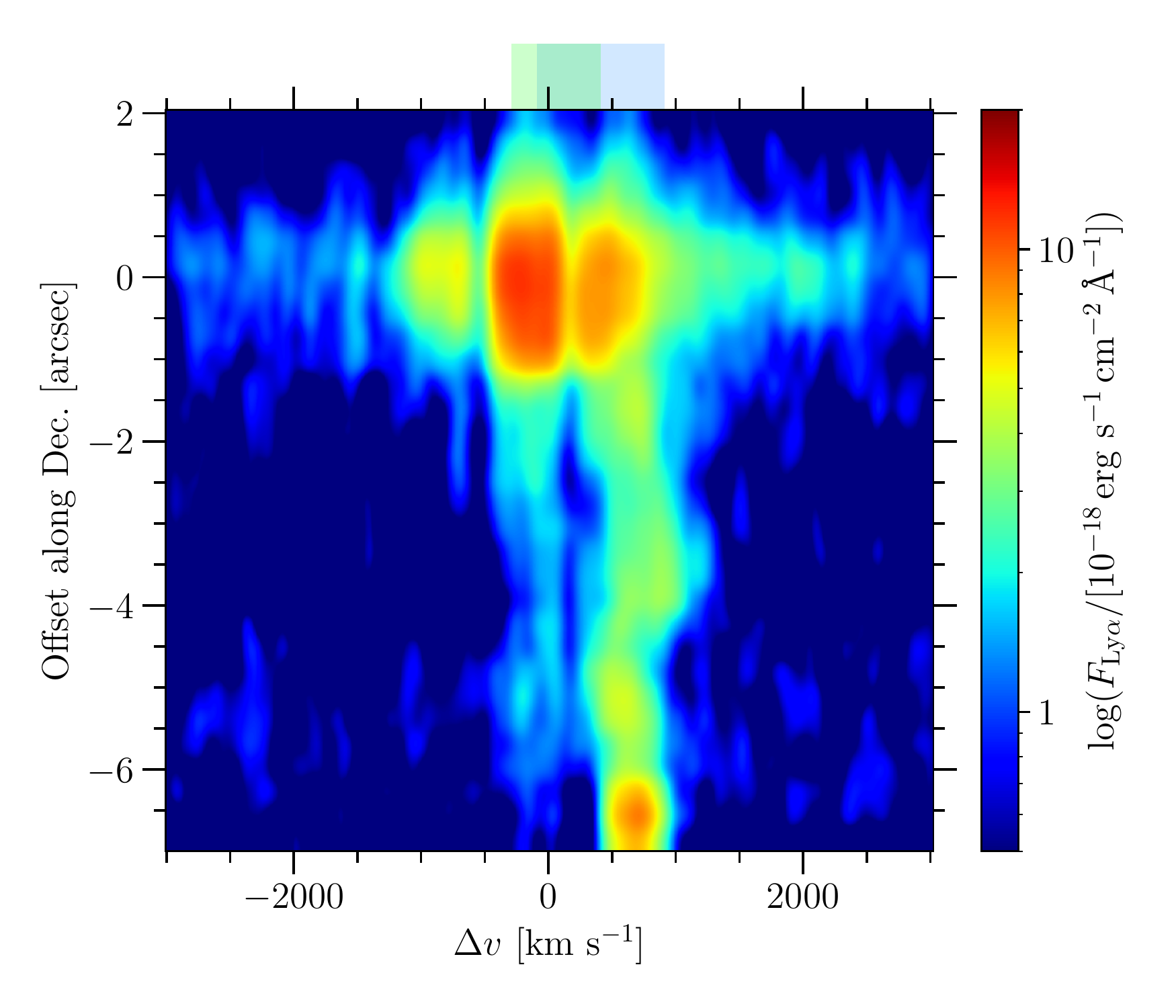}
		\includegraphics[width=\columnwidth]{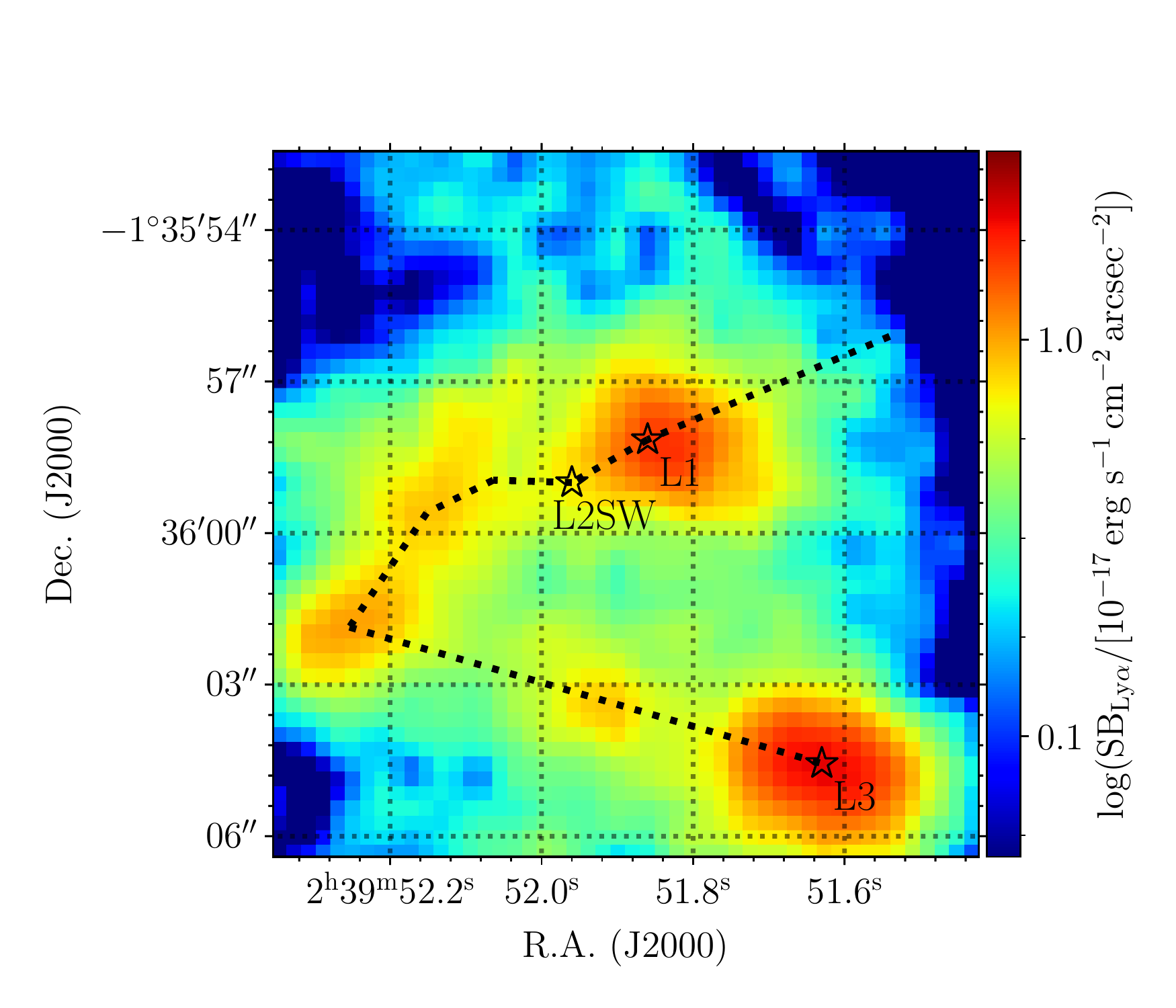}
    \caption{{\it Left:} Position-velocity map of the \Lya\ nebula taken along the track shown in the right panel (black dots). The origin of the velocity scale is $z_{\rm ref}=2.8041$. 
    The positions are labelled by their offset in declination with respect to L1. 
     Note that the extremely broad \Lya\ wings, visible from -3000 to +3000 \kms\ at the position of the BAL QSO L1 are unresolved by the Keck/KCWI observations because of the seeing. 
    The blue and green boxes at the top show the full velocity coverage of the \CHp\ absorption lines in front of L2SW and L1, respectively, as in Fig.  ~\ref{fig:Lyaspec}. {\it Right:} Another map of the continuum-subtracted \Lya\ emission  of the gas surrounding SMM\,J02399$-$0136 from \citet{Li2019} as Fig.~\ref{fig:Lyamap}, on which the positions used for the position-velocity plot of the left panel are shown. 
    The width of the band used to make this map is narrower than that used in Fig. ~\ref{fig:Lyamap} (5\,\AA\ over 4630--4635\AA\ instead of 30\AA), therefore encompassing only the redder wing of the \Lya\ line.}
    \label{fig:Lyapv}
\end{figure*}

The position-velocity diagram (Fig.~\ref{fig:Lyapv}) shows that the extremely broad \Lya\ lines (FWZI possibly larger than $\sim$~6000~\kms) are found only in the inner part of the nebula, indeed unresolved by the KCWI observations 
 which had a seeing of 1.5" \citep{Li2019}, i.e. the extent of the extremely broad \Lya\ emission in the position-velocity cut.
 
  In its more extended part, the lines are narrower and redshifted with respect 
 to $z_{\rm ref}=2.8041$ by up to $\sim 750$ \kms.
Remarkably, the average velocity across the southern part of the nebula does not change much.
At first sight, the \Lya\ lines cannot be interpreted simply in terms of outflows and inflows, as discussed in  \citet{Verhamme2006} and \citet{Dijkstra2017} except for the line in the direction of L1 
that has a redshifted centroid velocity, computed as its first moment, $C_{\rm 1,\Lya}=260 \pm 40 $\kms. This redshifted velocity centroid can be seen as a signature of outflowing gas 
due to the strong scattering of \Lya\ photons. However, an infalling CGM 
could contribute to the blue peak asymmetry as seen in the direction of L2 and to a lesser extent towards L1 and L2SW. The \Lya\ lines therefore bear the signatures of both outflows and inflows.
We show in the next section the unique power of combining  \CHp\ and \Lya\ spectroscopy to get some insight into the CGM gas dynamics on a 50\,kpc-scale.


 \section{Comparison of the \CHp(1-0) and \Lya\ perspectives}

Despite the vast differences between the gas phases emitting and absorbing the \CHp(1-0) and \Lya\ lines,
a number of straightforward and remarkable properties can be inferred from their comparison. 

\subsection{The \CHp\ absorptions and \Lya\ asymmetric line shape}

\subsubsection{Radius of the halo of diffuse molecular gas}

The radius inferred for the turbulent reservoir (TR) of diffuse molecular gas traced by \CHp(1-0) absorption depends linearly on the  duration of the starburst phase (Eq. (\ref{eq:radius})). 
This radius is drawn on Fig.~\ref{fig:Lyamap}
for two different values, $t_{\rm SB} = $50~Myr and 100~Myr  that bracket the value 66~Myr computed for SMM\,J02399$-$0136 (Sect 3.1.1) 
and correspond to the range 
inferred from the statistical studies of SMGs (e.g.~\citealt{Tacconi2008}). These timescales are uncertain, partly because of the unknown 
IMF in starburst galaxies \citep{Zhang2018}.
However, the broad agreement between the projected extent of the bright core of the \Lya\ nebula and that of the TR is remarkable.
 This agreement suggests that the CGM is (at least) biphase with a cool molecular phase of low density ($n_{\rm H} < 10^3$\cc) and $T \sim 10^2$\,K contributing 
 $\sim 3.5 \times 10^{10}$\msol\ (Table~\ref{tab:tabfitvalues}). 
The warm phase at $T \sim 10^4$\,K, the \Lya\ emitter, is possibly fully ionized.

 Even more significant, this agreement supports the theoretical basis of our size determination of a \CHp\ absorbing medium, relying on the turbulent dissipation rate and its scaling with size (Paper I). 
Indeed, this was the original motivation for imaging the  \Lya\ emission  of SMM\,J02399$-$0136 with the Keck/KCWI.

\subsubsection{Inflow signatures}

The range of velocities covered by the \CHp\ absorption lines against L2SW and L1 is highlighted in blue and green, respectively,  in Fig.~\ref{fig:Lyaspec}. All velocities have been computed 
with respect to $z_{\rm ref}= 2.8041$.
Together they cover most of the red side of the \Lya\ line.
As the \CHp\ absorption is redshifted, the cool molecular gas is inflowing towards the galaxies and is located between the galaxies and the observer.  
This diffuse molecular gas necessarily  contains a fraction of atomic hydrogen susceptible to absorb and scatter \Lya\ photons, most likely at the origin of the \Lya\ line asymmetry at the three positions, L1, L2 and L2SW.

This same velocity range is also highlighted in blue and green in the position-velocity cut of Fig.~\ref{fig:Lyapv}. It is very similar to that of the whole southern part of the \Lya\ nebula, to which L3 belongs.
Unless it is an unlikely coincidence, this means that the bulk of the imaged \Lya\ nebula is inflowing towards the galaxies because its velocity (and velocity dispersion) is the same as that of the gas causing the redshifted  \CHp(1-0) absorption. This redshifted gas, inflowing towards the galaxies,  scatters the \Lya\ photons back towards the galaxies, and contributes to the weakening of the observed red side of the \Lya\ line.

Remarkably, the \Lya\ absorption trough at $v\sim 150$ \kms\ falls at the same velocity as the \CHp\ absorption common to both L1 and L2SW sight-lines (Fig.~\ref{fig:Lyaspec}).
It thus likely corresponds to inflowing gas located between the SMGs and the observer.
The other narrow \Lya\ absorption at $v \sim -550$ \kms\  does not coincide with any \CHp\ absorption trough:
it might be caused by gas located {\it behind} the SMGs, and therefore also inflowing towards the galaxies, but we cannot rule out that 
it is \CHp-poor because metal-poor, reflecting inhomogeneities in the metallicity of inflowing matter.
In the former case, since the $-550$\kms\ trough has also to lie in front of the whole blue side of the \Lya\ profile, this blue side would be emitted by gas located not only behind the SMGs but also behind the gas causing the absorption at $-550$\kms, and its negative velocities would be the signature of the inflow of the rear part of the nebula onto the galaxies.
It is therefore plausible that the entire \Lya\ nebula be inflowing towards the galaxies at velocities of several $\times$ 100 \kms.

This redshift of the \CHp\ absorption lines, however, could be ascribed to a gas component unrelated to the galaxies.
This is unlikely because the \CHp\  profile in the direction of L2SW-em (Fig.~\ref{fig:3specGal}) has the characteristic shape of an inverse P-Cygni profile in which the emission and the absorption are linked by the radiative transfer of photons in infalling gas.  So, as long as \CHp\ is concerned, the gas responsible for the emission (i.e. from the UV-irradiated shocks) and that causing absorption (i.e. from the large-scale turbulent reservoir of diffuse molecular gas) are dynamically and radiatively coupled. 
The gas causing the \CHp\ absorption and the extended part of the \Lya\ 
nebula, at similar velocities and velocity dispersions, are most likely also dynamically linked to the system. 
We therefore are confident that the gas emitting the redshifted \Lya\ emission of the nebula is dynamically connected to the system, and is inflowing towards L1 and L2SW.

Last, the mere existence of this extended cool molecular phase moving at the same velocities as the warm phase emitting \Lya\ indicates that not only they are mixed together i.e. the CGM is multiphase, as said above, but also that they share the same dynamics. However, the \Lya\ emission probably extends beyond the limits of the KCWI image (Fig.~\ref{fig:Lyamap}) i.e. its radius is $> 9" \sim  50$ kpc in source plane. It is larger than the 
radius inferred for the diffuse molecular reservoir which suggests that the inflowing gas is molecular only within the inner regions, at distances $\lesssim 20$ kpc. 
The co-spatiality of these two phases further suggests that some thermal cooling radiation necessarily contributes to the \Lya\  line emission \citep{Gnat2012}.


\subsection{The broad \CHp\ and \Lya\ emission lines}

\begin{figure}
		\includegraphics[width=\columnwidth]{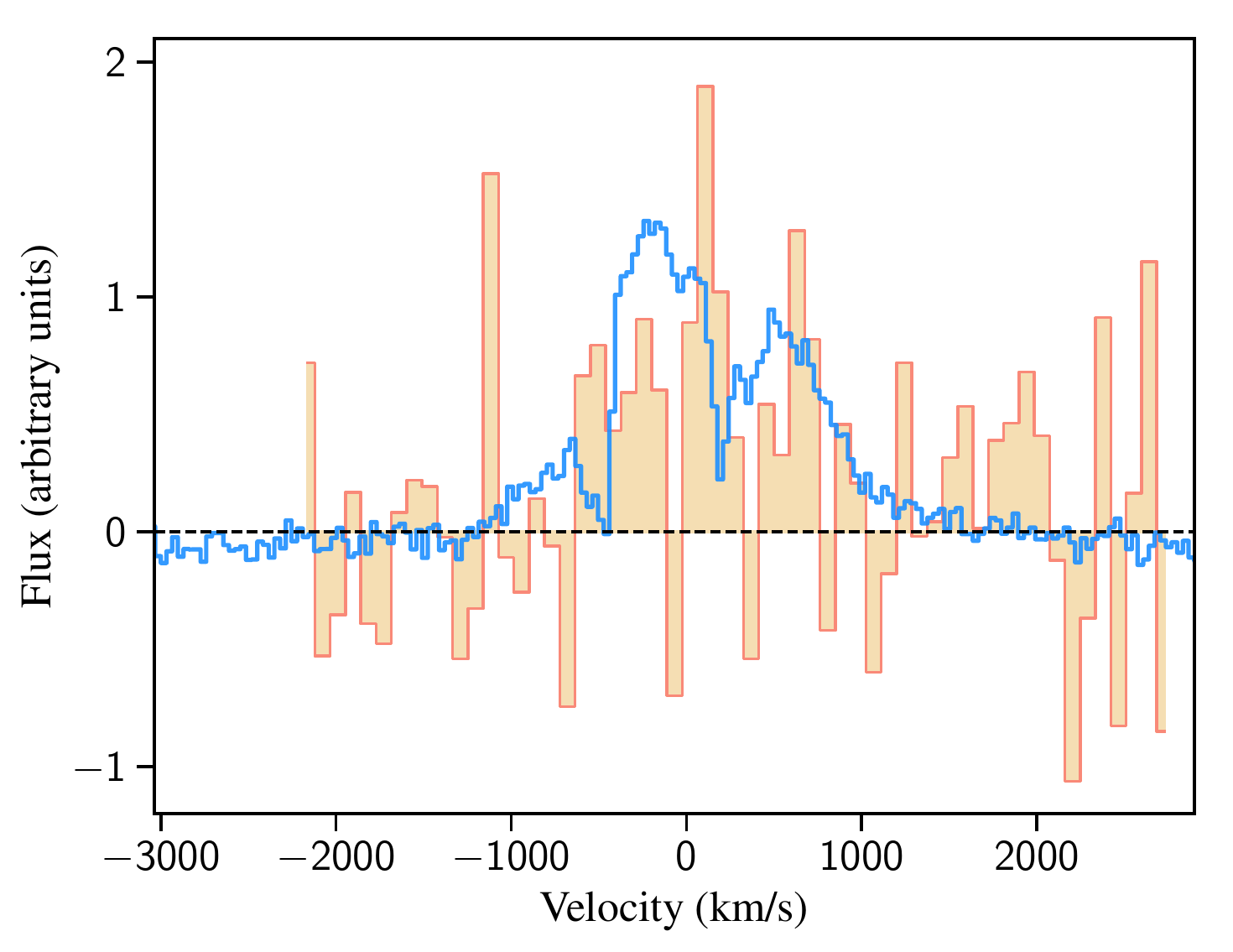}
    \caption{ Superposition (in arbitrary units) of the IRAM-30m \CHp(1-0) spectrum of SMM\,J02399$-$0136 at a spectral resolution of 64 MHz (or 87 \kms)  (beige)  and the KCWI \Lya\ spectrum integrated  over the IRAM-30m half-power beamwidth (HPBW=11.2") (blue). The common origin of the velocity scale is computed for  $z_{\rm ref}=2.8041$. }
    \label{fig:CHpLyaSpec}
\end{figure}

\begin{figure}
		\includegraphics[width=\columnwidth]{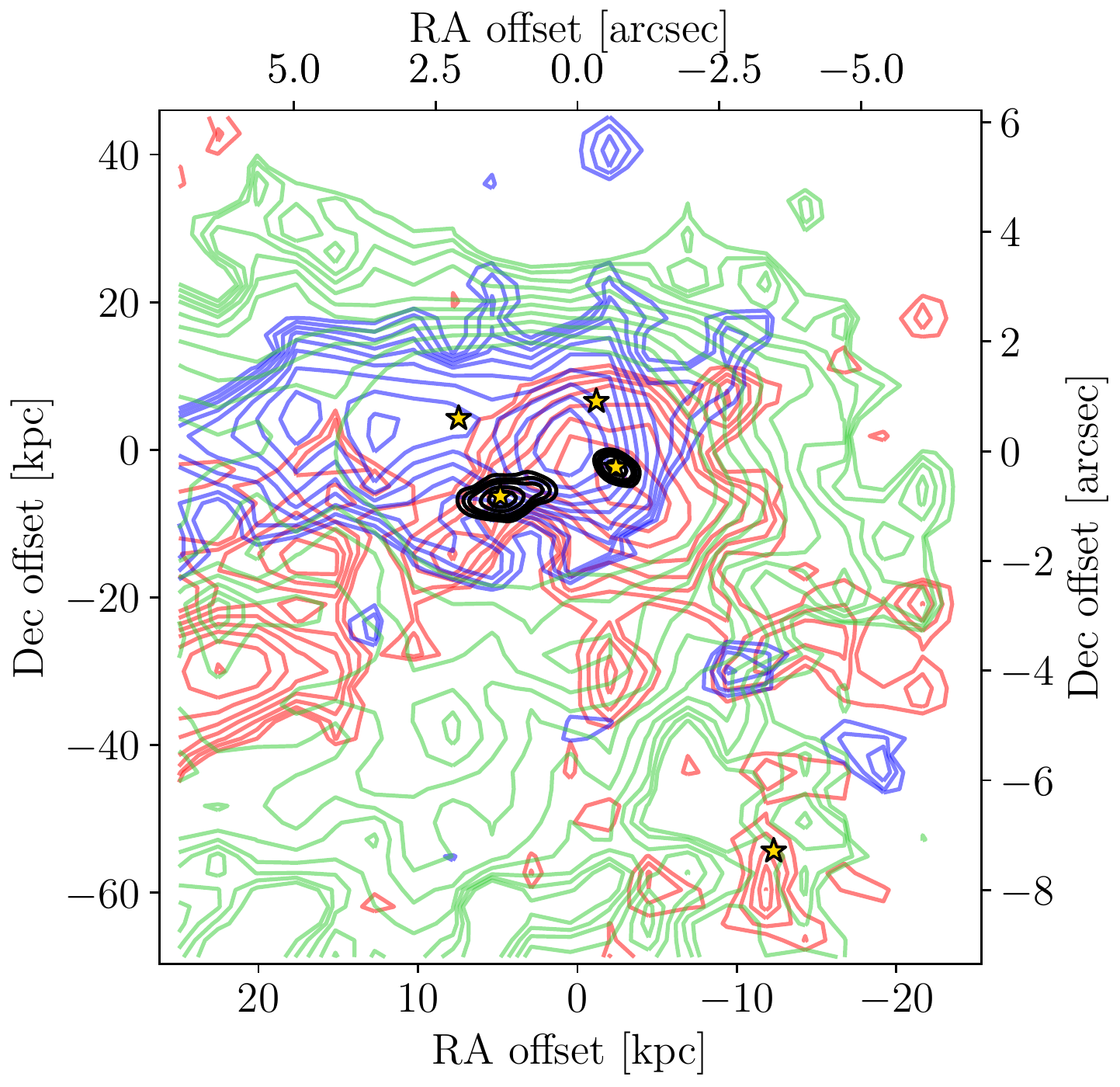}
    \caption{ Contours of the \Lya\ emission integrated over the velocity ranges [0,500] \kms\ (green), [1000,1500] \kms\ (red)  and [-1200,-700] \kms\ (blue). 
    The contours are at ${\rm log}_{10}(SB_\Lya/[10^{-17} {\rm erg} {\rm s}^{-1} {\rm cm}^{-2} {\rm arcsec}^{-2}])=-0.1\,(10-n)$ for the ten lowest levels, $n=$1 to 10, and 0.5 and 1 for the two highest, to emphasise the gradients at low brightness. 
     Black contours delineate the continuum sources and the four yellow stars indicate the positions of L1, L2SW  and L1N and L2. 
     }
    \label{fig:Lya-3vel}
\end{figure}

The \Lya\ emission has been integrated over the IRAM-30m HPBW=11.2" to compare its line profile to that of the \CHp\ emission (Fig.~\ref{fig:CHpLyaSpec}). 
As expected, the line cores differ because of the 
vastly different opacities of the two lines (the Einstein coefficients are respectively $A_{\rm 10}= 5.9 \times 10^{-3}$\,s$^{-1}$ for \CHp(1-0) and $A_{\rm \Lya}= 6.265 \times10^{8}$\,s$^{-1}$), 
but  the similarity of their full extent  (FWZI $\sim$ 3000\,\kms) is 
striking and suggests a shock contribution to the \Lya\ lines 
as done in the analysis of another conspicuous ELAN \citep{Cai2017} and as modeled in self-irradiated shocks \citep{Lehmann2020a,Lehmann2020b}. Extremely broad \Lya\  lines may therefore have a kinematic origin, in parallel to being due to opacity broadening \citep{Faucher-Giguere2010}.

 We detail below the rich facets of this similarity, in space and velocity space. 
The \Lya\ profiles are complex but their spatial distribution in three different velocity ranges (Fig.~\ref{fig:Lya-3vel}) is most informative. In the velocity range close to the \Lya\ line centroid ([0,500] \kms, green contours) there is a bright extended plateau with little spatial structure, encompassing the galaxies and co-located with the turbulent reservoir of diffuse gas causing the \CHp\ absorption (Fig. \ref{fig:Lyamap}).  
At the opposite, the contours in the high-velocity (HV) positive ([1000,1500]\,\kms, red) and negative ([-1200,-700]\,\kms, blue)  ranges are highly structured in space. 

Positive velocities are dominating the HV \Lya\ emission, consistent with primarily outflowing gas according to \Lya\ radiative transfer \citep{Verhamme2006,Dijkstra2017}. 
The geometry of these high positive velocities on each side of L1 and L2SW suggests high opening angles of the outflows.

\begin{figure*}
		\includegraphics[width=\textwidth]{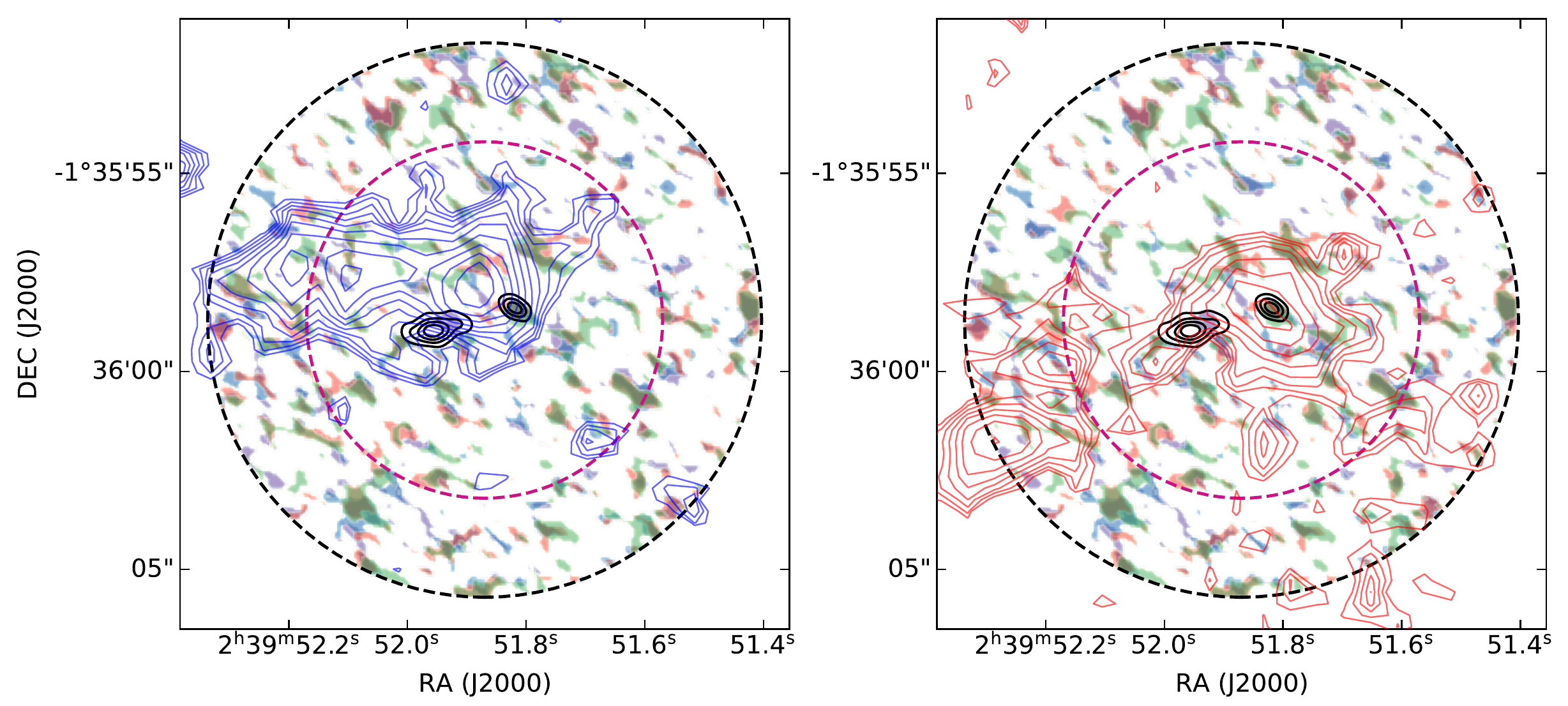}
    \caption{ Superposition of the  velocity-dependent shapes of the structures of \CHp\ emission (same as in Fig.\ref{fig:7specFar}) tentatively detected at the 
    1\,$\times \sigma_{\rm m0}$ level with contours of the \Lya\ emission integrated over the velocity ranges  [-1200,-700] \kms\ (blue) {\it (left panel)} and [1000,1500] \kms\ (red) {\it (right  panel)}. The  origin of the velocity scale is the systemic velocity of the galaxies at $z_{\rm ref}=2.8041$. The ALMA 1/3 primary beam is shown (black dashed circle) as well as the distance of 4.5 arcsec from the phase centre (dashed magenta circle).}
    \label{fig:11CHpLyaHV}
\end{figure*}

The overlay of the \Lya\ HV contours with the \CHp\ emission structures identified in the ALMA data suggests that these structures are not randomly distributed in space (Fig.~\ref{fig:11CHpLyaHV}). They follow the edges (i.e. brightness gradients) of the high-velocity contours, or are found at the tip of elongated \Lya\ structures. Half of those located within 4.5 arcsec of the phase centre lie in the environment of L1, at projected distances up to $\sim$ 15 kpc in the source plane (see Table\,\ref{tab:11spots}). Most of the least plausible candidates are clustered to the east of L2SW, where the positive and negative \Lya\ high-velocity contours overlap in projection, signposting complex dynamics.
Another view of this overlay is given in Appendix \,\ref{E}, where the spatial distribution of \Lya\ high-velocity emissions is compared to \CHp\ moment-0 maps computed on a spatially smoothed version of the ALMA data. Long and weak structures of \CHp\ line emission appear that exquisitely follow, in a few cases, the edges of the high-velocity \Lya\ emissions.

The spatial distribution of the \CHp\ emission structures with respect to the HV \Lya\ emissions suggests that they trace molecular shocks located at the thin and highly convoluted interface of the AGN- and starburst-driven outflows (i.e. the high-velocity components of  the \Lya\ line)  with the inflowing CGM.
 
\begin{figure*}
		\includegraphics[width=\textwidth]{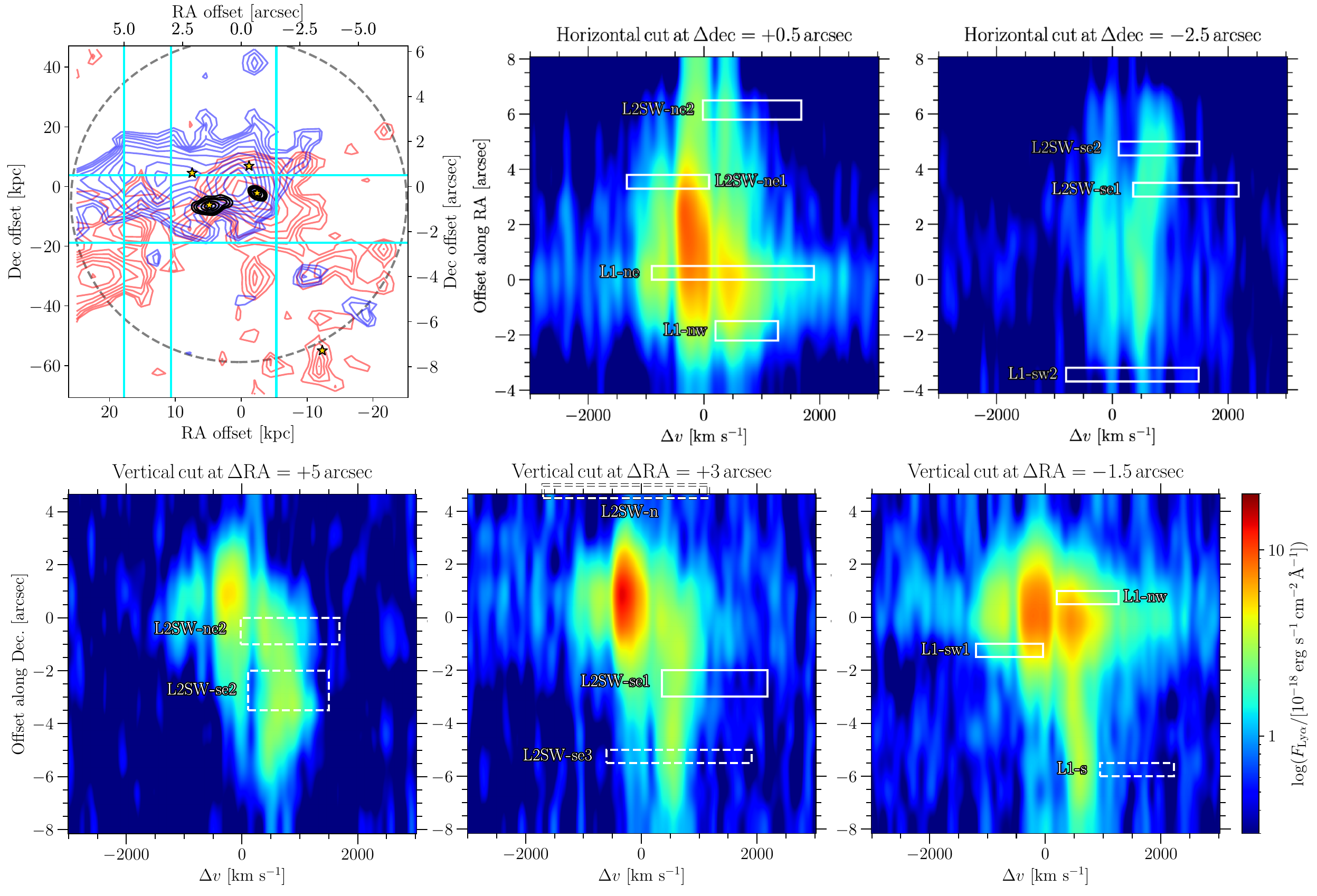}
    \caption{ Position-velocity cuts of the \Lya\ emission along five directions (shown in the upper left panel) crossing the scattered structures of \CHp\ emission. The position and extents in space (full size at the $\sigma_{\rm m0}$ level) and velocity (FWHM of the \CHp\ line) of each of them is displayed as a white box, next to each name. Dashed boxes mark the five least plausible structures listed in Table~\ref{tab:11spots}. Some of them are sampled in two different cuts. The common origin of the velocity scale is computed for  $z=2.8041$. }
    \label{fig:5LyaPV}
\end{figure*}

 The position-velocity cuts of Fig.\ref{fig:5LyaPV} offer another projection of the data. There, it is clear that the
structures of \CHp\ emission never share the position and velocity of \Lya\ peaks, but nevertheless always have a \Lya\ counterpart, albeit sometimes weak. This finding further supports the shock interpretation of the \CHp\ emission structures, shifted both in space and velocity from the \Lya\ peaks. It also highlights the weak contribution of these shocks to the \Lya\ emission of the nebula. The diversity of the relative positions of the shocks and \Lya\ emission is due mostly to projection effects. This comparison of the position-velocity distributions of \CHp\ and \Lya\ emission suggests that L2SW-ne2, L2SW-se2 and L2SW-se3 are plausible 
regions of \CHp\ emission, although located $\sim 5$\,arcsec (or $\sim$ 18 kpc) from the phase centre.  In the following, we adopt 18 kpc as the largest distance of the \CHp\ emission structures from the galaxies.

The observed fact that the broadest \Lya\ lines are found only in the unresolved brightest part of the nebula, in the vicinity of the BAL QSO L1, might appear in apparent contradiction with the findings of \citet{Laursen2009} who show that, in their simulations of forming galaxies, dust absorption primarily affects the wings of the \Lya\  lines. This is so because the wings are found to originate in the densest and most opaque parts of the galaxies.
It is not the case here and this suggests that the \Lya\  broad wings in SMM\,J02399$-$0136 are not originating in the most opaque regions of this system but in far less compact regions at distances $>3.5$ kpc from L1, at the interface of powerful galactic winds and the CGM, as suggested in Paper I.

 \section{Discussion}
 
 Thanks to the unique kinematic information carried out jointly by the \CHp\ and  \Lya\ lines, augmented by the ancillary CO data, we are now able to draw a coherent picture in which a large-scale inflow interacts with powerful AGN- and stellar-driven outflows, 
 and where the mass and energy injection rates from both the galactic winds and gravitational accretion sustain the co-existence over the starburst phase of the massive turbulent CGM and the high star formation rate.

 \subsection{What \CHp, \Lya\ and multi-transition CO observations tell us.}

 \begin{table*}
\begin{center}
\caption[]{\bf{Comparison of the kinematic properties of the low and high excitation phases seen by the \Lya, \CHp\ and CO lines set at the systemic redshift $z_{\rm ref}=2.8041$ (only the error bars relevant to the discussion are reported).}}
\bigskip
\smallskip
\label{tab:tab3}
\begin{tabular}{llllllllll}
\hline \noalign {\smallskip}
Line   & Telescope  & Beam &  $z_{\rm obs}$ &Source Component  & $v_{\rm abs}$ & $\Delta v_{\rm abs}$ & $v_{\rm em}$ & $\Delta v_{\rm em}$ & Reference \\
           &             &           arcsec     &                                            & \kms                   &   \kms      &  \kms          &  \kms &        \\
\hline \noalign {\smallskip}
 \Lya\     &     KCWI  &  1.5 (seeing) & 2.8048 &    L1              &                --          &    -- & $\approx$ 0  $^a$   &  $\approx$ 1500 &  b \\
             &             &                           &          &              L2SW                               &  --               &    --       &$\approx$ 0        &  $\approx$ 1500   &        \\
             &             &       &    &       L3                                      &     --          & --       &   {\bf 670$\pm$6}       &  {\bf  353$\pm$15 } &  b      \\
\CHp(1-0)     & ALMA    & 0.36$\times$0.59     &     2.803                   &L2SW  LV                  &        280             &   210     &       --         &   --      &     c     \\         
                   &                   &  &            & L2SW   HV                                        &     {\bf 660$\pm$70}           &   {\bf  290$\pm$160}      &    --            &    --     &          \\       
                    &           &       &         &   L2SW  LV+HV                                 &    430           &   580     &       --         &     --    &          \\       
                    &           &   &    &L2SW-w                                                       &    600           &  750      &   -10             &  1280       &          \\       
                      &     &    &  &   L1-sw                                                            &     80          &  290      &        10        &   1350      &          \\       
                   &       &     &   &  Scattered structures                                              &   --            &   -- &       -600 to 1600&  1000 to 3000 &  \\
                   &        &   & &      Average                                                       &    380$\pm$ 200           &  490      &         170 $\pm$ 540     &     1330$\pm$270   &          \\                 
                    & IRAM-30m & 11.2 &                                  &                      &       --        &    --    &        105$\pm$170        & 1300$\pm$500        &          \\   
  CO(1-0)    &   GBT     &  25 &  2.808 &        A                                       &        --       &   --     &   {\bf 637$\pm$15}              &    {\bf  260$\pm$35 }      &       d  \\       
                            &        &   &        &          B                                   &     --          &      --  &         57$\pm$35         &    660$\pm$100     &         \\       
   CO(7-6)                 & ALMA       & 0.6$\times$0.5   &                 &             L1, L2SW                                  &      --         &     --   &          0      &  $\approx$ 650      &     d,e     \\       
    CO(3-2)          & OVRO            &    7.4$\times$4.8             &     2.808                           &          &     --          &  --      &        307$\pm$158        &    710 $\pm$80    &       f  \\       
                           &  IRAM-PdBI      &  5.2$\times$2.4     &          2.808                               &           &     --          &    --    &      $\approx$     100      &  $\approx$900        &     g     \\        
                           &   IRAM-30m       & 22        &           2.803    &  &  -- & -- & 160$\pm$60 &  800$\pm$100 & b     \\   
                           & ALMA              &   0.72$\times$ 0.61 & 2.808 & W1 &         -- & -- &              {\bf  $\sim$650 }&   {\bf  $\sim$260} & d\\   
\hline \noalign {\smallskip}
\end{tabular}
\label{tab:tab3}
\end{center}
$^a$  The line is non-Gaussian and its extremely broad red wing makes its first moment positive (see text), $^b$ \cite{Li2019},
$^c$ This work,
$^d$ \cite{Frayer2018}, $^e$ From a sentence in \cite{Frayer2018} stating that the CO(7-6) lines detected by ALMA have a similar FWHM as the GBT component B in CO(1-0),  $^f$ \cite{Frayer1998}, $^g$ \cite{Genzel2003} 
\end{table*}

 The published CO observations of SMM\,J02399$-$01236 are a precious addition to the \CHp\ -- \Lya\ comparison  because they have been carried out in several transitions and at different angular resolutions.
For comparison, the velocities and FWHM of the \Lya, \CHp\ and CO lines are gathered in Table~\ref{tab:tab3} with the respective angular resolution of their observations.

The key kinematic properties of the different components seen in various tracers can be summarized as follows:

(1) The \Lya\ line in the direction of L3 and in the bulk of the \Lya\ nebula, the HV absorption line in the direction of L2SW, the CO(1-0) Green Bank Telescope (GBT) 
 component A and the CO(3-2) ALMA component W1, all have, within the error bars, the same redshifted velocity and about the same FWHM (in boldface in Table~\ref{tab:tab3}). These CO and \Lya\ components extend, up to 6"  and $\gtrsim 10"$ from the galaxies, respectively.    

(2) The extremely broad \Lya\ linewings in the direction of L1 and L2SW, the \CHp(1-0) emission in the close vicinity of the galaxies and the CO($J$=7-6) lines trace highly excited gas. 
They are centred at the same velocity, $\approx 0$ \kms\  suggesting that there is some shock contribution to the \Lya\ and CO(7-6) emissions associated with the close (kpc-scale) environment of the galaxies.
 We note that the high-$J$ CO line is about twice narrower than the \CHp\ emission lines. This can be understood by the fact that the abundance of \CHp\ is enhanced in strong UV 
 fields while CO is photo-dissociated \citep{Godard2019}.

(3) The  average velocity of the \CHp\ absorptions,  ${\overline v}_{\rm abs} = 380 \pm 200$ \kms\  is  commensurate with  those of the  \CHp\ emission lines 
 (Table~\ref{tab:11spots}), ${\overline v}_{\rm em} = 170 \pm 540 $\kms\  (resp. 412$\pm$ 520 \kms) for the  eight (resp. eleven) most plausible structures.

 As the similarities given in (1) are unlikely to be due to a coincidence, we infer that all these gas components, in addition to being projected over a similar large scale area, are sharing the same dynamics i.e. inflow towards the galaxies at $v_{\rm in} \sim 650$\kms\ because the redshifted HV \CHp\ absorption line in the direction of L2SW is seen in absorption. 
The inflowing stream is therefore multiphase within the inner $\sim$ 20 kpc, with a hint that the purely atomic (or ionized) hydrogen inflow extends even further than 50 kpc.
 A shock contribution to the \Lya\ emission is inferred from (2).
 
  Lastly, point (3), and the fact that the most plausible  \CHp\  emission structures are scattered within a large area around the galaxies 
  (up to distances $\sim$ 18\,kpc in the source frame), commensurate with the radius of the diffuse turbulent reservoir, $r_{\rm TR}=22\,  t_{\rm 66} \,{\rm kpc}$, 
 further support that the shocks traced by the \CHp\ emission lines and the diffuse, highly turbulent, molecular gas are dynamically coupled: the shocks are the signposts of the  mechanical fuelling of turbulence by the outflows within the whole CGM. 
 This was already suggested by the inverse-P Cygni profile tentatively detected 
 in the direction of L2SW-em (see Sect. 5.1).
 
The velocity coherence of the \Lya\ 
  extended component at $v_{\rm in} \sim 650$\kms\  over the full observed projected size of the \Lya\  nebula is striking (Figs. \ref{fig:Lyapv} and \ref{fig:5LyaPV}). Its linewidth varies however, mostly $\sim 500$ \kms, 
  much broader in the Eastern area with a linewidth $\sim 1000$ \kms\ (pv-cut at RA offset = +5 arcsec in Fig.~\ref{fig:5LyaPV}).
This velocity component is dominant in the western part of the galaxy environment seen in CO(3-2)  \citep{Genzel2003,Frayer2018}. 
 Lower velocity components, still inflowing (since redshifted  in absorption with respect to $z_{\rm ref}$), are also present,  e.g. the LV \CHp\ absorption component towards L2SW and the \CHp\ absorption in the direction of L1.
  These are also seen in \Lya\ emission  up to 5" (or $\approx 15$ kpc) eastward of L1 and L2SW (Fig. \ref{fig:Lyapv}). 
 This is why the average inflowing velocity is only $\overline{v}_{\rm in} = 380$ \kms\ (Table\,\ref{tab:tabfitvalues}).  
 
 These inflow velocities are large compared to what is expected in forming galaxies \citep{Goerdt2015a},  but recent improved numerical treatments of shock heating in the halo of massive galaxies show that large amounts of gas cool down and fall at such high velocities \citep{Bennett2020}. 
 We note that the escape velocity $v_{\rm esc}=(2GM_{\rm h}/R_{\rm h})^{1/2}$ at radius 200 kpc
 of a  halo of virial mass $M_{\rm h}= 5 \times 10^{12}$ \msol, is
 \begin{equation}
 v_{\rm esc}= 475 \kms\ \bigg( \frac{M_{\rm h}} {5 \times 10^{12} \msol}  \bigg)^{1/2} \bigg( \frac{R_{\rm h}} {200{\rm  kpc}}  \bigg)^{-1/2}.
 \end{equation}
So, the observed inflow velocites could then be the kinematic signature of large scale recycled winds, in which case outflows and inflows would be intimately mixed \citep{Herrera-Camus2020}. 
 Galactic winds do generate turbulence and a fraction of the expelled matter is susceptible, if enabled to cool down, to fall back onto the galaxies
 at large velocities ~\citep[e.g.][]{GaborBournaud2014}.
 
The turbulent velocity of the inflowing gas $\overline{v}_{\rm turb,TR}=340$\kms\  is also very large and traces highly supersonic turbulence, even in gas at $T\sim 10^4$\,K.
The processes susceptible to sustain this high level of large scale turbulence are AGN- and stellar-driven winds from L1 and L2SW  \citep{Faucher-Giguere2012,Costa2020,Nims2015} but also gravitational energy 
either from the  merger \citep{Oh2015,Blecha2018,Biernacki2018,Bustamante2018} -- the SMM\,J02399$-$0136 galaxy group being an archetypal example of a merger -- or recycled outflows or infalling cold streams \citep{Dekel2009,Narayanan2015}. This alternative may just be two facets of the same process since galaxies are already forming in accreting cold streams \citep[e.g.][]{Narayanan2015}. Not unexpectedly, \cite{Bennett2020} show that the improvement in the resolution of accretion shocks causes CGM turbulence to increase dramatically, up to values comparable to those observed here.

 The inflowing gas could be pristine gas infalling in the dark matter potential well of the galaxies, at large radii where only the \Lya\ line is detected i.e.  $\gtrsim$40 kpc from the galaxies. 
 The inner part, rich in CO and detected in \CHp\ is metal-enriched and could be part of recycled outflows.
We discuss these possibilities 
in the next sections, relying on the \CHp\ and \Lya\ line observations.

 \subsection{Where inflow meets outflows}

 The faster  \Lya\ outflow at  $v_{\rm out,1}>3000$ \kms, originating in L1, and to a lesser extent L2SW, is clearly visible in the position-velocity cut of Fig.~\ref{fig:Lyapv}. 
It is unresolved and is therefore confined within the inner 0.75" (or $\approx$ 3 kpc) environment of the galaxies.
We show below that the \CHp\ and \Lya\ emission lines make it possible to outline the fate of less powerful outflows, at  $v_{\rm out,2} < v_{\rm out,1}$ at distances at least as large as $\sim$ 20 kpc from the galaxies.
   
 The longest structures of broad and weak \CHp\ emission obtained in the smooth moment-0 maps (Fig.~\ref{fig:SmoothMom0}) are close to the maximum recoverable size of $\sim 5$~arcsec, i.e. $\sim$ 15 to $\sim$ 40 kpc in the source plane.  
Such large lengths and large velocity dispersions are reminiscent of the cold gas kinematic properties  observed in the 50-kpc long high-velocity shock (HVS) of the Stephan's Quintet (SQ) where the velocity 
 dispersion of the molecular gas in the post-shock layer of the HVS is that of  myriad low-velocity molecular shocks (LVMS).
 This large dispersion spans the full velocity difference between the colliders, i.e. the intruder galaxy and the HI tidal stream \citep{Guillard2012,Guillard2021}. 
 
 Here, the CGM into which the outflows penetrate is inflowing at the projected velocity ${\overline v}_{\rm in} = +380$ \kms. The HVS post-shock extent in velocity, 
 provided by the \CHp\ linewidths, ranges from 1000 to 3000 \kms\ (Table~\ref{tab:tab3}). From its average value, $\overline{\Delta v}_{\rm em} = 1330$ \kms, and 
 $\overline{\Delta v}_{\rm em} = \lvert v_{\rm in} -  v_{\rm out,2}\lvert $, we infer $v_{\rm out,2} \sim 1000$\,\kms\ with a large scatter due to projection effects. It is 
 remarkable that this estimate of a characteristic outflow velocity, based on \CHp\ line emission, falls in the ranges of the blue and red high velocity \Lya\ emissions (Fig.~\ref{fig:Lya-3vel}).
 Their complex geometry  prevents more accurate estimates. 
 The \CHp\   linewidths also yield the velocity of the large scale HVS that is 
 close to the relative velocity of the colliders, $v_{\rm HVS} \sim 1300$ \kms. 
 
Interestingly, the 122$\mu$m [NII] line was detected with the 10.4m telescope of the Caltech Submillimeter Observatory of same HPBW at that frequency as the IRAM-30m telescope \citep{Ferkinhoff2011} while almost resolved out in subsequent ALMA observations \citep{Ferkinhoff2015}. The similarity of the [NII] and \CHp\ velocities and velocity coverages over the same area supports an origin of the [NII] line in which nitrogen would be collisionally excited in the HVS shocks rather than photo-ionised by hard-UV photons.
  
 Now, the \CHp\  emission brings a new perspective on these kpc-scale HVS.
 Shocks at high Mach number, in which the magnetic field does not store a large fraction of the pre-shock kinetic energy, heat the post-shock medium to a temperature  
 \begin{equation} 
 T = \frac{3\mu_{\rm p} m_{\rm H}}{16 k_{\rm B}} v_{\rm sh}^2 \sim 5 \times 10^7 {\rm K} \bigg(\frac{v_{\rm sh}}{1300 \kms}\bigg)^2
 \end{equation}
  in a monoatomic gas with a mean molecular weight per particle $\mu_{\rm p}=1.3$ \citep{Spitzer1968}.
 This is equivalent to say that half the pre-shock kinetic energy density is transformed into thermal energy density in the post-shock gas. 
 
 It is not the case here since, as in the SQ shock, the molecules would not survive, nor reform fast enough, at such high temperatures. 
 The width of the \CHp\ lines is not thermal but has to trace fast turbulent motions of dense gas. 
Indeed, the mean turbulent velocity of the LVMS  within the post-shock layer of the HVS,  $0.7\, \overline{\Delta v}_{\rm em} \sim $ 900 \kms, 
 is commensurate with the sound velocity $c_{\rm S} \sim 750 $ \kms\  of the HVS post-shock layer heated at $\sim 5 \times 10^7$K
 indicating that these turbulent motions of the LVMS within this HVS post-shock layer are about trans-sonic.

The physics of such a turbulent cascade through gas phases of temperatures ranging from $\gtrsim 10^7$\,K to 100\,K  is still unknown although beautiful observations have started to 
stress the importance of turbulence, even in the hot intracluster gas \citep{Zhuravleva2014,Zhuravleva2019,Li2020}.   
The existence of dense gas within AGN-driven shocks is consistent with theoretical predictions \citep{Richings2018a,Richings2018b,Costa2020}. 
What is new here is that a fraction of the HVS kinetic energy 
is not thermalized at $T>10^7$ K, and therefore radiated in X-rays or highly excited lines of ions, but stored in the extremely fast trans-sonic motions of dense molecular gas  i.e. the post-shock layers of the LVMS, emitting the \CHp\ lines. 

The cascade of kinetic energy has to proceed up to the stage where the turbulent motions drive shocks at $v_{\rm sh} \sim 20 $\kms, enabled to efficiently form \CHp. It involves a broad distribution of shock velocities, from $v_{\rm sh} \sim 1300$ to 20 \kms, propagating in a variety of pre-shock densities and gas phases. 
   
In summary, although they are only individual tentative detections, 
the spatially scattered and broad \CHp\ emission lines,  when analysed in conjunction with the \Lya\ high-velocity field and morphology,  
 can be understood as tracing kpc-scale HVS at $\overline{v}_{\rm HVS}\sim$ 1300 \kms\  formed at  the interface of AGN- and stellar-driven outflows 
at  $\overline{v}_{\rm out,2} \sim 1000$ \kms\ and inflowing CGM at $\overline{v}_{\rm in} \sim 400$ \kms,  up to distances as large as $\sim 20$ kpc from the galaxies (in projection).  Interestingly, the standard deviation of their projected velocities, $\sigma_{\rm v_{em}} =$ 540 \kms\ (Table \ref{tab:11spots})  
corresponds to a mean turbulent velocity of their ensemble  $1.6 \sigma_{\rm v_{em}} \sim 900$ \kms, close to $\overline{v}_{\rm out,2}$.

 
\subsection{Following the energy trails: radiative, turbulent and gravitational  luminosities}

The above results confirm the multiphase nature of the cold CGM, including cool diffuse and cold dense molecular phases, in addition to the hot and warm phases, over large distances from the galaxies. 
They also disclose not only the high level of turbulence in the massive CGM but also how and where the stellar- and AGN-driven outflows inject kinetic energy into the CGM (i.e. the myriad shocks seen in \CHp\ emission). 

The two questions raised by these results are therefore: {\it (i)} are these outflows able to sustain 
the turbulence observed in such a massive cold CGM over the starburst phase duration? and 
{\it (ii)} are they able to compensate the mass drain of the cold CGM due to the high star formation rate? 

The cold phases of the CGM comprise the cool diffuse molecular gas of mass $M_{\rm TR}$ inferred from \CHp\ absorption and denser molecular gas 
of mass $M_{\rm CO}$ seen as extended CO(1-0) emission with the EVLA  \citep{Ivison2010a} and the GBT \citep{Frayer2018}. 
Its  turbulent luminosity is 
$L_{\rm turb,cCGM} = \frac{1}{2} M_{\rm TR} \,  \overline{v}_{\rm turb,TR}^2\, / \,  t_{\rm dyn} +  \frac{1}{2} M_{\rm CO} \,  \overline{v}_{\rm turb,CO}^2\, / \,  t_{\rm dyn}$,
where the dynamical time $ t_{\rm dyn}$ should be similar for the two components.

For the diffuse gas, $M_{\rm TR}=3.7\times 10^{10} \, t_{66}^2$ \msol\ and  $ \overline{v}_{\rm turb,TR}=340$ \kms\  (Table~\ref{tab:tabfitvalues}) so that 
$L_{\rm turb,TR}= 2.2 \times 10^{43}\,  t_{66}$  \ergs\  with  $t_{66}=t_{\rm dyn}/66$\,Myr. 
The mass of the denser component extending over $\sim$ 25 kpc, $M_{\rm CO}= 1.7 \times 10^{11}$\,\msol, is the mass of the low-excitation component that contributes 75\% of the total mass seen in CO(1-0). We adopt the linewidth measured by \cite{Frayer2018} for the low-excitation CO emission, 
$\Delta v_{\rm CO} =260$ \kms, so that $\overline{v}_{\rm turb,CO}= 182$\kms,  and 
 $L_{\rm turb,CO} \sim  2.8  \times 10^{43} t_{66}^{-1}$ \ergs\ . Note that $t_{\rm dyn}=68$~Myr for that low-excitation CO component.
The turbulent luminosity of the cool and cold CGM (hereafter cold CGM) is therefore  
$L_{\rm turb,cCGM} \sim 5 \times 10 ^{43} $ \ergs\ for $t_{66}=1$.

We now estimate the mechanical luminosity of the stellar- and AGN-driven outflows from the luminosity of the \CHp($J$=1-0) emission lines, relying on models of UV-irradiated molecular shocks \citep{Godard2019}.
The \CHp($J$=1-0) integrated line flux of all the shocks detected by IRAM-30m, $S_{\CHp} \,\Delta v= 4.7$ Jy \kms\ (see Sect. 3.2), corresponds to a 
line luminosity, $\mu \, L_{\CHp(1-0)}= 2.3\times 10^{42}$ \ergs,
at the luminosity distance of SMM\,J02399$-$0136, $D_{\rm L}= 24.0$ Gpc \citep{Solomon1997}. As in the UV-irradiated shock models of \cite{Godard2019}, the $J=1-0$ transition is found to contribute about 20$\%$ of the energy radiated by all the \CHp\  rotational transitions, 
the total \CHp\  line luminosity,  $L_{\CHp} \sim 6.5 \times 10^{42}$ \ergs\ (after correction for lensing), is as large as $\sim$ 5\% of the  \Lya\ luminosity, 
$L_\Lya = 1.4\times 10^{44}$~\ergs\ \citep{Li2019}. 
In these shock models, the total flux radiated in the \CHp\  lines follows the kinetic energy flux entering the shocks, with different ratios 
$f_{\CHp, {\rm kin}}=L_{\CHp}/L_{\rm kin,sh}$ depending  on the pre-shock densities, the shock velocities and 
UV-irradiations.  For shock velocities $v_{\rm sh} \geq 10$ \kms\ and an UV-irradiation $G_0 =10^4$, this ratio increases sharply between 5$\times 10^{-4}$ and $10^{-2}$ for pre-shock densities between $ 5 \times 10^4$ and $10^5$\,\cc\  that are the most appropriate conditions to explain the \CHp\ line intensities (Appendix \ref{D}).
Estimates of this ratio for molecular shocks at higher velocities will be published in a forthcoming paper (Lehmann et al., in prep.)  but these higher velocity shocks are J-shocks, much less efficient to form \CHp.  

The UV-irradiation of the CGM at distances of several $\times$10 kpc from the galaxies is difficult to estimate. However, 
values as high as $G_0 = 10^4$ or larger, even at distances $d \sim 20$\,kpc from L1, are expected in the case of  SMM\,J02399$-$0136 given the illumination 
of the dark-cloud L3 located at $\sim 60$\,kpc from L1 \citep{Li2019}.
The bolometric  luminosity of SMM\,J02399$-$0136 is $L_{\rm bol}=1.2 \times 10^{13}$ \lsol\ \citep{Ivison2010a}.
As the AGN contributes 42\% of the total FIR luminosity and assuming that the difference between $L_{\rm bol}$ and $L_{\rm FIR}$ is entirely due to the AGN, we infer $L_{\rm bol,AGN}=2.8\times 10^{46}$ \ergs. The UV-luminosity below  912\AA\ is $\sim 0.25 \times L_{\rm bol}$ \citep{Li2019}, so that the flux of UV photons  expressed in Habing units\footnote{Habing flux integrated between 912\AA\ and 2066\AA, $F_{\rm \lambda} \Delta  \lambda=1.3\times10^{-4}$ \ergs\,cm$^{-2}$ sr $^{-1}$} at  a distance $d_{\rm 20}$ of 20\,kpc from L1 depends only on the solid angle $\Omega_{\rm free}$ of the quasi dust-free lanes opened up by the AGN-driven winds and responsible from the large anisotropy of UV photons around QSOs: 
\begin{equation} 
G_0=1.5 \times 10^5 \bigg( \frac{\Omega _{\rm free}}{0.1} \bigg)^{-1} \, d_{\rm 20}^{-2}.
\end{equation}
We note that \cite{Li2019} find $\Omega_{\rm free} \sim 0.07$ for the UV-irradiation in the direction of the L3 dark-cloud at 60\,kpc from L1.
 The UV-luminosity of the AGN alone is therefore sufficient to provide $G_0 \geq10^4$ over several  regions located at the tip of dust-free lanes around L1, to which the contribution of the L2SW starburst should be added. The anisotropy of the UV field at large scale around these galaxies is supported by the 
scattered structure of the visible spots in Fig.\ref{fig:HST}.
 
In the following, we therefore adopt $G_0=10^4$ as a lower limit of the shock UV-irradiation in the CGM at distances $\sim 20$\,kpc from the galaxies 
and the upper limit 
$f_{\CHp, {\rm kin}}=10^{-2}$ that provides a lower limit to the kinetic luminosity of the ensemble of LVMS,  
$ L_{\rm kin,sh} =L_{\CHp}/ f_{\CHp, {\rm kin}} >  6.5 \times 10^{44}$ \ergs.
 This is an immense kinetic luminosity that signposts the terminal dissipative step of the energy cascade triggered, within their post-shock layers, by the HVS. Assuming mass and energy flux conservation between HVS and the myriad LVMS and neglecting the radiative losses in the warm gas that are much smaller that the  total \Lya\ luminosity, 
 a massive molecular outflow rate is inferred from  $L_{\rm kin,sh}= \frac{1}{2}\dot{M}_{\rm out} \, v_{\rm out ,2}^2$: 
 \begin{equation}
 \dot{M}_{\rm out} > 2 \times 10^3 \msol\ {\rm yr}^{-1} \bigg(\frac{v_{\rm out,2}}{10^3 \kms}\bigg)^{-2} 
 \end{equation}
 leading to a mass loading factor $\dot{M}_{\rm out}/{\rm SFR} >2.4$ in the range of those observed for high-$z$ starburst galaxies \citep{Veilleux2020}.
 The spatial pattern of the high-velocity \Lya\  emissions (Fig.~\ref{fig:Lya-3vel}) supports a joint AGN and starburst driving of the galactic outflows.  Indeed, \cite{Biernacki2018} show that  only the non-linear coupling of AGN and supernovae feedbacks is able to produce massive and extended outflows.
 
As the AGN contribution to the kinetic luminosity is necessarily smaller than the above value, 
we find that $L_{\rm kin}^{\rm AGN} <  0.02 \, L_{\rm bol,AGN}$, an upper limit broadly consistent with the values  
inferred from numerical simulations of AGN-driven large-scale outflows, $L_{\rm kin}=0.004\, L_{\rm AGN}$ \citep{Costa2020}. 

Even taking into account the $\sim 10$\% of the shock energy dissipated in the LVMS and radiated in the \HH\ lines \citep{Godard2019}, the kinetic luminosity estimated above, $ L_{\rm kin,sh}>   6.5 \times 10^{44}$ \ergs, is more than 10 $\times$  larger than that needed to sustain the cold CGM turbulence within a radius $\sim 20$ kpc and over the starburst lifetime.  
The mass outflow rate sufficient to feed the cold CGM turbulent luminosity, $L_{\rm turb,cCGM} = 5 \times 10 ^{43}$ \,\ergs, 
is therefore much smaller than $\dot{M}_{\rm out}$:
 \begin{equation}
\dot{M}_{\rm out,cCGM} \sim  150 \msol\ {\rm yr}^{-1} \bigg( \frac{v_{\rm out ,2}}{10^3 \kms} \bigg)^{-2}.
 \end{equation}
 
 A net mass inflow rate into the cold CGM,  $ \dot M_{\rm in,cCGM} = {\rm SFR} - \dot{M}_{\rm out,cCGM} \sim  720 \msol {\rm yr }^{-1}$
 is required to reach a steady-state where the cold CGM mass drain due to star formation is compensated jointly by the outflows and the gas inflow.  
This net mass inflow to the CGM of the galaxy group could be due to the gas tidal streams of the galaxy merger.
 However, in numerical simulations, tidal streams due to mergers are not 
 found as massive and extended as what is observed here, and are most often seen as outflows, not inflows  \citep{Narayanan2006}.
 The net mass inflow could also be due to gas expelled further away than $\sim $ 20 kpc from the galaxies by the powerful galactic winds, falling back onto the inner CGM if it succeeds at cooling efficiently. Last, it could be due to cold stream accretion onto the host halo of this galaxy group. 
  
 We therefore speculate that the missing mass contribution, $\dot M_{\rm in,CGM}  \sim 720 \msol {\rm yr }^{-1}$, whatever its origin, is due to net gas accretion into the inner gravitational potential well of the galaxy group and that the gravitational cooling rate from this infalling material is an additional energy source of the cold CGM turbulence. 
 The gravitational luminosity estimated at a representative distance $d$ from the galaxies 
\begin{equation}
L_{\rm grav} = \frac{G \dot M_{\rm in,CGM} M_{\rm stars}}{d}
\end{equation}
for $M_{\rm stars}=7\times 10^{11}$ \msol\ \citep{Aguirre2013}, 
appears to be also comparable to all the other luminosities: 
\begin{equation}
L_{\rm grav} = 3.7\times 10^{43} \ergs\ \frac{\dot M_{\rm in,CGM}}{720 \msol\ {\rm yr}^{-1}} \frac{M_{\rm stars}}{7 \times 10^{11} \msol} \bigg(\frac{d}{40 {\rm kpc}}\bigg)^{-1}
\end{equation}
where the infalling time from distance $\sim$ 40 kpc at $v_{\rm in}= 400 $\kms\ is still  commensurate with the estimated duration of the starburst phase.
 The gravitational energy is therefore another contribution to the feeding of  turbulence of the massive CGM, and as such, it is also eventually lost by radiation in the cool phase, locally heated by turbulent dissipation and producing \CHp\ in the diffuse CGM.  

These results reveal the critical role of turbulence in reprocessing the different sources of energy, were they AGN- and starburst-driven outflows or large-scale gravitational energy of infalling streams, down to the kinetic and thermal energy of the cool phase, in which this energy is eventually radiated and lost.

All the above rates are order-of-magnitude estimates in agreement with the model of \cite{Bouche2010}. They show that, 
 although a merger is  accelerating the process of star formation in this system, 
 the star formation and black-hole growth are  most likely fuelled, ultimately, by large-scale accretion. 
It is not possible, at this stage to distinguish the origin of the infalling gas, namely cold stream accretion or recycled outflows.
Any infalling gas of low metallicity may just be indistinguishable in the inner tens of kpc of 
the CGM because it is mixed with powerful outflows and metal-enriched by small-scale turbulent mixing, 
 as shown in numerical simulations \cite[e.g.][]{Shen2013}. 
 As a final note, it is fascinating to observers that the theoretical predictions for the total mass inflow rate onto the host halo \citep{Dekel2009}, and the numerical results for both the smooth and merger-based baryon accretion efficiencies at redshift $z=2.8$  \citep{Wright2020}, $\dot M_{\rm grav} = 740 \msol\ {\rm yr}^{-1} $
be so close to our estimated infall rate $\dot M_{\rm in,cCGM}$ for a halo mass $M_{\rm h} = 4.5 \times 10^{12}$\,\msol, and a cosmological baryon fraction
$f_b =0.165$.  If not due to chance, this would suggest that the contribution of recycled outflows to the gas infall in this galaxy group is low.


 
\section{Summary and open questions}

Unexpectedly, the 
\CHp\ and  \Lya\  lines observed, at high angular and spectral resolution, towards the galaxy group SMM\,J02399$-$0136, 
provide coherent and complementary information on the dynamics of the multiphase CGM. 
The main results are summarized as follows:

\begin{itemize}
\item
The co-existence of the two cool and warm thermal phases up to distances $\sim$ 20\,kpc in the CGM of the galaxies validates, in this case, the critical assumptions 
made to interpret the \CHp\ absorption lines detected against the continuum emission of several starburst galaxies (Paper I). It supports the turbulent framework itself and the link between 
the \CHp\ abundance in diffuse gas and the turbulent dissipation rate. 
\item
Inflow of the massive, multiphase, highly turbulent CGM towards the galaxies  at an average velocity $\overline{v}_{\rm in} \sim 380$ \kms\ is ascertained, from both the redshifted \CHp\ absorption 
with respect to the velocity of the galaxies and the detailed comparison of the \Lya, \CHp\ and CO line profiles. 
\item 
Multiple kpc-scale shocks are tentatively detected as broad \CHp\ emission lines 
($\Delta v_{\rm em} = 1300$ \kms\ on average) at the interface of the inflowing CGM and the high-velocity \Lya\ emission of galactic winds from the AGN and the starburst galaxy. 
They are scattered in the CGM, up to $\sim 20 $~kpc from the galaxies.
They trigger turbulent cascades in the post-shock gas, generating a hierarchy of shocks, from the high-velocity shocks (HVS) at $\gtrsim$ 1000 \kms\ down to the low-velocity molecular shocks at $\gtrsim 20$ \kms,  where \CHp\ can form efficiently.
The HVS kinetic energy density is not fully thermalized in the post-shock layer.
\item
 While the immense kinetic luminosity of the shock ensemble signposts powerful AGN- and starburst-driven outflows 
likely enabled to eject considerable amounts of gas out of the galaxies potential well, only $\sim 10$\% 
of this kinetic luminosity is required to sustain turbulence in the cold CGM. However, a net mass accretion, at a rate 
$\dot M_{\rm in,cCGM} \sim 720$ \msol\ yr$^{-1}$ is required to balance the mass
drain of the cold CGM due to star formation. The loss of gravitational energy of this infalling gas is an additional energy source for the CGM turbulence. 
\end{itemize}

 These results, obtained in the galaxy group of SMM\,J02399$-$0136, illustrate the invaluable complementarity of the \Lya\ lines and molecular line tracers, as different as high-$J$ CO and \CHp\ lines, 
 in unraveling not only the complex physics of the turbulent, multiphase CGM, but also the disheartening intricacy of the \Lya\ radiative transfer.
 They suggest that there is thermal cooling and shock contributions to the \Lya\ emission.

Because  of a conjunction of three unrelated properties, i.e. its highly endothermic formation, its UV-driven abundance enhancement and its high dipole moment, \CHp\ is a very unique molecular species that provides 
a wealth of clues inaccessible to other tracers. The \CHp\ lines do not only trace the presence of diffuse molecular gas, of density marginally able to excite the low-$J$ CO lines, but they also highlight the bursts of dissipation 
of mechanical energy, either in emission in UV-irradiated shocks where CO is photo-dissociated, or in absorption, on the sites of intermittent dissipation of turbulence in the cool diffuse gas. 

These results  reveal the critical role of turbulence in channeling energy and momentum across gas phases and scales, drastically modifying the timescales over which the energy is lost by radiation.
The amount of energy stored in fast turbulent motions of cool gas and eventually radiated in low-excitation lines has probably been broadly underestimated. 

 Lastly, some fine tuning in terms of mass and energy seems to be at work in this galaxy group between the AGN and stellar feedback, the gravitational infall and the star formation rate. 
 While the starburst phase is triggered by the galaxy merger, the steady-state suggested by the existence of the massive turbulent reservoir of cool gas, i.e. the cold CGM,  requires large-scale accretion
 
 How is the gravitational potential energy shared between the turbulent and thermal energy of the 
warm and cool gas in the CGM remains an open issue that could be addressed with future observations of the cooling radiation of the diffuse molecular gas (i.e. the [CII] line,  -- unfortunately not accessible 
from ground-based telescopes in this source -- and the 
pure rotational lines of \HH. A similar question opens up for the prodigious shock kinetic energy at the interface of galactic outflows and the CGM,  with the additional complexity borne by the hot gas and the multiphase nature of the post-shock gas.


\section*{Acknowledgements}
We are grateful to the referee for their thorough comments that led us to explore new facets of this study.
We thank Dr Vinod Arumugam for creating Fig. 1.
This paper makes use of the data under project numbers: ADS/JAO.ALMA$\#$2016.1.00282.S (ALMA), U128 (Keck/KCWI) and 086-19 (IRAM 30-m). They are all publicly available in the corresponding archives. ALMA is a partnership of ESO (representing its member states), NSF (USA) and NINS (Japan), together with NRC (Canada) and NSC and ASIAA (Taiwan), in cooperation with the Republic of Chile. The Joint ALMA Observatory is operated by ESO, AUI/NRAO and NAOJ. NRAO is a facility of the NSF operated under cooperative agreement by Associated Universities, Inc..
IRAM is supported by INSU/CNRS (France), MPG (Germany) and IGN (Spain).
W. M. Keck Observatory is operated as a scientific partnership among the California Institute of Technology, the University of California and the National Aeronautics and Space Administration.
A.V.-G., E.F. and B.G. acknowledge support from the European Research Council Advanced Grant MIST (No 742719, PI: E. Falgarone).  



\bibliographystyle{mnras}
\bibliography{MyBiblio2010} 

\begin{thebibliography}{}
\makeatletter
\relax
\def\mn@urlcharsother{\let\do\@makeother \do\$\do\&\do\#\do\^\do\_\do\%\do\~}
\def\mn@doi{\begingroup\mn@urlcharsother \@ifnextchar [ {\mn@doi@}
  {\mn@doi@[]}}
\def\mn@doi@[#1]#2{\def\@tempa{#1}\ifx\@tempa\@empty \href
  {http://dx.doi.org/#2} {doi:#2}\else \href {http://dx.doi.org/#2} {#1}\fi
  \endgroup}
\def\mn@eprint#1#2{\mn@eprint@#1:#2::\@nil}
\def\mn@eprint@arXiv#1{\href {http://arxiv.org/abs/#1} {{\tt arXiv:#1}}}
\def\mn@eprint@dblp#1{\href {http://dblp.uni-trier.de/rec/bibtex/#1.xml}
  {dblp:#1}}
\def\mn@eprint@#1:#2:#3:#4\@nil{\def\@tempa {#1}\def\@tempb {#2}\def\@tempc
  {#3}\ifx \@tempc \@empty \let \@tempc \@tempb \let \@tempb \@tempa \fi \ifx
  \@tempb \@empty \def\@tempb {arXiv}\fi \@ifundefined
  {mn@eprint@\@tempb}{\@tempb:\@tempc}{\expandafter \expandafter \csname
  mn@eprint@\@tempb\endcsname \expandafter{\@tempc}}}

\bibitem[\protect\citeauthoryear{{Aguirre}, {Baker}, {Menanteau}, {Lutz}  \&
  {Tacconi}}{{Aguirre} et~al.}{2013}]{Aguirre2013}
{Aguirre} P.,  {Baker} A.~J.,  {Menanteau} F.,  {Lutz} D.,   {Tacconi} L.~J.,
  2013, \mn@doi [\apj] {10.1088/0004-637X/768/2/164}, \href
  {https://ui.adsabs.harvard.edu/abs/2013ApJ...768..164A} {768, 164}

\bibitem[\protect\citeauthoryear{{Ag{\'u}ndez}, {Goicoechea}, {Cernicharo},
  {Faure}  \& {Roueff}}{{Ag{\'u}ndez} et~al.}{2010}]{Agundez2010}
{Ag{\'u}ndez} M.,  {Goicoechea} J.~R.,  {Cernicharo} J.,  {Faure} A.,
  {Roueff} E.,  2010, \mn@doi [\apj] {10.1088/0004-637X/713/1/662}, \href
  {http://adsabs.harvard.edu/abs/2010ApJ...713..662A} {713, 662}

\bibitem[\protect\citeauthoryear{{Amano}}{{Amano}}{2010}]{Amano2010}
{Amano} T.,  2010, \mn@doi [\apjl] {10.1088/2041-8205/716/1/L1}, \href
  {http://adsabs.harvard.edu/abs/2010ApJ...716L...1A} {716, L1}

\bibitem[\protect\citeauthoryear{{Appleton} et~al.,}{{Appleton}
  et~al.}{2013}]{Appleton2013}
{Appleton} P.~N.,  et~al., 2013, \mn@doi [\apj] {10.1088/0004-637X/777/1/66},
  \href {http://adsabs.harvard.edu/abs/2013ApJ...777...66A} {777, 66}

\bibitem[\protect\citeauthoryear{{Appleton} et~al.,}{{Appleton}
  et~al.}{2017}]{Appleton2017}
{Appleton} P.~N.,  et~al., 2017, \mn@doi [\apj] {10.3847/1538-4357/836/1/76},
  \href {http://cdsads.u-strasbg.fr/abs/2017ApJ...836...76A} {836, 76}

\bibitem[\protect\citeauthoryear{{Arrigoni Battaia}, {Prochaska}, {Hennawi},
  {Obreja}, {Buck}, {Cantalupo}, {Dutton}  \& {Macci{\`o}}}{{Arrigoni Battaia}
  et~al.}{2018}]{Arrigoni-Battaia2018}
{Arrigoni Battaia} F.,  {Prochaska} J.~X.,  {Hennawi} J.~F.,  {Obreja} A.,
  {Buck} T.,  {Cantalupo} S.,  {Dutton} A.~A.,   {Macci{\`o}} A.~V.,  2018,
  \mn@doi [\mnras] {10.1093/mnras/stx2465}, \href
  {http://adsabs.harvard.edu/abs/2018MNRAS.473.3907A} {473, 3907}

\bibitem[\protect\citeauthoryear{{Bennett} \& {Sijacki}}{{Bennett} \&
  {Sijacki}}{2020}]{Bennett2020}
{Bennett} J.~S.,  {Sijacki} D.,  2020, \mn@doi [\mnras]
  {10.1093/mnras/staa2835}, \href
  {https://ui.adsabs.harvard.edu/abs/2020MNRAS.499..597B} {499, 597}

\bibitem[\protect\citeauthoryear{{Berta} et~al.,}{{Berta}
  et~al.}{2021}]{Berta2021}
{Berta} S.,  et~al., 2021, \mn@doi [\aap] {10.1051/0004-6361/202039743}, \href
  {https://ui.adsabs.harvard.edu/abs/2021A&A...646A.122B} {646, A122}

\bibitem[\protect\citeauthoryear{{Bielby} et~al.,}{{Bielby}
  et~al.}{2020}]{Bielby2020}
{Bielby} R.~M.,  et~al., 2020, \mn@doi [\mnras] {10.1093/mnras/staa546}, \href
  {https://ui.adsabs.harvard.edu/abs/2020MNRAS.493.5336B} {493, 5336}

\bibitem[\protect\citeauthoryear{{Biernacki} \& {Teyssier}}{{Biernacki} \&
  {Teyssier}}{2018}]{Biernacki2018}
{Biernacki} P.,  {Teyssier} R.,  2018, \mn@doi [\mnras] {10.1093/mnras/sty216},
  \href {https://ui.adsabs.harvard.edu/abs/2018MNRAS.475.5688B} {475, 5688}

\bibitem[\protect\citeauthoryear{{Blecha}, {Snyder}, {Satyapal}  \&
  {Ellison}}{{Blecha} et~al.}{2018}]{Blecha2018}
{Blecha} L.,  {Snyder} G.~F.,  {Satyapal} S.,   {Ellison} S.~L.,  2018, \mn@doi
  [\mnras] {10.1093/mnras/sty1274}, \href
  {https://ui.adsabs.harvard.edu/abs/2018MNRAS.478.3056B} {478, 3056}

\bibitem[\protect\citeauthoryear{{Bouch{\'e}} et~al.,}{{Bouch{\'e}}
  et~al.}{2010}]{Bouche2010}
{Bouch{\'e}} N.,  et~al., 2010, \mn@doi [\apj] {10.1088/0004-637X/718/2/1001},
  \href {http://adsabs.harvard.edu/abs/2010ApJ...718.1001B} {718, 1001}

\bibitem[\protect\citeauthoryear{{Bouch{\'e}}, {Murphy}, {Kacprzak},
  {P{\'e}roux}, {Contini}, {Martin}  \& {Dessauges-Zavadsky}}{{Bouch{\'e}}
  et~al.}{2013}]{Bouche2013}
{Bouch{\'e}} N.,  {Murphy} M.~T.,  {Kacprzak} G.~G.,  {P{\'e}roux} C.,
  {Contini} T.,  {Martin} C.~L.,   {Dessauges-Zavadsky} M.,  2013, \mn@doi
  [Science] {10.1126/science.1234209}, \href
  {https://ui.adsabs.harvard.edu/abs/2013Sci...341...50B} {341, 50}

\bibitem[\protect\citeauthoryear{{Bussmann} et~al.,}{{Bussmann}
  et~al.}{2013}]{Bussmann2013}
{Bussmann} R.~S.,  et~al., 2013, \mn@doi [\apj] {10.1088/0004-637X/779/1/25},
  \href {http://adsabs.harvard.edu/abs/2013ApJ...779...25B} {779, 25}

\bibitem[\protect\citeauthoryear{{Bustamante}, {Sparre}, {Springel}  \&
  {Grand}}{{Bustamante} et~al.}{2018}]{Bustamante2018}
{Bustamante} S.,  {Sparre} M.,  {Springel} V.,   {Grand} R. J.~J.,  2018,
  \mn@doi [\mnras] {10.1093/mnras/sty1692}, \href
  {https://ui.adsabs.harvard.edu/abs/2018MNRAS.479.3381B} {479, 3381}

\bibitem[\protect\citeauthoryear{{Cai} et~al.,}{{Cai} et~al.}{2017}]{Cai2017}
{Cai} Z.,  et~al., 2017, \mn@doi [\apj] {10.3847/1538-4357/aa5d14}, \href
  {http://adsabs.harvard.edu/abs/2017ApJ...837...71C} {837, 71}

\bibitem[\protect\citeauthoryear{{Carilli} \& {Walter}}{{Carilli} \&
  {Walter}}{2013}]{Carilli2013}
{Carilli} C.~L.,  {Walter} F.,  2013, \mn@doi [\araa]
  {10.1146/annurev-astro-082812-140953}, \href
  {http://cdsads.u-strasbg.fr/abs/2013ARA%26A..51..105C} {51, 105}

\bibitem[\protect\citeauthoryear{{Carter} et~al.,}{{Carter}
  et~al.}{2012}]{Carter2012}
{Carter} M.,  et~al., 2012, \mn@doi [\aap] {10.1051/0004-6361/201118452}, \href
  {https://ui.adsabs.harvard.edu/abs/2012A&A...538A..89C} {538, A89}

\bibitem[\protect\citeauthoryear{{Costa}, {Pakmor}  \& {Springel}}{{Costa}
  et~al.}{2020}]{Costa2020}
{Costa} T.,  {Pakmor} R.,   {Springel} V.,  2020, \mn@doi [\mnras]
  {10.1093/mnras/staa2321}, \href
  {https://ui.adsabs.harvard.edu/abs/2020MNRAS.497.5229C} {497, 5229}

\bibitem[\protect\citeauthoryear{{Dav{\'e}}, {Crain}, {Stevens}, {Narayanan},
  {Saintonge}, {Catinella}  \& {Cortese}}{{Dav{\'e}} et~al.}{2020}]{Dave2020}
{Dav{\'e}} R.,  {Crain} R.~A.,  {Stevens} A. R.~H.,  {Narayanan} D.,
  {Saintonge} A.,  {Catinella} B.,   {Cortese} L.,  2020, \mn@doi [\mnras]
  {10.1093/mnras/staa1894}, \href
  {https://ui.adsabs.harvard.edu/abs/2020MNRAS.497..146D} {497, 146}

\bibitem[\protect\citeauthoryear{{Dekel} et~al.,}{{Dekel}
  et~al.}{2009}]{Dekel2009}
{Dekel} A.,  et~al., 2009, \mn@doi [\nat] {10.1038/nature07648}, \href
  {http://adsabs.harvard.edu/abs/2009Natur.457..451D} {457, 451}

\bibitem[\protect\citeauthoryear{{Diego}, {Broadhurst}, {Wong}, {Silk}, {Lim},
  {Zheng}, {Lam}  \& {Ford}}{{Diego} et~al.}{2016}]{Diego2016}
{Diego} J.~M.,  {Broadhurst} T.,  {Wong} J.,  {Silk} J.,  {Lim} J.,  {Zheng}
  W.,  {Lam} D.,   {Ford} H.,  2016, \mn@doi [\mnras] {10.1093/mnras/stw865},
  \href {https://ui.adsabs.harvard.edu/abs/2016MNRAS.459.3447D} {459, 3447}

\bibitem[\protect\citeauthoryear{{Dijkstra}}{{Dijkstra}}{2017}]{Dijkstra2017}
{Dijkstra} M.,  2017, arXiv e-prints, \href
  {http://cdsads.u-strasbg.fr/abs/2017arXiv170403416D} {}

\bibitem[\protect\citeauthoryear{{Douglas} \& {Herzberg}}{{Douglas} \&
  {Herzberg}}{1941}]{Douglas1941}
{Douglas} A.~E.,  {Herzberg} G.,  1941, \mn@doi [\apj] {10.1086/144342}, \href
  {http://adsabs.harvard.edu/abs/1941ApJ....94..381D} {94, 381}

\bibitem[\protect\citeauthoryear{{Draine}}{{Draine}}{1986}]{Draine1986}
{Draine} B.~T.,  1986, \mn@doi [\apj] {10.1086/164694}, \href
  {http://adsabs.harvard.edu/abs/1986ApJ...310..408D} {310, 408}

\bibitem[\protect\citeauthoryear{{Eales} et~al.,}{{Eales}
  et~al.}{2010}]{Eales2010}
{Eales} S.,  et~al., 2010, \mn@doi [\pasp] {10.1086/653086}, \href
  {http://adsabs.harvard.edu/abs/2010PASP..122..499E} {122, 499}

\bibitem[\protect\citeauthoryear{{Faisst} et~al.,}{{Faisst}
  et~al.}{2020}]{Faisst2020}
{Faisst} A.~L.,  et~al., 2020, \mn@doi [\apjs] {10.3847/1538-4365/ab7ccd},
  \href {https://ui.adsabs.harvard.edu/abs/2020ApJS..247...61F} {247, 61}

\bibitem[\protect\citeauthoryear{{Falgarone}, {Hily-Blant}  \&
  {Levrier}}{{Falgarone} et~al.}{2004}]{Falgarone2004}
{Falgarone} E.,  {Hily-Blant} P.,   {Levrier} F.,  2004, \mn@doi [\apss]
  {10.1023/B:ASTR.0000045004.70345.21}, \href
  {https://ui.adsabs.harvard.edu/abs/2004Ap&SS.292...89F} {292, 89}

\bibitem[\protect\citeauthoryear{{Falgarone} et~al.,}{{Falgarone}
  et~al.}{2017}]{Falgarone2017}
{Falgarone} E.,  et~al., 2017, \mn@doi [\nat] {10.1038/nature23298}, \href
  {http://adsabs.harvard.edu/abs/2017Natur.548..430F} {548, 430}

\bibitem[\protect\citeauthoryear{{Faucher-Gigu{\`e}re} \&
  {Quataert}}{{Faucher-Gigu{\`e}re} \& {Quataert}}{2012}]{Faucher-Giguere2012}
{Faucher-Gigu{\`e}re} C.-A.,  {Quataert} E.,  2012, \mn@doi [\mnras]
  {10.1111/j.1365-2966.2012.21512.x}, \href
  {https://ui.adsabs.harvard.edu/abs/2012MNRAS.425..605F} {425, 605}

\bibitem[\protect\citeauthoryear{{Faucher-Gigu{\`e}re}, {Kere{\v{s}}},
  {Dijkstra}, {Hernquist}  \& {Zaldarriaga}}{{Faucher-Gigu{\`e}re}
  et~al.}{2010}]{Faucher-Giguere2010}
{Faucher-Gigu{\`e}re} C.-A.,  {Kere{\v{s}}} D.,  {Dijkstra} M.,  {Hernquist}
  L.,   {Zaldarriaga} M.,  2010, \mn@doi [\apj] {10.1088/0004-637X/725/1/633},
  \href {https://ui.adsabs.harvard.edu/abs/2010ApJ...725..633F} {725, 633}

\bibitem[\protect\citeauthoryear{{Federman}, {Rawlings}, {Taylor}  \&
  {Williams}}{{Federman} et~al.}{1996}]{Federman1996}
{Federman} S.~R.,  {Rawlings} J.~M.~C.,  {Taylor} S.~D.,   {Williams} D.~A.,
  1996, \mnras, \href {http://adsabs.harvard.edu/abs/1996MNRAS.279L..41F} {279,
  L41}

\bibitem[\protect\citeauthoryear{{Ferkinhoff} et~al.,}{{Ferkinhoff}
  et~al.}{2011}]{Ferkinhoff2011}
{Ferkinhoff} C.,  et~al., 2011, \mn@doi [\apjl] {10.1088/2041-8205/740/1/L29},
  \href {https://ui.adsabs.harvard.edu/abs/2011ApJ...740L..29F} {740, L29}

\bibitem[\protect\citeauthoryear{{Ferkinhoff}, {Brisbin}, {Nikola}, {Stacey},
  {Sheth}, {Hailey-Dunsheath}  \& {Falgarone}}{{Ferkinhoff}
  et~al.}{2015}]{Ferkinhoff2015}
{Ferkinhoff} C.,  {Brisbin} D.,  {Nikola} T.,  {Stacey} G.~J.,  {Sheth} K.,
  {Hailey-Dunsheath} S.,   {Falgarone} E.,  2015, \mn@doi [\apj]
  {10.1088/0004-637X/806/2/260}, \href
  {https://ui.adsabs.harvard.edu/abs/2015ApJ...806..260F} {806, 260}

\bibitem[\protect\citeauthoryear{{Flower}, {Roueff}  \& {Zeippen}}{{Flower}
  et~al.}{1998}]{Flower1998}
{Flower} D.~R.,  {Roueff} E.,   {Zeippen} C.~J.,  1998, \mn@doi [Journal of
  Physics B Atomic Molecular Physics] {10.1088/0953-4075/31/5/017}, \href
  {http://adsabs.harvard.edu/abs/1998JPhB...31.1105F} {31, 1105}

\bibitem[\protect\citeauthoryear{{Frayer}, {Ivison}, {Scoville}, {Yun},
  {Evans}, {Smail}, {Blain}  \& {Kneib}}{{Frayer} et~al.}{1998}]{Frayer1998}
{Frayer} D.~T.,  {Ivison} R.~J.,  {Scoville} N.~Z.,  {Yun} M.,  {Evans} A.~S.,
  {Smail} I.,  {Blain} A.~W.,   {Kneib} J.-P.,  1998, \mn@doi [\apjl]
  {10.1086/311639}, \href {http://cdsads.u-strasbg.fr/abs/1998ApJ...506L...7F}
  {506, L7}

\bibitem[\protect\citeauthoryear{{Frayer} et~al.,}{{Frayer}
  et~al.}{1999}]{Frayer1999}
{Frayer} D.~T.,  et~al., 1999, \mn@doi [\apjl] {10.1086/311940}, \href
  {https://ui.adsabs.harvard.edu/abs/1999ApJ...514L..13F} {514, L13}

\bibitem[\protect\citeauthoryear{{Frayer}, {Maddalena}, {Ivison}, {Smail},
  {Blain}  \& {Vanden Bout}}{{Frayer} et~al.}{2018}]{Frayer2018}
{Frayer} D.~T.,  {Maddalena} R.~J.,  {Ivison} R.~J.,  {Smail} I.,  {Blain}
  A.~W.,   {Vanden Bout} P.,  2018, \mn@doi [\apj] {10.3847/1538-4357/aac49a},
  \href {http://cdsads.u-strasbg.fr/abs/2018ApJ...860...87F} {860, 87}

\bibitem[\protect\citeauthoryear{{Fu}, {Xue}, {Prochaska}, {Stockton},
  {Ponnada}, {Lau}, {Cooray}  \& {Narayanan}}{{Fu} et~al.}{2021}]{Fu2021}
{Fu} H.,  {Xue} R.,  {Prochaska} J.~X.,  {Stockton} A.,  {Ponnada} S.,  {Lau}
  M.~W.,  {Cooray} A.,   {Narayanan} D.,  2021, \mn@doi [\apj]
  {10.3847/1538-4357/abdb32}, \href
  {https://ui.adsabs.harvard.edu/abs/2021ApJ...908..188F} {908, 188}

\bibitem[\protect\citeauthoryear{{Fujimoto} et~al.,}{{Fujimoto}
  et~al.}{2019}]{Fujimoto2019}
{Fujimoto} S.,  et~al., 2019, arXiv e-prints, \href
  {https://ui.adsabs.harvard.edu/abs/2019arXiv190206760F} {p. arXiv:1902.06760}

\bibitem[\protect\citeauthoryear{{Gabor} \& {Bournaud}}{{Gabor} \&
  {Bournaud}}{2014}]{GaborBournaud2014}
{Gabor} J.~M.,  {Bournaud} F.,  2014, \mn@doi [\mnras] {10.1093/mnras/stu677},
  \href {https://ui.adsabs.harvard.edu/abs/2014MNRAS.441.1615G} {441, 1615}

\bibitem[\protect\citeauthoryear{{Gaspari}, {Ruszkowski}  \& {Oh}}{{Gaspari}
  et~al.}{2013}]{Gaspari2013}
{Gaspari} M.,  {Ruszkowski} M.,   {Oh} S.~P.,  2013, \mn@doi [\mnras]
  {10.1093/mnras/stt692}, \href
  {https://ui.adsabs.harvard.edu/abs/2013MNRAS.432.3401G} {432, 3401}

\bibitem[\protect\citeauthoryear{{Gaspari}, {Temi}  \& {Brighenti}}{{Gaspari}
  et~al.}{2017}]{Gaspari2017}
{Gaspari} M.,  {Temi} P.,   {Brighenti} F.,  2017, \mn@doi [\mnras]
  {10.1093/mnras/stw3108}, \href
  {https://ui.adsabs.harvard.edu/abs/2017MNRAS.466..677G} {466, 677}

\bibitem[\protect\citeauthoryear{{Gavazzi} et~al.,}{{Gavazzi}
  et~al.}{2011}]{Gavazzi2011}
{Gavazzi} R.,  et~al., 2011, \mn@doi [\apj] {10.1088/0004-637X/738/2/125},
  \href {https://ui.adsabs.harvard.edu/abs/2011ApJ...738..125G} {738, 125}

\bibitem[\protect\citeauthoryear{{Genzel}, {Baker}, {Tacconi}, {Lutz}, {Cox},
  {Guilloteau}  \& {Omont}}{{Genzel} et~al.}{2003}]{Genzel2003}
{Genzel} R.,  {Baker} A.~J.,  {Tacconi} L.~J.,  {Lutz} D.,  {Cox} P.,
  {Guilloteau} S.,   {Omont} A.,  2003, \mn@doi [\apj] {10.1086/345718}, \href
  {http://cdsads.u-strasbg.fr/abs/2003ApJ...584..633G} {584, 633}

\bibitem[\protect\citeauthoryear{{Gerin}, {Neufeld}  \& {Goicoechea}}{{Gerin}
  et~al.}{2016}]{Gerin2016}
{Gerin} M.,  {Neufeld} D.~A.,   {Goicoechea} J.~R.,  2016, \mn@doi [\araa]
  {10.1146/annurev-astro-081915-023409}, \href
  {http://cdsads.u-strasbg.fr/abs/2016ARA%26A..54..181G} {54, 181}

\bibitem[\protect\citeauthoryear{{Gnat} \& {Ferland}}{{Gnat} \&
  {Ferland}}{2012}]{Gnat2012}
{Gnat} O.,  {Ferland} G.~J.,  2012, \mn@doi [\apjs]
  {10.1088/0067-0049/199/1/20}, \href
  {https://ui.adsabs.harvard.edu/abs/2012ApJS..199...20G} {199, 20}

\bibitem[\protect\citeauthoryear{{Godard} \& {Cernicharo}}{{Godard} \&
  {Cernicharo}}{2013}]{Godard2013}
{Godard} B.,  {Cernicharo} J.,  2013, \mn@doi [\aap]
  {10.1051/0004-6361/201220151}, \href
  {http://adsabs.harvard.edu/abs/2013A%26A...550A...8G} {550, A8}

\bibitem[\protect\citeauthoryear{{Godard}, {Falgarone}  \& {Pineau des
  For{\^e}ts}}{{Godard} et~al.}{2009}]{Godard2009}
{Godard} B.,  {Falgarone} E.,   {Pineau des For{\^e}ts} G.,  2009, \mn@doi
  [\aap] {10.1051/0004-6361:200810803}, \href
  {http://adsabs.harvard.edu/abs/2009A%26A...495..847G} {495, 847}

\bibitem[\protect\citeauthoryear{{Godard}, {Falgarone}  \& {Pineau des
  For{\^e}ts}}{{Godard} et~al.}{2014}]{Godard2014}
{Godard} B.,  {Falgarone} E.,   {Pineau des For{\^e}ts} G.,  2014, \mn@doi
  [\aap] {10.1051/0004-6361/201423526}, \href
  {http://adsabs.harvard.edu/abs/2014A%26A...570A..27G} {570, A27}

\bibitem[\protect\citeauthoryear{{Godard}, {Pineau des For{\^e}ts}, {Lesaffre},
  {Lehmann}, {Gusdorf}  \& {Falgarone}}{{Godard} et~al.}{2019}]{Godard2019}
{Godard} B.,  {Pineau des For{\^e}ts} G.,  {Lesaffre} P.,  {Lehmann} A.,
  {Gusdorf} A.,   {Falgarone} E.,  2019, \mn@doi [\aap]
  {10.1051/0004-6361/201834248}, \href
  {http://adsabs.harvard.edu/abs/2019A%26A...622A.100G} {622, A100}

\bibitem[\protect\citeauthoryear{{Goerdt} \& {Ceverino}}{{Goerdt} \&
  {Ceverino}}{2015}]{Goerdt2015a}
{Goerdt} T.,  {Ceverino} D.,  2015, \mn@doi [\mnras] {10.1093/mnras/stv786},
  \href {https://ui.adsabs.harvard.edu/abs/2015MNRAS.450.3359G} {450, 3359}

\bibitem[\protect\citeauthoryear{{Gredel}, {van Dishoeck}  \& {Black}}{{Gredel}
  et~al.}{1993}]{Gredel1993}
{Gredel} R.,  {van Dishoeck} E.~F.,   {Black} J.~H.,  1993, \aap, \href
  {http://adsabs.harvard.edu/abs/1993A%26A...269..477G} {269, 477}

\bibitem[\protect\citeauthoryear{{Greve} et~al.,}{{Greve}
  et~al.}{2005}]{Greve2005}
{Greve} T.~R.,  et~al., 2005, \mn@doi [\mnras]
  {10.1111/j.1365-2966.2005.08979.x}, \href
  {https://ui.adsabs.harvard.edu/abs/2005MNRAS.359.1165G} {359, 1165}

\bibitem[\protect\citeauthoryear{{Gronke} \& {Oh}}{{Gronke} \&
  {Oh}}{2020}]{Gronke2020b}
{Gronke} M.,  {Oh} S.~P.,  2020, \mn@doi [\mnras] {10.1093/mnrasl/slaa033},
  \href {https://ui.adsabs.harvard.edu/abs/2020MNRAS.494L..27G} {494, L27}

\bibitem[\protect\citeauthoryear{{Guillard} et~al.,}{{Guillard}
  et~al.}{2012}]{Guillard2012}
{Guillard} P.,  et~al., 2012, \mn@doi [\apj] {10.1088/0004-637X/747/2/95},
  \href {http://adsabs.harvard.edu/abs/2012ApJ...747...95G} {747, 95}

\bibitem[\protect\citeauthoryear{{Guillard} et~al.,}{{Guillard}
  et~al.}{2021}]{Guillard2021}
{Guillard} P.,  et~al., 2021, arXiv e-prints, \href
  {https://ui.adsabs.harvard.edu/abs/2021arXiv210206843G} {p. arXiv:2102.06843}

\bibitem[\protect\citeauthoryear{{Hafen} et~al.,}{{Hafen}
  et~al.}{2019}]{Hafen2019}
{Hafen} Z.,  et~al., 2019, \mn@doi [\mnras] {10.1093/mnras/stz1773}, \href
  {https://ui.adsabs.harvard.edu/abs/2019MNRAS.488.1248H} {488, 1248}

\bibitem[\protect\citeauthoryear{{Herrera-Camus} et~al.,}{{Herrera-Camus}
  et~al.}{2020}]{Herrera-Camus2020}
{Herrera-Camus} R.,  et~al., 2020, \mn@doi [\aap]
  {10.1051/0004-6361/201937109}, \href
  {https://ui.adsabs.harvard.edu/abs/2020A&A...633L...4H} {633, L4}

\bibitem[\protect\citeauthoryear{{Hodge} \& {da Cunha}}{{Hodge} \& {da
  Cunha}}{2020}]{Hodge2020}
{Hodge} J.~A.,  {da Cunha} E.,  2020, \mn@doi [Royal Society Open Science]
  {10.1098/rsos.200556}, \href
  {https://ui.adsabs.harvard.edu/abs/2020RSOS....700556H} {7, 200556}

\bibitem[\protect\citeauthoryear{{Hopkins} et~al.,}{{Hopkins}
  et~al.}{2018}]{Hopkins2018}
{Hopkins} P.~F.,  et~al., 2018, \mn@doi [\mnras] {10.1093/mnras/sty1690}, \href
  {https://ui.adsabs.harvard.edu/abs/2018MNRAS.480..800H} {480, 800}

\bibitem[\protect\citeauthoryear{{Ikarashi} et~al.,}{{Ikarashi}
  et~al.}{2017}]{Ikarashi2017}
{Ikarashi} S.,  et~al., 2017, \mn@doi [\apjl] {10.3847/2041-8213/aa9572}, \href
  {https://ui.adsabs.harvard.edu/abs/2017ApJ...849L..36I} {849, L36}

\bibitem[\protect\citeauthoryear{{Indriolo}, {Bergin}, {Falgarone}, {Godard},
  {Zwaan}, {Neufeld}  \& {Wolfire}}{{Indriolo} et~al.}{2018}]{Indriolo2018}
{Indriolo} N.,  {Bergin} E.~A.,  {Falgarone} E.,  {Godard} B.,  {Zwaan} M.~A.,
  {Neufeld} D.~A.,   {Wolfire} M.~G.,  2018, \mn@doi [\apj]
  {10.3847/1538-4357/aad7b3}, \href
  {https://ui.adsabs.harvard.edu/abs/2018ApJ...865..127I} {865, 127}

\bibitem[\protect\citeauthoryear{{Ivison}, {Smail}, {Le Borgne}, {Blain},
  {Kneib}, {Bezecourt}, {Kerr}  \& {Davies}}{{Ivison}
  et~al.}{1998}]{Ivison1998}
{Ivison} R.~J.,  {Smail} I.,  {Le Borgne} J.-F.,  {Blain} A.~W.,  {Kneib}
  J.-P.,  {Bezecourt} J.,  {Kerr} T.~H.,   {Davies} J.~K.,  1998, \mn@doi
  [\mnras] {10.1046/j.1365-8711.1998.01677.x}, \href
  {http://cdsads.u-strasbg.fr/abs/1998MNRAS.298..583I} {298, 583}

\bibitem[\protect\citeauthoryear{{Ivison}, {Smail}, {Papadopoulos}, {Wold},
  {Richard}, {Swinbank}, {Kneib}  \& {Owen}}{{Ivison}
  et~al.}{2010a}]{Ivison2010a}
{Ivison} R.~J.,  {Smail} I.,  {Papadopoulos} P.~P.,  {Wold} I.,  {Richard} J.,
  {Swinbank} A.~M.,  {Kneib} J.~P.,   {Owen} F.~N.,  2010a, \mn@doi [\mnras]
  {10.1111/j.1365-2966.2010.16322.x}, \href
  {https://ui.adsabs.harvard.edu/abs/2010MNRAS.404..198I} {404, 198}

\bibitem[\protect\citeauthoryear{{Ivison} et~al.,}{{Ivison}
  et~al.}{2010b}]{Ivison2010b}
{Ivison} R.~J.,  et~al., 2010b, \mn@doi [\aap] {10.1051/0004-6361/201014548},
  \href {http://cdsads.u-strasbg.fr/abs/2010A%26A...518L..35I} {518, L35}

\bibitem[\protect\citeauthoryear{{Ivison}, {Papadopoulos}, {Smail}, {Greve},
  {Thomson}, {Xilouris}  \& {Chapman}}{{Ivison} et~al.}{2011}]{Ivison2011}
{Ivison} R.~J.,  {Papadopoulos} P.~P.,  {Smail} I.,  {Greve} T.~R.,  {Thomson}
  A.~P.,  {Xilouris} E.~M.,   {Chapman} S.~C.,  2011, \mn@doi [\mnras]
  {10.1111/j.1365-2966.2010.18028.x}, \href
  {https://ui.adsabs.harvard.edu/abs/2011MNRAS.412.1913I} {412, 1913}

\bibitem[\protect\citeauthoryear{{Johnson}, {Sharon}, {Bayliss}, {Gladders},
  {Coe}  \& {Ebeling}}{{Johnson} et~al.}{2014}]{Johnson2014}
{Johnson} T.~L.,  {Sharon} K.,  {Bayliss} M.~B.,  {Gladders} M.~D.,  {Coe} D.,
   {Ebeling} H.,  2014, \mn@doi [\apj] {10.1088/0004-637X/797/1/48}, \href
  {https://ui.adsabs.harvard.edu/abs/2014ApJ...797...48J} {797, 48}

\bibitem[\protect\citeauthoryear{{Kennicutt}}{{Kennicutt}}{1998}]{Kennicutt1998}
{Kennicutt} Robert~C. J.,  1998, \mn@doi [\apj] {10.1086/305588}, \href
  {https://ui.adsabs.harvard.edu/abs/1998ApJ...498..541K} {498, 541}

\bibitem[\protect\citeauthoryear{{Laursen}, {Sommer-Larsen}  \&
  {Andersen}}{{Laursen} et~al.}{2009}]{Laursen2009}
{Laursen} P.,  {Sommer-Larsen} J.,   {Andersen} A.~C.,  2009, \mn@doi [\apj]
  {10.1088/0004-637X/704/2/1640}, \href
  {https://ui.adsabs.harvard.edu/abs/2009ApJ...704.1640L} {704, 1640}

\bibitem[\protect\citeauthoryear{{Lehmann}, {Godard}, {Pineau des For{\^e}ts}
  \& {Falgarone}}{{Lehmann} et~al.}{2020a}]{Lehmann2020a}
{Lehmann} A.,  {Godard} B.,  {Pineau des For{\^e}ts} G.,   {Falgarone} E.,
  2020a, in {da Cunha} E.,  {Hodge} J.,  {Afonso} J.,  {Pentericci} L.,
  {Sobral} D.,  eds,  IAU Symposium Vol. 352, IAU Symposium. pp 73--74,
  \mn@doi{10.1017/S1743921320000654}

\bibitem[\protect\citeauthoryear{{Lehmann}, {Godard}, {Pineau des For{\^e}ts}
  \& {Falgarone}}{{Lehmann} et~al.}{2020b}]{Lehmann2020b}
{Lehmann} A.,  {Godard} B.,  {Pineau des For{\^e}ts} G.,   {Falgarone} E.,
  2020b, \mn@doi [\aap] {10.1051/0004-6361/202038644}, \href
  {https://ui.adsabs.harvard.edu/abs/2020A&A...643A.101L} {643, A101}

\bibitem[\protect\citeauthoryear{{Lesaffre}, {Pineau des For{\^e}ts}, {Godard},
  {Guillard}, {Boulanger}  \& {Falgarone}}{{Lesaffre}
  et~al.}{2013}]{Lesaffre2013}
{Lesaffre} P.,  {Pineau des For{\^e}ts} G.,  {Godard} B.,  {Guillard} P.,
  {Boulanger} F.,   {Falgarone} E.,  2013, \mn@doi [\aap]
  {10.1051/0004-6361/201219928}, \href
  {http://adsabs.harvard.edu/abs/2013A%26A...550A.106L} {550, A106}

\bibitem[\protect\citeauthoryear{{Li} et~al.,}{{Li} et~al.}{2019}]{Li2019}
{Li} Q.,  et~al., 2019, \mn@doi [\apj] {10.3847/1538-4357/ab0e6f}, \href
  {https://ui.adsabs.harvard.edu/abs/2019ApJ...875..130L} {875, 130}

\bibitem[\protect\citeauthoryear{{Li} et~al.,}{{Li} et~al.}{2020}]{Li2020}
{Li} Y.,  et~al., 2020, \mn@doi [\apjl] {10.3847/2041-8213/ab65c7}, \href
  {https://ui.adsabs.harvard.edu/abs/2020ApJ...889L...1L} {889, L1}

\bibitem[\protect\citeauthoryear{{Lochhaas}, {Bryan}, {Li}, {Li}  \&
  {Fielding}}{{Lochhaas} et~al.}{2020}]{Lochhaas2020}
{Lochhaas} C.,  {Bryan} G.~L.,  {Li} Y.,  {Li} M.,   {Fielding} D.,  2020,
  \mn@doi [\mnras] {10.1093/mnras/staa358}, \href
  {https://ui.adsabs.harvard.edu/abs/2020MNRAS.493.1461L} {493, 1461}

\bibitem[\protect\citeauthoryear{{Lotz} et~al.,}{{Lotz}
  et~al.}{2017}]{Lotz2017}
{Lotz} J.~M.,  et~al., 2017, \mn@doi [\apj] {10.3847/1538-4357/837/1/97}, \href
  {https://ui.adsabs.harvard.edu/abs/2017ApJ...837...97L} {837, 97}

\bibitem[\protect\citeauthoryear{{Madau} \& {Dickinson}}{{Madau} \&
  {Dickinson}}{2014}]{Madau2014}
{Madau} P.,  {Dickinson} M.,  2014, \mn@doi [\araa]
  {10.1146/annurev-astro-081811-125615}, \href
  {http://adsabs.harvard.edu/abs/2014ARA%26A..52..415M} {52, 415}

\bibitem[\protect\citeauthoryear{{Magnelli} et~al.,}{{Magnelli}
  et~al.}{2012}]{Magnelli2012}
{Magnelli} B.,  et~al., 2012, \mn@doi [\aap] {10.1051/0004-6361/201118312},
  \href {https://ui.adsabs.harvard.edu/abs/2012A&A...539A.155M} {539, A155}

\bibitem[\protect\citeauthoryear{{Miville-Desch{\^e}nes}, {Murray}  \&
  {Lee}}{{Miville-Desch{\^e}nes} et~al.}{2017}]{Miville-Deschenes2017}
{Miville-Desch{\^e}nes} M.-A.,  {Murray} N.,   {Lee} E.~J.,  2017, \mn@doi
  [\apj] {10.3847/1538-4357/834/1/57}, \href
  {https://ui.adsabs.harvard.edu/abs/2017ApJ...834...57M} {834, 57}

\bibitem[\protect\citeauthoryear{{Narayanan} et~al.,}{{Narayanan}
  et~al.}{2006}]{Narayanan2006}
{Narayanan} D.,  et~al., 2006, \mn@doi [\apjl] {10.1086/504846}, \href
  {https://ui.adsabs.harvard.edu/abs/2006ApJ...642L.107N} {642, L107}

\bibitem[\protect\citeauthoryear{{Narayanan} et~al.,}{{Narayanan}
  et~al.}{2015}]{Narayanan2015}
{Narayanan} D.,  et~al., 2015, \mn@doi [\nat] {10.1038/nature15383}, \href
  {https://ui.adsabs.harvard.edu/abs/2015Natur.525..496N} {525, 496}

\bibitem[\protect\citeauthoryear{{Nims}, {Quataert}  \&
  {Faucher-Gigu{\`e}re}}{{Nims} et~al.}{2015}]{Nims2015}
{Nims} J.,  {Quataert} E.,   {Faucher-Gigu{\`e}re} C.-A.,  2015, \mn@doi
  [\mnras] {10.1093/mnras/stu2648}, \href
  {https://ui.adsabs.harvard.edu/abs/2015MNRAS.447.3612N} {447, 3612}

\bibitem[\protect\citeauthoryear{{Oh}, {Kim}  \& {Lee}}{{Oh}
  et~al.}{2015}]{Oh2015}
{Oh} S.~H.,  {Kim} W.-T.,   {Lee} H.~M.,  2015, \mn@doi [\apj]
  {10.1088/0004-637X/807/1/73}, \href
  {https://ui.adsabs.harvard.edu/abs/2015ApJ...807...73O} {807, 73}

\bibitem[\protect\citeauthoryear{{Oliver} et~al.,}{{Oliver}
  et~al.}{2010}]{Oliver2010}
{Oliver} S.~J.,  et~al., 2010, \mn@doi [\aap] {10.1051/0004-6361/201014697},
  \href {http://cdsads.u-strasbg.fr/abs/2010A%26A...518L..21O} {518, L21}

\bibitem[\protect\citeauthoryear{{Pearson} \& {Drouin}}{{Pearson} \&
  {Drouin}}{2006}]{Pearson2006}
{Pearson} J.~C.,  {Drouin} B.~J.,  2006, \mn@doi [\apjl] {10.1086/506522},
  \href {http://adsabs.harvard.edu/abs/2006ApJ...647L..83P} {647, L83}

\bibitem[\protect\citeauthoryear{{Planck Collaboration} et~al.,}{{Planck
  Collaboration} et~al.}{2020}]{Planck-Collaboration2020}
{Planck Collaboration} et~al., 2020, \mn@doi [\aap]
  {10.1051/0004-6361/201833910}, \href
  {https://ui.adsabs.harvard.edu/abs/2020A&A...641A...6P} {641, A6}

\bibitem[\protect\citeauthoryear{{Richard}, {Kneib}, {Limousin}, {Edge}  \&
  {Jullo}}{{Richard} et~al.}{2010}]{Richard2010}
{Richard} J.,  {Kneib} J.~P.,  {Limousin} M.,  {Edge} A.,   {Jullo} E.,  2010,
  \mn@doi [\mnras] {10.1111/j.1745-3933.2009.00796.x}, \href
  {https://ui.adsabs.harvard.edu/abs/2010MNRAS.402L..44R} {402, L44}

\bibitem[\protect\citeauthoryear{{Richard} et~al.,}{{Richard}
  et~al.}{2014}]{Richard2014}
{Richard} J.,  et~al., 2014, \mn@doi [\mnras] {10.1093/mnras/stu1395}, \href
  {https://ui.adsabs.harvard.edu/abs/2014MNRAS.444..268R} {444, 268}

\bibitem[\protect\citeauthoryear{{Richings} \&
  {Faucher-Gigu{\`e}re}}{{Richings} \&
  {Faucher-Gigu{\`e}re}}{2018a}]{Richings2018a}
{Richings} A.~J.,  {Faucher-Gigu{\`e}re} C.-A.,  2018a, \mn@doi [\mnras]
  {10.1093/mnras/stx3014}, \href
  {https://ui.adsabs.harvard.edu/abs/2018MNRAS.474.3673R} {474, 3673}

\bibitem[\protect\citeauthoryear{{Richings} \&
  {Faucher-Gigu{\`e}re}}{{Richings} \&
  {Faucher-Gigu{\`e}re}}{2018b}]{Richings2018b}
{Richings} A.~J.,  {Faucher-Gigu{\`e}re} C.-A.,  2018b, \mn@doi [\mnras]
  {10.1093/mnras/sty1285}, \href
  {https://ui.adsabs.harvard.edu/abs/2018MNRAS.478.3100R} {478, 3100}

\bibitem[\protect\citeauthoryear{{Riechers} et~al.,}{{Riechers}
  et~al.}{2013}]{Riechers2013}
{Riechers} D.~A.,  et~al., 2013, \mn@doi [\nat] {10.1038/nature12050}, \href
  {https://ui.adsabs.harvard.edu/abs/2013Natur.496..329R} {496, 329}

\bibitem[\protect\citeauthoryear{{Rupke}}{{Rupke}}{2018}]{Rupke2018}
{Rupke} D.,  2018, \mn@doi [Galaxies] {10.3390/galaxies6040138}, \href
  {http://cdsads.u-strasbg.fr/abs/2018Galax...6..138R} {6, 138}

\bibitem[\protect\citeauthoryear{{Salpeter}}{{Salpeter}}{1955}]{Salpeter1955}
{Salpeter} E.~E.,  1955, \mn@doi [\apj] {10.1086/145971}, \href
  {https://ui.adsabs.harvard.edu/abs/1955ApJ...121..161S} {121, 161}

\bibitem[\protect\citeauthoryear{{Schaye} et~al.,}{{Schaye}
  et~al.}{2015}]{Schaye2015}
{Schaye} J.,  et~al., 2015, \mn@doi [\mnras] {10.1093/mnras/stu2058}, \href
  {http://cdsads.u-strasbg.fr/abs/2015MNRAS.446..521S} {446, 521}

\bibitem[\protect\citeauthoryear{{Schroetter} et~al.,}{{Schroetter}
  et~al.}{2019}]{Schroetter2019}
{Schroetter} I.,  et~al., 2019, \mn@doi [\mnras] {10.1093/mnras/stz2822}, \href
  {https://ui.adsabs.harvard.edu/abs/2019MNRAS.490.4368S} {490, 4368}

\bibitem[\protect\citeauthoryear{{Shen}, {Madau}, {Guedes}, {Mayer},
  {Prochaska}  \& {Wadsley}}{{Shen} et~al.}{2013}]{Shen2013}
{Shen} S.,  {Madau} P.,  {Guedes} J.,  {Mayer} L.,  {Prochaska} J.~X.,
  {Wadsley} J.,  2013, \mn@doi [\apj] {10.1088/0004-637X/765/2/89}, \href
  {https://ui.adsabs.harvard.edu/abs/2013ApJ...765...89S} {765, 89}

\bibitem[\protect\citeauthoryear{{Solomon}, {Downes}, {Radford}  \&
  {Barrett}}{{Solomon} et~al.}{1997}]{Solomon1997}
{Solomon} P.~M.,  {Downes} D.,  {Radford} S.~J.~E.,   {Barrett} J.~W.,  1997,
  \mn@doi [\apj] {10.1086/303765}, \href
  {https://ui.adsabs.harvard.edu/abs/1997ApJ...478..144S} {478, 144}

\bibitem[\protect\citeauthoryear{{Somerville} \& {Dav{\'e}}}{{Somerville} \&
  {Dav{\'e}}}{2015}]{Somerville2015}
{Somerville} R.~S.,  {Dav{\'e}} R.,  2015, \mn@doi [\araa]
  {10.1146/annurev-astro-082812-140951}, \href
  {https://ui.adsabs.harvard.edu/abs/2015ARA&A..53...51S} {53, 51}

\bibitem[\protect\citeauthoryear{{Spitzer}}{{Spitzer}}{1968}]{Spitzer1968}
{Spitzer} Lyman J.,  1968, {Dynamics of Interstellar Matter and the Formation
  of Stars}.
p.~1

\bibitem[\protect\citeauthoryear{{Tacconi} et~al.,}{{Tacconi}
  et~al.}{2008}]{Tacconi2008}
{Tacconi} L.~J.,  et~al., 2008, \mn@doi [\apj] {10.1086/587168}, \href
  {http://adsabs.harvard.edu/abs/2008ApJ...680..246T} {680, 246}

\bibitem[\protect\citeauthoryear{{Tacconi}, {Genzel}  \& {Sternberg}}{{Tacconi}
  et~al.}{2020}]{Tacconi2020}
{Tacconi} L.~J.,  {Genzel} R.,   {Sternberg} A.,  2020, arXiv e-prints, \href
  {https://ui.adsabs.harvard.edu/abs/2020arXiv200306245T} {p. arXiv:2003.06245}

\bibitem[\protect\citeauthoryear{{Tumlinson}, {Peeples}  \& {Werk}}{{Tumlinson}
  et~al.}{2017}]{Tumlinson2017}
{Tumlinson} J.,  {Peeples} M.~S.,   {Werk} J.~K.,  2017, \mn@doi [\araa]
  {10.1146/annurev-astro-091916-055240}, \href
  {https://ui.adsabs.harvard.edu/abs/2017ARA&A..55..389T} {55, 389}

\bibitem[\protect\citeauthoryear{{Valdivia}, {Godard}, {Hennebelle}, {Gerin},
  {Lesaffre}  \& {Le Bourlot}}{{Valdivia} et~al.}{2017}]{Valdivia2017}
{Valdivia} V.,  {Godard} B.,  {Hennebelle} P.,  {Gerin} M.,  {Lesaffre} P.,
  {Le Bourlot} J.,  2017, \mn@doi [\aap] {10.1051/0004-6361/201629905}, \href
  {https://ui.adsabs.harvard.edu/abs/2017A&A...600A.114V} {600, A114}

\bibitem[\protect\citeauthoryear{{Veilleux}, {Maiolino}, {Bolatto}  \&
  {Aalto}}{{Veilleux} et~al.}{2020}]{Veilleux2020}
{Veilleux} S.,  {Maiolino} R.,  {Bolatto} A.~D.,   {Aalto} S.,  2020, \mn@doi
  [\aapr] {10.1007/s00159-019-0121-9}, \href
  {https://ui.adsabs.harvard.edu/abs/2020A&ARv..28....2V} {28, 2}

\bibitem[\protect\citeauthoryear{{Verhamme}, {Schaerer}  \&
  {Maselli}}{{Verhamme} et~al.}{2006}]{Verhamme2006}
{Verhamme} A.,  {Schaerer} D.,   {Maselli} A.,  2006, \mn@doi [\aap]
  {10.1051/0004-6361:20065554}, \href
  {http://cdsads.u-strasbg.fr/abs/2006A%26A...460..397V} {460, 397}

\bibitem[\protect\citeauthoryear{{Vernet} \& {Cimatti}}{{Vernet} \&
  {Cimatti}}{2001}]{Vernet2001}
{Vernet} J.,  {Cimatti} A.,  2001, \mn@doi [\aap] {10.1051/0004-6361:20011442},
  \href {http://cdsads.u-strasbg.fr/abs/2001A%26A...380..409V} {380, 409}

\bibitem[\protect\citeauthoryear{{Voit} et~al.,}{{Voit}
  et~al.}{2020}]{Voit2020}
{Voit} G.~M.,  et~al., 2020, \mn@doi [\apj] {10.3847/1538-4357/aba42e}, \href
  {https://ui.adsabs.harvard.edu/abs/2020ApJ...899...70V} {899, 70}

\bibitem[\protect\citeauthoryear{{Walter} et~al.,}{{Walter}
  et~al.}{2020}]{Walter2020}
{Walter} F.,  et~al., 2020, arXiv e-prints, \href
  {https://ui.adsabs.harvard.edu/abs/2020arXiv200911126W} {p. arXiv:2009.11126}

\bibitem[\protect\citeauthoryear{{Wright}, {Lagos}, {Power}  \&
  {Mitchell}}{{Wright} et~al.}{2020}]{Wright2020}
{Wright} R.~J.,  {Lagos} C. d.~P.,  {Power} C.,   {Mitchell} P.~D.,  2020,
  \mn@doi [\mnras] {10.1093/mnras/staa2359}, \href
  {https://ui.adsabs.harvard.edu/abs/2020MNRAS.498.1668W} {498, 1668}

\bibitem[\protect\citeauthoryear{{Yang} et~al.,}{{Yang}
  et~al.}{2017}]{Yang2017}
{Yang} C.,  et~al., 2017, \mn@doi [\aap] {10.1051/0004-6361/201731391}, \href
  {https://ui.adsabs.harvard.edu/abs/2017A&A...608A.144Y} {608, A144}

\bibitem[\protect\citeauthoryear{{Zabl} et~al.,}{{Zabl}
  et~al.}{2019}]{Zabl2019}
{Zabl} J.,  et~al., 2019, \mn@doi [\mnras] {10.1093/mnras/stz392}, \href
  {https://ui.adsabs.harvard.edu/abs/2019MNRAS.485.1961Z} {485, 1961}

\bibitem[\protect\citeauthoryear{{Zhang}, {Romano}, {Ivison}, {Papadopoulos}
  \& {Matteucci}}{{Zhang} et~al.}{2018}]{Zhang2018}
{Zhang} Z.-Y.,  {Romano} D.,  {Ivison} R.~J.,  {Papadopoulos} P.~P.,
  {Matteucci} F.,  2018, \mn@doi [\nat] {10.1038/s41586-018-0196-x}, \href
  {https://ui.adsabs.harvard.edu/abs/2018Natur.558..260Z} {558, 260}

\bibitem[\protect\citeauthoryear{{Zhuravleva} et~al.,}{{Zhuravleva}
  et~al.}{2014}]{Zhuravleva2014}
{Zhuravleva} I.,  et~al., 2014, \mn@doi [\nat] {10.1038/nature13830}, \href
  {https://ui.adsabs.harvard.edu/abs/2014Natur.515...85Z} {515, 85}

\bibitem[\protect\citeauthoryear{{Zhuravleva}, {Churazov}, {Schekochihin},
  {Allen}, {Vikhlinin}  \& {Werner}}{{Zhuravleva}
  et~al.}{2019}]{Zhuravleva2019}
{Zhuravleva} I.,  {Churazov} E.,  {Schekochihin} A.~A.,  {Allen} S.~W.,
  {Vikhlinin} A.,   {Werner} N.,  2019, \mn@doi [Nature Astronomy]
  {10.1038/s41550-019-0794-z}, \href
  {https://ui.adsabs.harvard.edu/abs/2019NatAs...3..832Z} {3, 832}

\bibitem[\protect\citeauthoryear{{van der Tak}, {Black}, {Sch{\"o}ier},
  {Jansen}  \& {van Dishoeck}}{{van der Tak} et~al.}{2007}]{van-der-Tak2007}
{van der Tak} F.~F.~S.,  {Black} J.~H.,  {Sch{\"o}ier} F.~L.,  {Jansen} D.~J.,
   {van Dishoeck} E.~F.,  2007, \mn@doi [\aap] {10.1051/0004-6361:20066820},
  \href {http://adsabs.harvard.edu/abs/2007A%26A...468..627V} {468, 627}

\makeatother
\end{thebibliography}



\appendix

\section{The \CHp\ manna}\label{A}

\CHp\ is a most fragile species: it has a highly endoenergic formation ($E_{\rm form} \sim 0.4$\,eV) 
 and a fast destruction rate, by collisions with H and \HH. In diffuse and weakly irradiated gas, a warm chemistry activated by 
dissipation of turbulence in shocks and/or intense velocity shears has been proposed to overcome 
its fast destruction \citep{Flower1998,Lesaffre2013, Godard2014}. Turbulent transport in a biphase medium comprising a cold ($T\sim 100$\,K) and a warm ($T\sim 10^4$\,K) phase
may also contribute to the \CHp\ production in diffuse gas \citep{Valdivia2017}. 

Once formed, its lifetime is so short,  
\begin{equation}
t_{\CHp}=1 \,{\rm yr} \, f(\HH)^{-1} \bigg(\frac{n_{\rm H}}{50\cc}\bigg)^{-1}, 
\end{equation}
that, unlike the CO molecule, \CHp\ is always observed where it formed, i.e. it cannot be transported. 
Here, $f_{\HH}=2n(\HH)/n_{\rm H}$ is the \HH\ fraction.

\CHp\ being a light hydride with high dipole moment, the critical density ($\sim 10^7$\,\cc) of its $J=$1--0 transition is  almost 10$^5\times$ larger than that of CO(1--0), so the line appears in absorption in low-density gas ($n_{\rm H} < 10^3$~\cc).
In emission, the \CHp\ $J$= 1--0 line is detected only
in dense and UV-illuminated gas, either photon-dominated
regions (PDRs) or UV-irradiated shocks because, unlike most molecules, the
abundance of \CHp\ is enhanced in intense UV fields \citep{Agundez2010,Godard2013,Godard2019}.

We recall here the method we devised in Paper I to infer the radius and mass of the reservoir of diffuse molecular gas  from the observables:  the radius $r_{\rm SMG}$ and flux
of the sub-mm background continuum source and the width $\Delta v_{\rm abs}$ and opacity $\tau_0$ of the \CHp\ absorption line.

The dominant driver of \CHp\ production in the diffuse gas causing the absorptions is assumed to be turbulent energy dissipation. As in 
a steady turbulent cascade,  the energy dissipation and transfer rates balance each other, the dissipation rate per unit volume is
$\overline{\epsilon} \propto \overline{\rho} \,\overline{v}_{\rm turb}^3/l$,  where $l \sim  r_{\rm TR}$ and $\overline{\rho}$ are the unknown sizescale and mean mass density 
of the part of the turbulent reservoir sampled by absorption and  $\overline{v}_{\rm turb}$ is its mean turbulent velocity inferred from the width of the absorption line.  
Absorption lines provide the column density of \CHp\ molecules and therefore the {\it number} of \CHp\ molecules in the volume $\pi r_{\rm SMG}^2 r_{\rm TR}$ subtended by the 
background continuum source of radius $r_{\rm SMG}$ and the unknown length of the line-of-sight $r_{\rm TR}$ across the diffuse CGM. 
Because we know that these molecules formed in that volume (i.e. they cannot be transported), we know that the energy injection 
rate due to turbulent dissipation required to sustain the observed {\it number} of \CHp\ 
molecules in that volume has to be: 
\begin{equation}
\dot{E}_{\CHp}= N(\CHp) \,\pi \,r_{\rm SMG}^2\, E_{\rm form}\,/\,t_{\CHp}.
\end{equation}
It only depends on the observed absorption line (see Eq. (2)) and the known values $E_{\rm form}$ and $t_{\CHp}$.
The injection rate of turbulent energy in that same volume is:
\begin{equation}
\dot{E}_{\rm turb}= {\overline \epsilon} \times  \pi r_{\rm SMG}^2  \,r_{\rm TR}.
\end{equation}
and a fraction $\alpha$ of this energy flux  is feeding the warm chemistry:
\begin{equation}
\dot{E}_{\CHp}= \alpha \,\dot{E}_{\rm turb}
\end{equation}

An estimate $\alpha \sim 10^{-3}$ is inferred from \CHp\ observations in the Milky Way as follows.
The \CHp\ abundances measured in the  Solar Neighbourhood are about 3$\times$ smaller than those measured in the inner Milky Way, at galactic radii between 4 and 6 kpc 
 \citep{Godard2014}. An exponential decrease of the turbulent energy transfer rate $\epsilon$ is found between the Inner Galaxy and the Solar Neigborhood from the analysis of CO(1-0) surveys, providing a similar ratio of the values of $\epsilon$ at these two galactic radii \citep{Miville-Deschenes2017}.
Since the turbulent cascade has been found to pervade the atomic medium and the molecular clouds down to the smallest non-star forming cores with a similar
 energy transfer rate $\epsilon \sim  2 \times 10^{-25}$ \ergccs\ in the Solar Neighborhood  \citep{Falgarone2004}, we assume that the same radial dependence applies to the entire cold interstellar medium (ISM):
\begin{equation}
X(\CHp) =N(\CHp)/N_{\rm H,cool}=\kappa \epsilon
\end{equation}
where  $N_{\rm H,cool}=N({\rm H}) + 2 N(\HH)$ is the total column density of the cool hydrogen along the \CHp\ lines-of-sight, $N({\rm H})$ being provided by the 21cm absorption optical depth and $N(\HH)$ by CH and HF absorptions over the same velocity intervals \citep{Godard2014}. 
We find $\kappa=10^{17}$ (\ergccs)$^{-1}$ from the observed variations of \CHp\ abundances and $\epsilon$ wiih the Galactic radius.
From Eq. (A4) adapted to the Galactic environment, we infer:
\begin{equation}
\alpha= \kappa \, \overline{n}_{\rm H,cool} \, \frac{E_{\rm form}}{t_{\CHp}} \sim 10^{-3}\, \bigg(\frac{ \overline{n}_{\rm H,cool}}{1 \cc}\bigg)
\end{equation}
where $\overline{n}_{\rm H,cool}$ is the mean density of the cool diffuse ISM.
An important assumption is made here: the same mean density is relevant for  
the turbulent energy transfer rate 
and the mean density of the gas in which \CHp\ is observed in absorption.
This is because the bulk of  the gas mass is assumed to be that of the  
cool molecular phase ($T \sim 100$ K) while the bulk of the volume is  filled with the warm phase  
($T \sim 10^4$ K). This mean density  is not the local density $n_{\rm H}$, involved in the collisional destruction rate of \CHp\  (Eq. (A1)) .

The gas mass contributing to the absorption is therefore:
 \begin{equation}
 M_{\rm abs}=\mu_{\rm p} \, m_{H} \, N_{\rm H,cool} \,\pi \,r_{\rm SMG}^2
 \end{equation}
 where $\mu_{\rm p}$ is the mean mass per particle, 
 and the turbulent energy flux pervading that mass, and eventually dissipated in the same volume, is:
  \begin{equation}
  \dot{E}_{\rm turb}=\frac{1}{2} \,M_{\rm abs} \,\overline{v}_{\rm turb}^3\,/  \,r_{TR}.
  \end{equation}
Using Eq.(A4), we then infer:
\begin{equation}
X(\CHp)= \alpha \,\frac{\mu_{\rm p} \,m_{\rm H}}{2} \frac{t_{\CHp}}{E_{\rm form}} \frac{\overline{v}_{\rm turb}^3}{r_{\rm TR}} 
\label{eq:abund}
\end{equation}

The mass of the turbulent reservoir is then computed by assuming mass conservation across the different shells, regardless of the global inward/outward 
or turbulent motions, so that its average density decreases with radius as $\overline{n}_{\rm H} \propto r^{-2}$, for an assumed spherical geometry. 
 In that case the hydrogen column density at impact parameter $b$ decreases as $N_{\rm H}(b) \propto b^{-1}$ and the total gas mass is
 \begin{equation} 
 M_{\rm TR}=2\,\pi\ \mu_{\rm p} \, m_{\rm H} \, r_{\rm TR} \, r_{\rm SMG} \, N_{\rm H,cool}. 
\label{eq:mass}
\end{equation}
where $N_{\rm H,cool}$ inferred from the \CHp\ absorption lines is half the average hydrogen column density within the radius $r_{\rm SMG}$.

Introducing $r_{\rm TR}= \overline{v}_{\rm turb} \, t_{\rm SB}$, the total gas mass depends only on the absorption opacity, as shown in Paper I:
\begin{equation}
M_{\rm TR}= 3.8 \times 10^{10} \msol\ \tau_0 \, t_{50}^2 \, r_{\rm SMG,kpc} \, \bigg( f_{\rm \HH} \frac{n_{50}}{\alpha_3} \bigg)
\label{eq:massnum}
\end{equation}
where we have introduced $\alpha_{3}=\alpha/10^{-3}$, $t_{50}=t_{\rm SB}/50$\,Myr for the lifetime of the turbulent reservoirs, 
the local gas density $n_{50}=n_{\rm H}/50\cc$ that determines the \CHp\ lifetime, 
A mean mass per particle $\mu_{\rm p}=1.27$ is adopted.


\section{The lensing model} \label{B}

SMM\,J02399$-$0136 is gravitationally lensed by the cluster Abell $370$ \citep{Ivison2010a}. The mass distribution of this cluster has been studied in~\citet{Richard2010} from the constraints given by $10$ multiply lensed background sources observed with bands F475W/F625W and F814W from {\it HST}/ACS. These authors assumed for their lens model a bimodal pseudo-isothermal elliptical mass distribution. More recently~\citet{Diego2016} have found an alternative way of deriving a free-form mass distribution by analysing the latest {\it HST} Frontier images and GLASS spectroscopy. In general, different authors find a good consistency between different models taking into account two cluster clumps and taking independent subsamples of lensed sources as constraints. We recall that the goal of this work is to study the background source rather than to obtain a very detailed modelling of the lens. Since SMM\,J02399$-$0136 is located far from the centres of these two cluster clumps, it is not strongly lensed and a simplification of the lensing model is justified.

In order to derive some of the intrinsic properties of the background source we correct from the distortion and magnification of A$370$ using sl$\_$fit \citep{Gavazzi2011}. This code adopts the standard Metropolis-Hastings algorithm, a standard Markov chain Monte Carlo (MCMC) method, to explore the space of parameters which best reproduces the observed data in the source plane. Given a set of parameters describing the lens model and the luminosity profile(s) of the background source(s), the code builds samples of the posterior probability distribution function of these parameters. We describe the lens as a singular isothermal ellipsoid (SIE) mass distribution centred on the cluster position obtained by~\citet{Richard2010}. We also fix the minor-to-major axis ratio $q_{\rm L}$ and orientation of the major axis in the plane of the sky ($PA_{\rm L}$) of the lens to 0.4 and 90.0 degrees respectively. To model the data in the background plane we assume two sources and adopt an elliptical Sersic profile for their luminosity, leaving the rest of the parameters free. These include for each source $q$, PA, its relative position with respect to the lens centre, its effective radius and its intrinsic flux.  

We obtain a magnification factor $\mu=1.9\pm0.1$ for L2SW and $\mu=1.7\pm0.1$ for L1, in agreement with the range of values obtained in the literature (Sect.~1). We derive an effective radius of $0.22\pm 0.01$\, arcsec corresponding to a physical size of 1.8 $\pm 0.1$\,kpc for L2SW and a radius of $0.8 \pm 0.1$\,kpc for L1, adopting for both the same redshift and the {\it Planck} cosmology ($H_0=67.4$, $\Omega_{\rm M}=0.315$).

\section{Statistical significance of the \CHp\ emission structures } \label{C}

As illustrated in Sect. 3.1,  the tentative structures of \CHp\ line emission are complex and not well suited to state-of-the-art statistical analysis 
of e.g. continuum point-sources observed by ALMA. The statistical confirmation of their existence has to be conducted in space and 
velocity-space jointly, because the structures are highly dynamical (see Sect. 6.2) and their boundary and shape depend on the velocity range over which the line emission is integrated. 
They are spatially extended and often elongated with the shortest dimensions in some cases barely resolved. 
Their moment-0 values are comparable to the negative values of the minima visible in the moment-0 maps.

These negative spots can be understood as combined sidelobes of multiple extended weak sources, scattered in the ALMA primary beam. As an example, the conspicuous negative feature 
close to the phase centre (2 arcsec South of L2SW) and extending over $\sim 2 $ ALMA synthesised beam, is due to a very broad (FWHM = 2600 \kms) line with 
a peak intensity of -0.3 mJy. Such a large linewidth suggests that it results from the chance combination of negative sidelobes of 
multiple emissions of narrower width.
The order of magnitude of its intensity can be recovered for a total extent of the eight central structures of average size $\overline{\Omega}=0.49$ arcsec$^2$,
equivalent to $8 \times 0.49/0.17=23$  synthesised beams, an average linewidth 
of 1330 \kms\ and an average flux of 0.2 mJy (Table 2): the combination of the sidelobes of 16 of these synthesised beams would be enough.

\begin{figure*}
 \begin{minipage}{0.45\linewidth}
	\includegraphics[width=\textwidth]{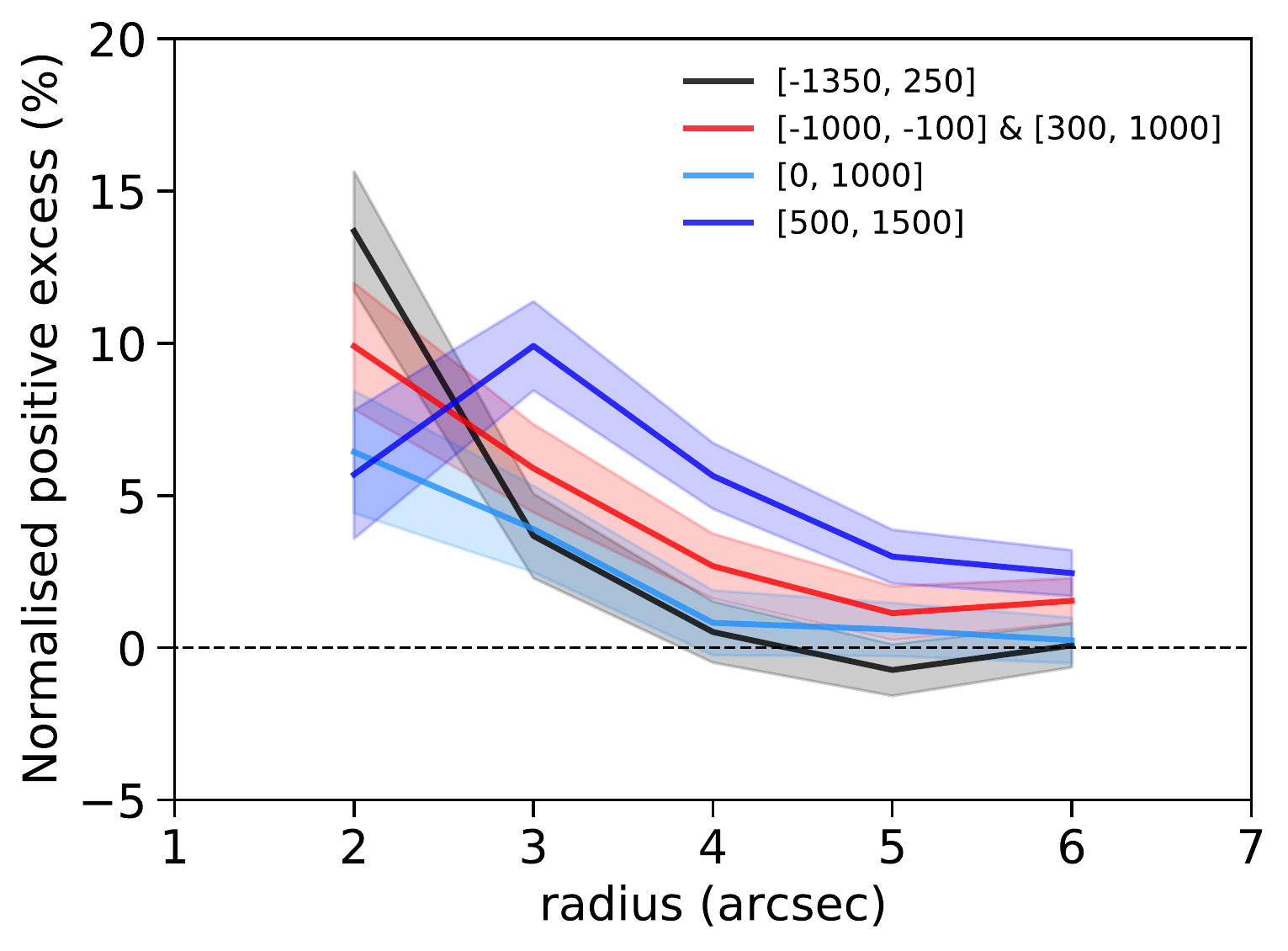}
\end{minipage}
 \begin{minipage}{0.45\linewidth}
	 \includegraphics[width=\textwidth]{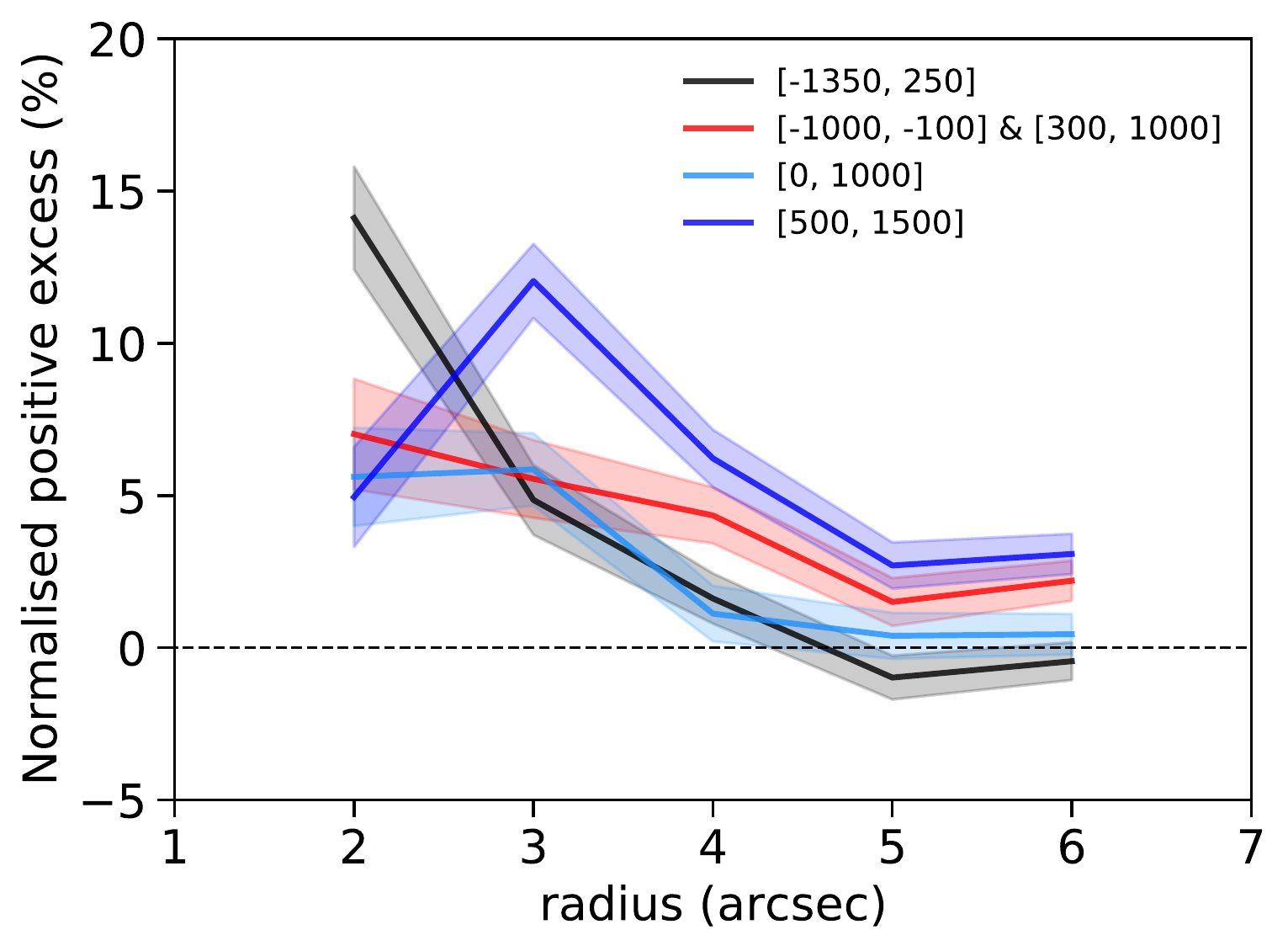}
 \end{minipage}	
    \caption{ Radial dependence of the excess of positive moment-0 pixels over those with negative values as a function of the radius (i.e. distance to the phase centre), for all the structures ({\it left}) and those larger than one synthesised beam ({\it right}). The difference is normalised to the total number of pixels within a given radius. Four moment-0 maps have been analysed, characterised by their velocity coverage (upper right corners). The shaded areas correspond to $\pm 2 \sigma$ r.m.s..  }
\label{fig:stat}
\end{figure*}

To assess the reality of an excess of extended positive structures in the moment-0 maps and its dependence on the distance to the centre of phase, we have 
identified in four moment-0 maps ensembles of neighbouring pixels brighter
than $\sigma_{\rm m0}$ (or weaker than $- \sigma_{\rm m0}$) 
and connected over areas more extended that one ALMA synthesised beam. 
In Fig.~\ref{fig:stat}, the excess number of pixels with positive moment-0 values over those with negative values, normalised to the total number of pixels within a given radius,  is displayed as a function of this radius (i.e. the distance to the phase centre). The left panel displays this excess for all the connected structures above $\sigma_{\rm m0}$ (and weaker than $-\sigma_{\rm m0}$) , whatever their size, and the right panel considers only the pixels in structures larger than one synthesised beam area. In the latter case, the excess of positive structures above those negative is statistically significant within 4.5 arcsec from the phase centre. This limit depends on the moment-0 map, as the positive excess computed over [500,1500] \kms\ remains significant up to a distance of 6 arcsec. This velocity interval corresponds to the velocity range where the \CHp\ emission is the most intense (Fig.~\ref{fig:SumALMA}).

 \begin{figure*}
	\includegraphics[width=0.9\textwidth]{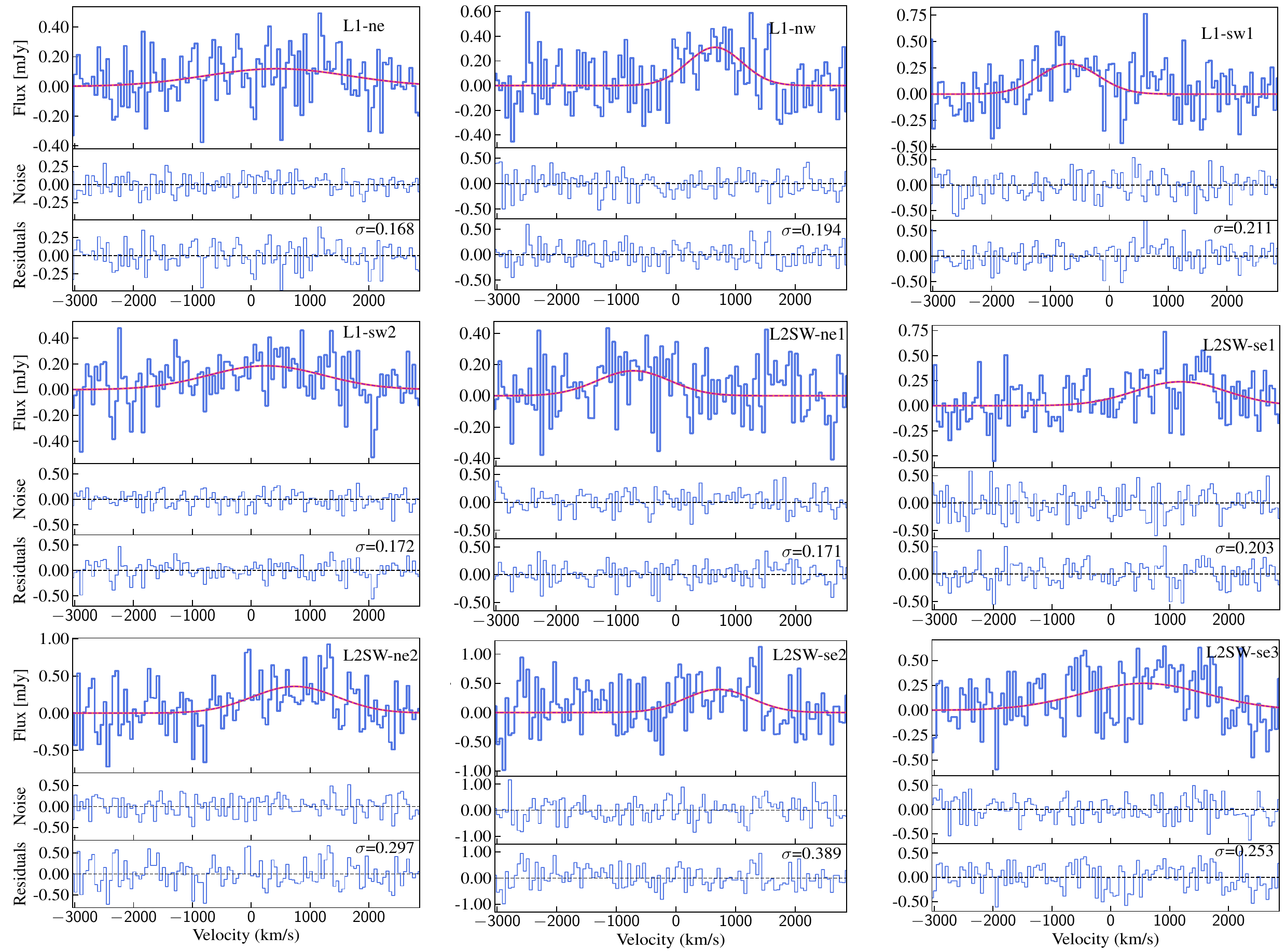}
    \caption{ Same  spectra as those of Fig.\ref{fig:4specL1} with the additional noise spectrum taken within the same shape as that of each \CHp\ emission structure, but localized in its vicinity. Note that the 
    flux scales of the noise spectra and the residuals are the same for a given structure.  }
\label{fig:9specnoise}
\end{figure*}

Last, Fig.\ref{fig:9specnoise} displays the same spectra as in Figs.\ref{fig:4specL1} and \ref{fig:7specFar}, but those of L1-s and L2SW-n,  
with an additional spectrum taken within a similar contour as that of each 
\CHp\ emission structure, localized in its vicinity. These spectra illustrate the excellent quality of the ALMA baselines.


\section{Results of \CHp\ non-LTE radiative transfer}\label{D}

 \begin{figure*}
	\includegraphics[width=\columnwidth]{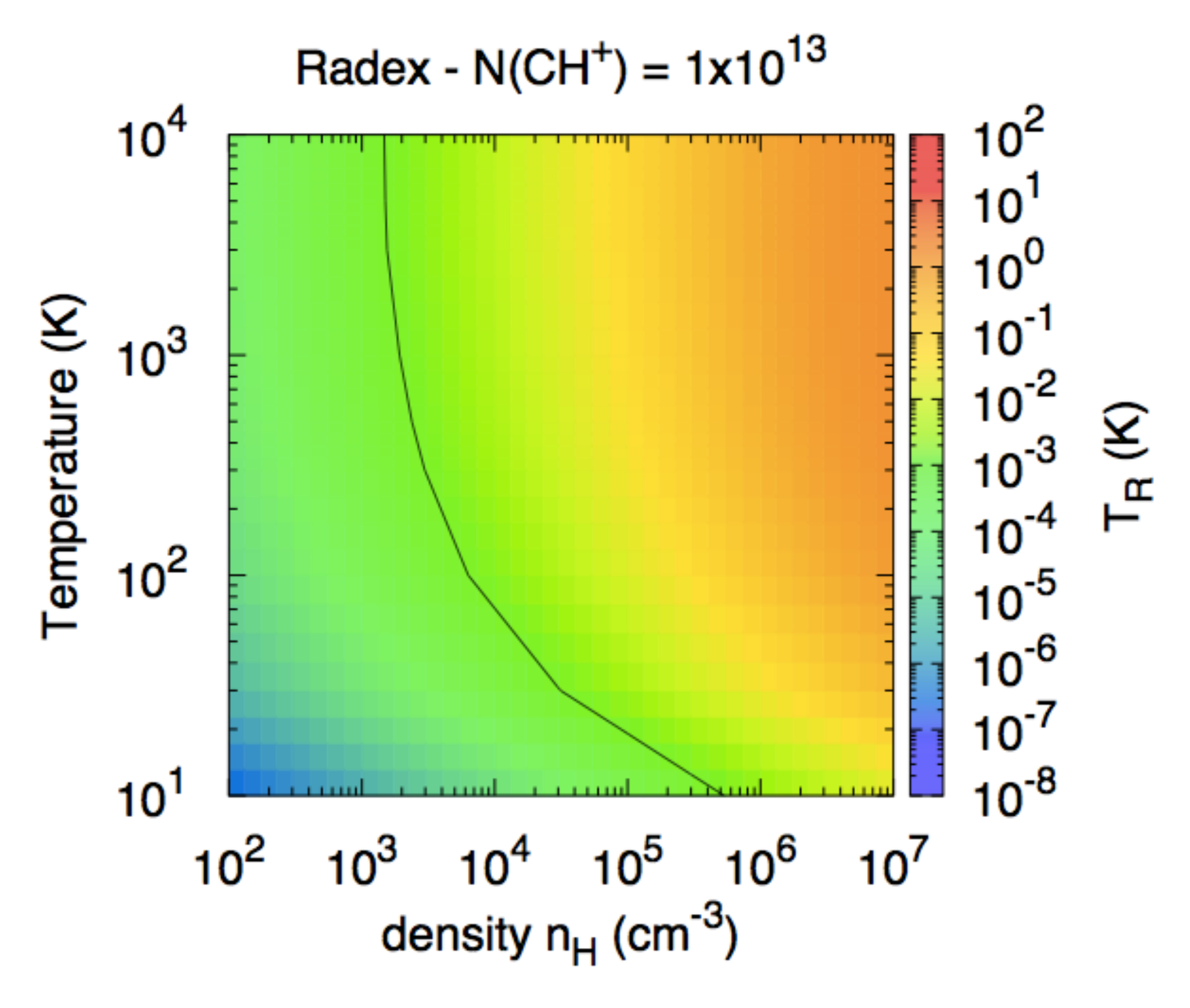}
      \includegraphics[width=\columnwidth]{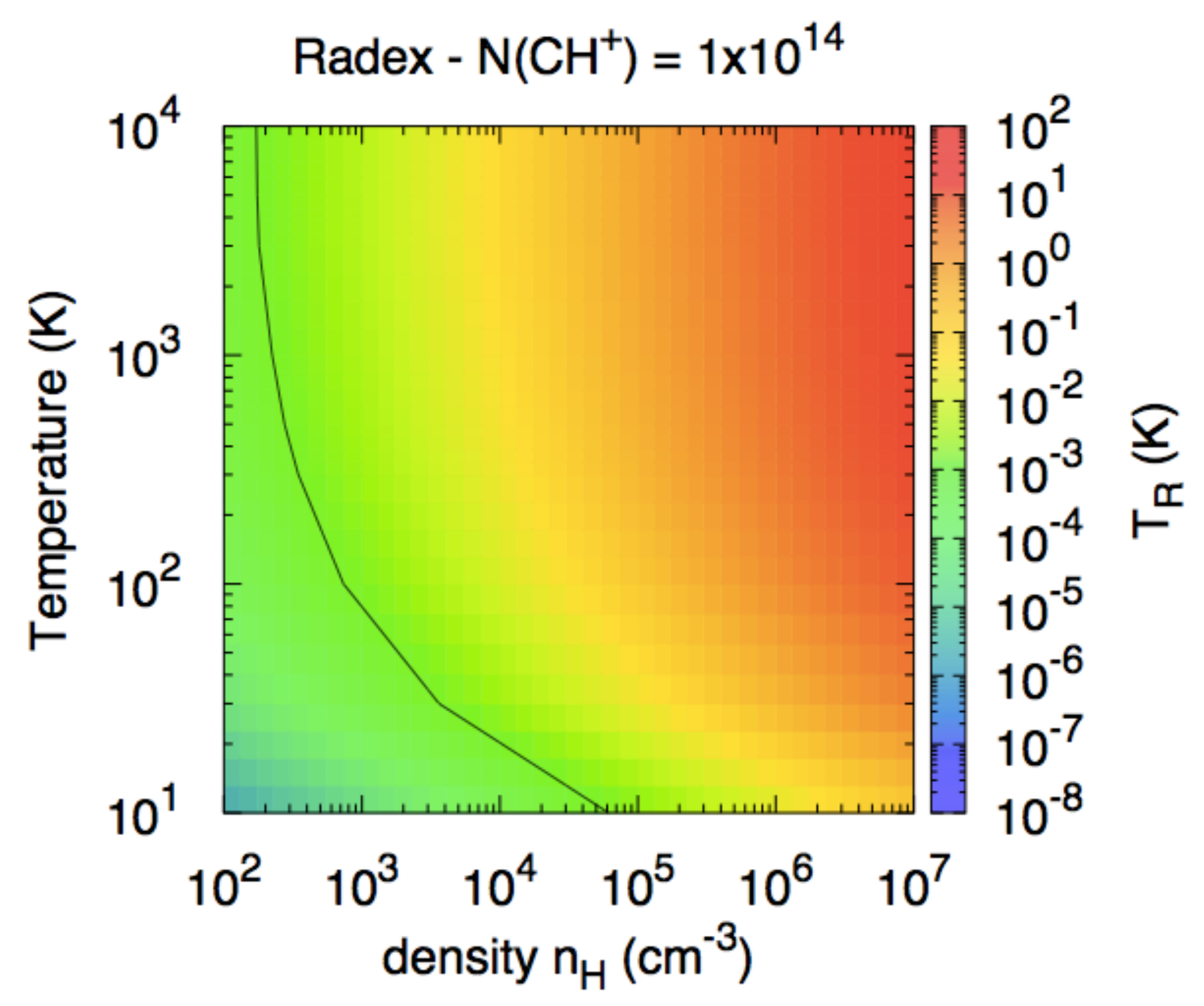}
    \caption{ \CHp(1-0) line temperatures, $T_{\rm R}$ from non-LTE radiative transfer using RADEX \citep{van-der-Tak2007} for two \CHp\ column densities 
    $N(\CHp)= 10^{13}$ (left) and $N(\CHp)=  10^{14}$ \cq\ (right) and broad ranges of local hydrogen density $n_{\rm H}$ and gas temperature $T$. The black lines at  $T_{\rm R} =1$\,mK 
    separate the parameter domains where the line is seen in emission  from that where it is seen in absorption against a continuum of temperature $T_{\rm cont}=2$K, a value 
    relevant for the bright sources of Paper I. For the fainter SMG SMM\,J02399$-$0136, this boundary is at  $T_{\rm R} \approx 0.1$\,mK. 
 }
\label{fig:radex}
\end{figure*}

The average \CHp\ peak emission line brightness of the thirteen kpc-scale shocks identified in the field is 0.07 mJy/beam inferred from their average flux density, 0.25 mJy,  and 
solid angle 0.75 arcsec$^2$, or $\sim$ 3.6 beams (Table \ref{tab:11spots}.   
We express this line brightness in mean brightness temperature, following the basic expressions of the line luminosity in the distant universe \citep{Solomon1997}.
At the redshift of the source $z=2.8041$, we find that 
 \begin{equation}
1 {\rm mJy/beam} = 1.4 \times 10^7 \frac{\lambda^2[{\rm m}] }{b_1 ["]\times b_2["]}\, {\rm mK}
\end{equation}
where $\lambda$ is the observed wavelength and $b_1 ["]\times b_2["]$ the synthesised beam.
The mean \CHp(1-0) line brightness of the extended sources detected by ALMA is therefore $T_{\rm R}\sim 10$ mK.

The results of non-LTE radiative transfer computed with RADEX \citep{van-der-Tak2007} are displayed in Fig.\ref{fig:radex} for two \CHp\ column densities and a velocity dispersion 
$\sigma_v =20$\kms\ representative of the intermediate velocity molecular shock models we consider \citep{Godard2019,Lehmann2020a}.
Densities larger than $n_{\rm H}=10^4$ \cc\ are required to produce the line intensities, even at high temperatures. 
The black curves correspond to the line temperature at which $T_{\rm ex} =T_{\rm cont}$, where  $T_{\rm ex}$ is the line excitation temperature and $T_{\rm cont}$ is the temperature of the continuum background source. These lines  separate the parameter domain where the line is seen in emission from that where it is seen in absorption  against 
the continuum background. The lines drawn here, $T_{\rm R}= 1$mK,  are computed for $T_{\rm cont}=2$K. For the smaller continuum fluxes of L1 and L2SW, the curves separating absorption from emission are 
at $\sim 0.1$mK, in the middle of the green area. 

\section{Further comparison of the \CHp\ and high-velocity \Lya\ emissions}\label{E}

 \begin{figure*}
	\includegraphics[width=0.8\textwidth]{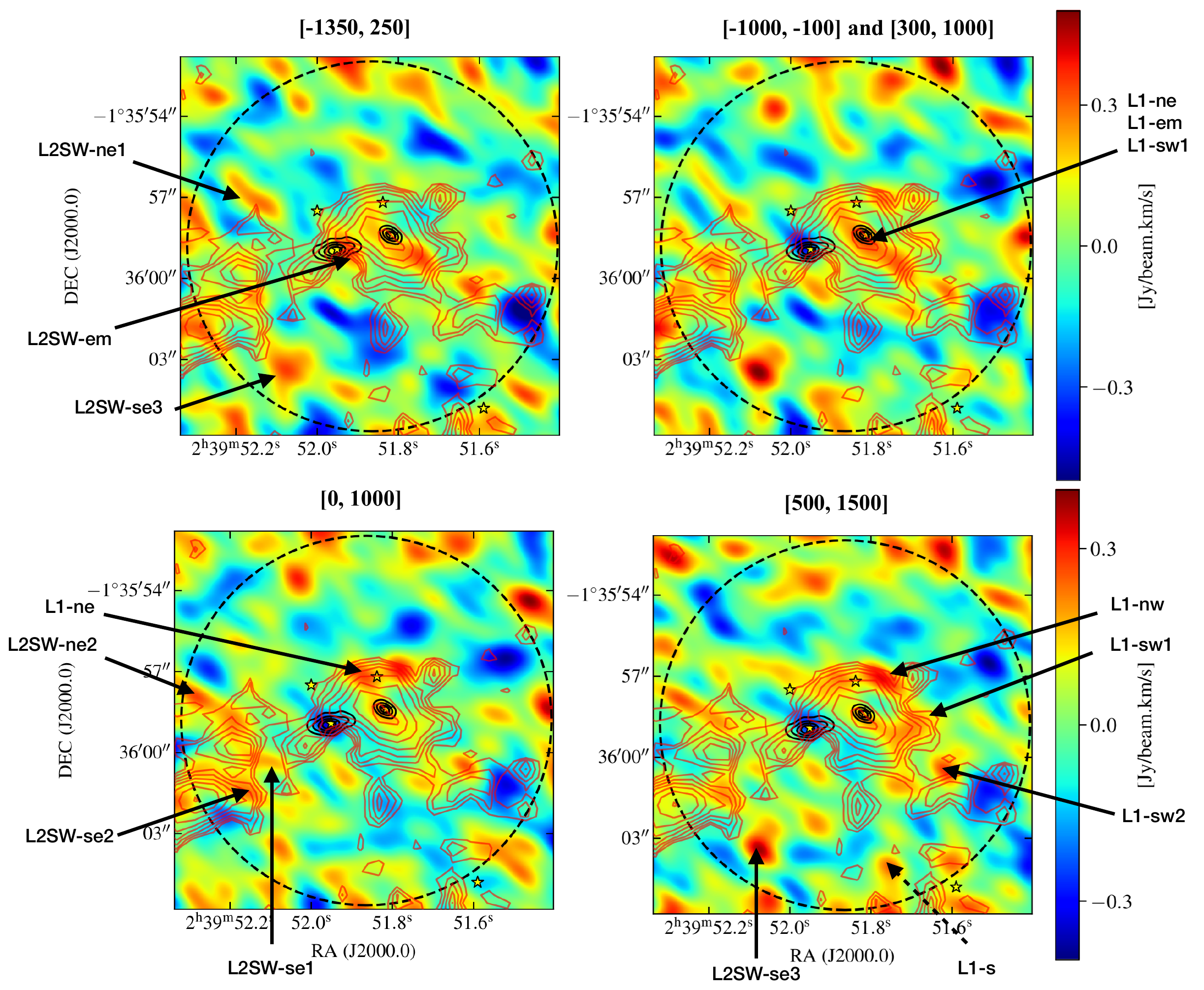}
	\includegraphics[width=0.8\textwidth]{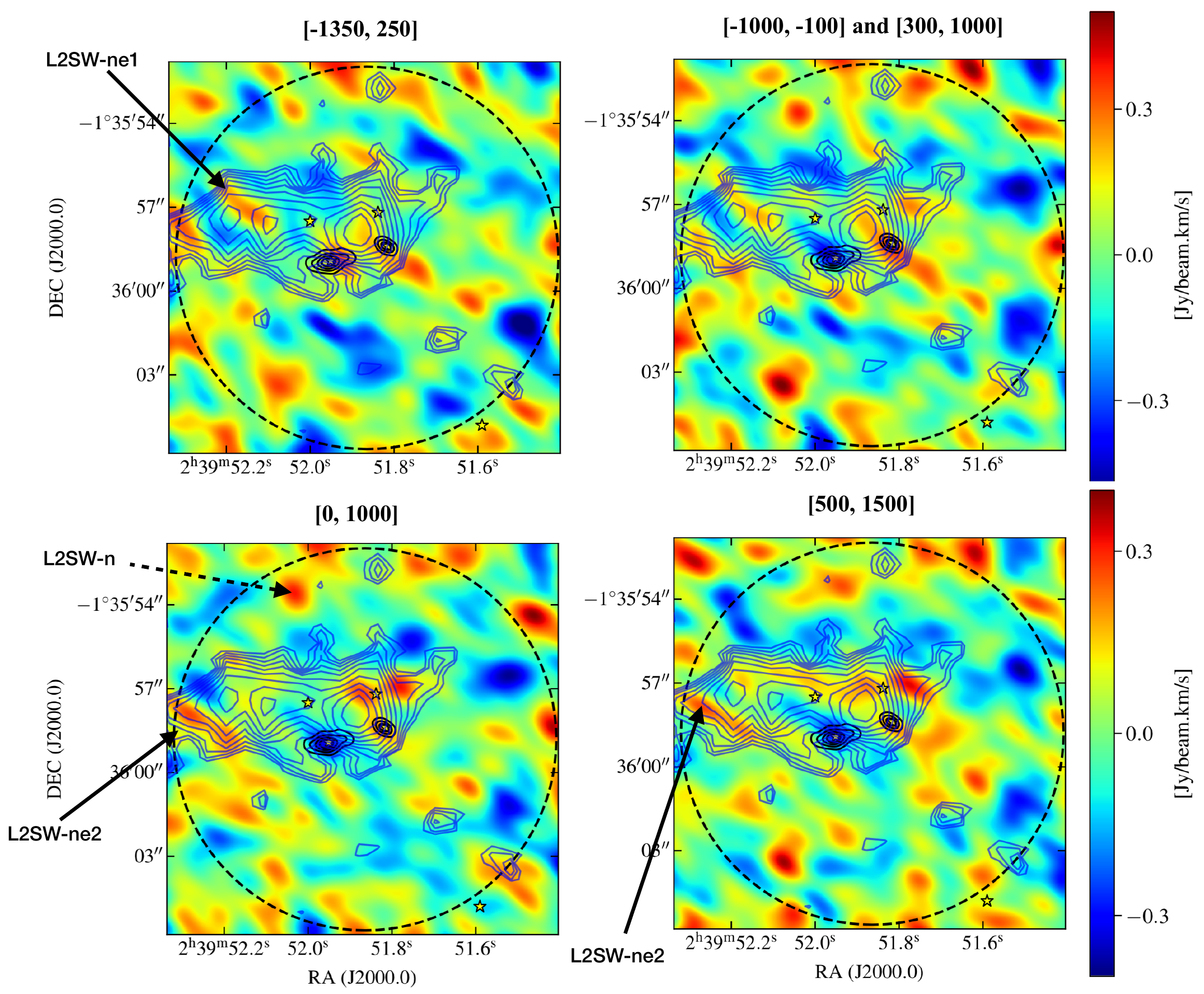}
    \caption{ Smoothed moment-0 maps computed over the velocity range indicated at the top of each panel, superposed to the \Lya\  brightness integrated over [1000,1500] \kms, (red contours) {\it (top panels)} and over [-1200, -700] (blue contours) {\it (bottom panels)}. }
   \label{fig:SmoothMom0}
\end{figure*}

Smoothed versions of selected moment-0 maps are compared with the high-velocity \Lya\ contours in Fig.~\ref{fig:SmoothMom0}. The smoothed moment-0 maps  are not computed as a spatial convolution of the moment-0 maps, with a beam twice larger than the original resolution, but on the original continuum-subtracted spectra. This is why the maps of  Fig.~\ref{fig:SmoothMom0} do not look like a spatially-smoothed version of those of Figs.~\ref{fig:3specGal} to \ref{fig:7specFar}. 
These figures show that the \CHp\ emission structures identified in the native resolution moment-0 maps are indeed connected by large and weaker structures that exquisitely follow the contours of the high-velocity \Lya\ maps. We stress the following features: \\
(1) the elongated NE-SW feature across L1 in the [-1000,-100] and [300,1000] \kms\ moment-0 maps comprises L1-ne, L1-em, L1-sw1 and L1-sw2 and is parallel to the broad orientation of the [1000,1500] \kms\ \Lya\ contours,\\
(2) L1-nw, L1-ne and L1-sw1 are parts of a ring in the [500,1500] \kms\ moment-0 map that closely follows the edge of the red HV \Lya\ contours,  \\
(3) same  kind of  coincidence for L2SW-se2 in the [0,1000]  \kms\ moment-0 map and L2SW-em in the [-1350,250] \kms\ moment-0 map with  the red HV \Lya\ contours,\\
(4) L2SW-se3, L2SW-se2, L2SW-ne1  located  at the tip of elongated structures in  the red HV \Lya\ map,\\
(5) a weak and elongated lane in the [500,1500] moment-0 map follows the northern edge of the blue HV \Lya\ contours as does L2SW-ne2 at their South-Eastern edge.

There are a few additional structures that have not been selected, such as that 2" south of L2SW in the [-1000,-100] and [300,1000] \kms\ moment-0 map, but as said in Sect. 3.1, the goal of this selection is not to make a complete census of all the  \CHp\ emission spots detected by ALMA but to shed light on a specific process, namely the molecular shocks at the interface of the CGM and outflows, seen as positive and negative high-velocity \Lya\  emission. 


\label{lastpage}

\bsp	

\end{document}